\documentclass[12pt]{article}

\usepackage{graphicx, fullpage}
\usepackage{longtable}
\usepackage[hypcap=false]{caption}
\usepackage{setspace}
\usepackage{multirow}
\usepackage{amsmath,amssymb,amsfonts}
\usepackage{amsthm}
\usepackage{mathrsfs}
\usepackage[title]{appendix}
\usepackage{xcolor}
\usepackage{textcomp}
\usepackage{booktabs}
\usepackage{algorithm}
\usepackage{algorithmicx}
\usepackage{algpseudocode}
\usepackage{listings}
\usepackage{comment}
\usepackage{natbib}
\usepackage{url}
\begin{document}

\title{FDR controlling procedures with dimension reduction and their application to GWAS with Linkage Disequilibrium score}







\author{Dayeon Jung  \thanks{Department of Statistics, Seoul National University, Gwanak-ro, Gwanak-gu, Seoul 08826, Korea}, Yewon Kim \thanks{{Cancer and Data Science Laboratory, National Cancer Institute, Bethesda, MD 20892, U.S.A.}} and Junyong Park \thanks{Department of Statistics, Seoul National University, Gwanak-ro, Gwanak-gu, Seoul 08826, Korea \texttt{junyongpark@snu.ac.kr}}}
\date{}
\maketitle

\begin{abstract}
Genome-wide association studies (GWAS) have led to the discovery of numerous single nucleotide polymorphisms (SNPs) associated with various phenotypes and complex diseases. 
However, the identified genetic variants do not fully explain the heritability of complex traits, known as the missing heritability problem. 
To address this challenge and accurately control false positives while maximizing true associations, we propose two approaches involving linkage disequilibrium (LD) scores as covariates.
We apply principal component analysis (PCA), one of the dimensionality reduction techniques, to control the False Discovery Rate (FDR) in the presence of high-dimensional covariates. 
This method not only provides a convenient interpretation of how multiple covariates in high dimensions affect the control of FDR but also offers higher statistical power compared to cases where covariates are not used. 
Using real-world datasets, including GWAS with Body Mass Index (BMI) as the phenotype, we evaluate the performance of our proposed approaches. 
Our methods alleviate computational burden and enhance interpretability while retaining essential information from LD scores. 
In general, our study contributes to the advancement of statistical methods in GWAS and provides practical guidance for researchers looking to improve the precision of genetic association analyses.

\end{abstract}

\section{Introduction}\label{sec1}

Genome-wide association studies (GWAS), initiated in 2005, have successfully identified numerous single nucleotide polymorphisms (SNPs) associated with a wide range of phenotypes and complex diseases (\cite{GWAS}).
Here, a SNP is a genomic variant at a single base position in DNA. 
SNPs have been considered biomarkers in genetics, depending on their location in the genome, for example, in coding regions, non-coding regions, or gene regulatory regions (\cite{SNP}).
Despite the considerable achievements of GWAS, the genetic variants discovered through these studies are insufficient to fully elucidate the heritability of complex traits, commonly called the missing heritability problem (\cite{SNP2}).
Consequently, it is crucial to accurately control for false positives while maximizing the discovery of true associations.

{
\cite{BH} introduced a widely adopted approach for multiple testing by controlling the false discovery rate (FDR). Since then, several variations of FDR-controlling procedures have been developed that incorporate covariates.
For example, \cite{Boca-Leek} proposed a method for using multiple covariates to estimate the null proportion, while \cite{IHW} introduced the independent hypothesis weighting (IHW) method, which allows for a single covariate to influence $p$-value adjustment. 
Depending on the data structure, one method can achieve more power than the other, as we demonstrate later in our simulations and real data examples.
}

In genomic studies, incorporating covariates that capture the underlying genomic structure can enhance the power and accuracy of GWAS analyses (\cite{GWAS2}).
{One such covariate is the linkage disequilibrium (LD) score, defined as the sum of pairwise correlations between a SNP and neighboring variants. LD scores reflect essential genetic dynamics such as recombination and demographic factors like population history and inbreeding (\cite{LD}). They are also used to account for population stratification and confounding effects in association testing (\cite{LDSC1, LDSC2}).
However, when incorporating LD scores as covariates, researchers face the challenge of high dimensionality and multicollinearity, which can complicate the stability and computational aspects of FDR control. 
To address these challenges, we propose two approaches for selecting informative LD scores.
The first approach involves assessing each LD score separately and identifying those with significant relationships to the phenotype of interest.
The second approach utilizes principal component analysis (PCA) to reduce high-dimensional LD scores into a smaller set of uncorrelated principal components that summarize the main sources of variation.
Such dimension reduction allows us to manage computational burden and avoid overfitting while retaining essential information embedded in LD scores.
}

{
While our primary focus is on human genetic studies, GWAS approaches have also been widely and successfully applied in plant and crop genetics to identify marker-trait associations and enhance selection strategies.  
For example, \cite{Shariatipur2021} conducted a comparative genomic analysis of quantitative trait loci (QTL) in wheat for micronutrient contents, grain quality, and agronomic traits. 
In rapeseed,  \cite{Salami2023} and \cite{Salami2024} integrated GWAS with other high-throughput datasets to pinpoint trait-associated regions and specific SNP markers under drought stress. 
In addition, \cite{Ravi2021}  provided a practical example in sugar beet, illustrating how SNP-based marker-trait associations were filtered and validated for Rhizoctonia solani resistance. 
Together, these plant-focused investigations underscore the broad utility of QTL and GWAS methodologies across diverse species, reinforcing the general applicability and adaptability of the approaches discussed in this study.
}

In this paper, we present the details of our proposed approaches and evaluate their performance using real-world datasets such as GWAS with a phenotype of body mass index (BMI) obtained from the genetic investigation of anthropometric traits (GIANT) consortium.
Traditional methods involve applying multiple testing corrections to $p$-values based on statistical measures evaluating the association between BMI and each SNP. 
We demonstrate the benefits of incorporating LD scores as covariates in FDR-controlled GWAS analyses and showcase the efficiency and effectiveness of our methods for selecting informative LD scores.
In particular  we identify how the covariates are used in our real data example in two methods in \cite{IHW} and \cite{Boca-Leek} through comparing the real data analysis and simulation studies.
The results of our study contribute to the advancement of statistical methods in GWAS and provide practical guidance for researchers aiming to enhance the accuracy of their genetic association analyses.

This paper is organized as follows:
In Section \ref{sec2}, we provide a brief review of the false discovery rate and the methods controlling FDR, especially when there are covariates. 
Section \ref{factor} discusses the effect of LD scores in multiple testing of SNPs  with controlling FDR.   
Section \ref{method} includes our proposed procedures incorporating high dimensional covariates using principal component analysis. 
We provide intensive simulation studies in Section \ref{sec4} showing the characteristics of two methods in the presence of covariates. 
In section \ref{sec5}, we analyze the real data set and compare the results with those from simulation studies. 
Concluding remarks are presented in Section \ref{sec6}.

\section{Review of FDR Controlling Methods with Covariates}\label{sec2}
In this section, we introduce methods for controlling FDR. 
There are several FDR controlling procedures available for GWAS data analysis, and in this paper, we will focus on comparing and analyzing the performance of two methods that control the FDR in the presence of covariates, such as the Independent Hypotheses Weighting (IHW) procedure, proposed by \cite{IHW}, and the Boca-Leek method, proposed by \cite{Boca-Leek}. 
These FDR controlling methods that incorporate informative covariates are known to be more powerful than classical approaches and do not underperform classic approaches.
Although these two methods reflect the covariates in multiple testing, they have different ways of using them in FDR controlling procedures.
In particular, the method in \cite{Boca-Leek} uses the covariates in estimating the null proportion, so it may not be effective when the null proportion is irrelevant to those covariates. 
In practice, it is not clear how the covariates affect the multiple testing procedures. 
We provide numerical studies of simulations and real-world data examples of GWAS data in Sections \ref{sec4} and \ref{sec5} to give some insight on the structure of the data.
We will compare these methods with traditional FDR controlling methods such as the Benjamini-Hochberg (BH) procedure in \cite{BH} and the Storey $q$-value method by \cite{qvalue}, which do not consider the covariates as well as the family-wise error rate (FWER). 
 
We briefly review these methods as well as the definition of the FDR as follows. 
Assume we are interested in testing $m$ hypotheses: 
$$H_{0i} \; \text{vs}  \; H_{1i} \text{,} \; i=1,2,\ldots,m$$
If a statistical test results in a significant result, we reject the null hypothesis.  
Let $P$ be the number of rejected hypotheses, which is also called ``discoveries" and 
$FP$ be the number of falsely rejected null hypotheses.

There are two typical types of Type I error rate: $FWER$ and $FDR$
which are defined as 
\begin{eqnarray}
 FWER = P( FP \geq 1 ), \\
 FDR = E\left[\frac{FP}{P \vee 1} \right].   
\end{eqnarray}
However, the Bonferroni method is an example of methods controlling $FWER$, when testing a large number of hypotheses or features, one can have difficulty in identifying potentially important discoveries. 
On the other hand, FDR-controlling procedures are more relevant in testing a large number of hypotheses that may occur in genomic research.  
Controlling FDR is a less stringent criterion than the FWER since FDR allows more significant results to be identified while still controlling the rate of false positives. 
This has led to increased power to detect true positive findings.

\subsection{Independent Hypothesis Weighting} 
The traditional FDR control methods, such as the BH procedure and the Storey $q$-value method, are widely used and effective in many cases.
However, they have some limitations and potential drawbacks, particularly in cases where the tested hypotheses are correlated or the null distribution is not well-understood. 
Then these traditional methods have suboptimal power.

The IHW method in \cite{IHW} is a data-driven approach for controlling the FDR in multiple hypothesis testing. 
The basic idea behind IHW is to assign weights to individual hypotheses from the data as in \cite{wpvalue}, and then use these weights to adjust the $p$-values of the individual tests. 
The IHW approach requires a two-column table of $p$-values and covariates as input. 
A covariate is a variable, either continuous or categorical, that is believed to provide relevant information about the statistical characteristics of a hypothesis test. 
When the null hypothesis is assumed to be true, it must be independent of the $p$-value.

The algorithm of the naive IHW method can be explained as follows:
\begin{enumerate}
    \item Divide hypothesis tests into $G$ different groups based on a relevant covariate
    \item A weight $w_g$ is assigned to each group $g$ ($g = 1, 2, \ldots, G$)
    \item Apply the weighted BH procedure (replace $p_i$ with $p_i/w_i$) at level $\alpha$ and calculate the total number of discoveries
    \item Find the weight vector $\textbf{w} = (w_1, \ldots, w_G)$ that maximizes the number of discoveries
\end{enumerate}

However, developing an optimization approach to determine weights that maximize the number of rejections in the above Step 4 is complex, 
and applying it to millions of hypotheses can pose computational difficulties. 
Furthermore, when dealing with cases where all hypotheses are true or when the statistical power to detect false hypotheses is insufficient, the lack of type \uppercase\expandafter{\romannumeral1} error control is caused by this algorithm. 
To address these drawbacks, the full IHW method is developed as a modification of the naive IHW method, with three significant extensions: convex relaxation using Grenander estimators, data splitting into $k$ folds, and regularization to have a restricted set of weights.

\subsection{Boca-Leek}
The Boca-Leek method in \cite{Boca-Leek} proposes a novel way for estimating and controlling FDR when multiple hypotheses are being tested and covariate information is available.
The use of a regression model to estimate the proportion of null hypotheses $\pi_0$ conditional on observed variables is the main novelty of the Boca-Leek approach. 
Then the adjusted $p$-values by the Benjamini-Hochberg method are multiplied by the estimated proportion to create a plug-in FDR estimator.
The main presumption for progressing is that the $p$-values are independent of the covariates given the null and alternative hypotheses.

Let us consider $m$ null hypotheses: $H_{01}, H_{02}, \ldots , H_{0m}$ and let $\theta_i$ be a binary parameter denoting whether $H_{0i}$ is true, defined as $\theta_i = \mathbb{I}(H_{0i} \; \text{true})$.
Then we have to estimate the null proportion $\pi_0(\textbf{x}_i) = Pr(\theta_i = 1 | \textbf{X}_i = \textbf{x}_i)$ 
and $\text{FDR}(\textbf{x}_i) = \mathbb{E}[\frac{FP}{P}| \textbf{X}_i = \textbf{x}_i]  = \mathbb{E}[\frac{FP}{P}| P>0, \textbf{X}_i = \textbf{x}_i] \cdot Pr(P>0 | \textbf{X}_i = \textbf{x}_i)$. 
Given a $p$-value $p_i$ for each hypothesis $i$ and the threshold of $p$-values $\lambda \in (0,1)$, 
we produce binary indicators $Y_i = \mathbb{I}(p_i > \lambda)$ and these are then used in a regression model to estimate $\pi_0(\textbf{x}_i)$ and $\text{FDR}(\textbf{x}_i)$. 
Since the $p$-values from the null follow a Uniform(0,1) distribution and the $p$-values from the alternative follow a distribution $G(\lambda) = Pr(p_i \le \lambda | \theta_i = 0)$, 
we can derive that $\mathbb{E}[Y_i|\textbf{X}_i = \textbf{x}_i] = (1-\lambda)\pi_0(\textbf{x}_i) + \{1-G(\lambda)\}\{1-\pi_0(\textbf{x}_i)\}.$ 
And we can estimate $\pi_0(\textbf{x}_i)$ using a regression framework.

The entire algorithm of the Boca-Leek method can be expressed mathematically as follows:
\begin{enumerate}
    \item Obtaion the $p$-values $p_1, p_2, \ldots, p_m$
    \item For a given threshhold $\lambda$, obtain $Y_i = \mathbb{I}(p_i \ge \lambda)$ for $1 \le i \le m$
    \item Estimate $\mathbb{E}[Y_i|\textbf{X}_i = \textbf{x}_i]$ via logistic regression by:
    $$\hat{\pi}_0^{\lambda}(\textbf{x}_i) = \frac{\hat{\mathbb{E}}[Y_i|\textbf{X}_i = \textbf{x}_i]}{1-\lambda}$$
    \item Smooth $\hat{\pi}_0^{\lambda}(\textbf{x}_i)$ over $\lambda \in (0, 1)$ to obtain $\hat{\pi}_0(\textbf{x}_i)$, at the largest threshold
    \item Multiply the BH adjusted $p$-values by $\hat{\pi}_0(\textbf{x}_i)$ to obtain $\widehat{FDR}(\textbf{x}_i)$
\end{enumerate}

\section{Factors Affecting Marker-Trait Association Detection and FDR Control}
\label{factor}

One of the most critical aspects in controlling the FDR and identifying reliable marker-trait associations in GWAS is understanding how various genomic, statistical, and biological factors interact to influence detection power and false discovery control. 
In GWAS, the number of SNPs included in the analysis is a key factor affecting FDR. A denser SNP array increases the multiple testing burden and, when combined with strong linkage disequilibrium, may lead to high correlation across tests.
Although our method does not use SNPs themselves as covariates, the LD scores serve as informative covariates that capture these dependencies. 
The covariate information used in our method is assigned per SNP but reflects local genomic structure rather than the SNP's own statistical significance, which may clarify potential confusion regarding our modeling strategy. 
The number of SNPs tested defines the scale of the multiple testing problem, while the LD scores help refine the weighting or prioritization of these tests. Statistically, sample size is another critical determinant, as small sample sizes reduce the stability of p-value estimation and limit the power to detect modest effect sizes. 
Additionally, biological factors further complicate FDR control. 
Traits with low heritability tend to exhibit weaker genetic signals, making it harder to detect associations. 
Measurement errors in phenotypes introduce noise that can obscure true effects, and pleiotropy—where one SNP affects multiple traits—can introduce hidden dependencies across tests. 
By integrating LD structure through dimension-reduced covariates, our method offers resilience against such confounding factors.
\section{Proposed Methodology}
\label{method}
We introduce an approach to controlling the FDR within the context of multiple covariates through dimensionality reduction.
Principal component analysis (PCA), a widely adopted technique in statistics, machine learning, and signal processing, serves as a powerful tool for dimensionality reduction and data visualization. 
PCA seeks to identify principal components, orthogonal directions along which data exhibits maximum variability. 
These components are sequentially orthogonal, with the first capturing the most variance.

Through the application of PCA to the covariate matrix, we illustrate the efficacy of dimensionality reduction in improving FDR control. 
PCA becomes instrumental in capturing essential information within high-dimensional covariate data.

One major significance of PCA in our proposed methodology lies in its ability to unveil the underlying structure of the covariate matrix by identifying principal components—orthogonal directions capturing maximum data variance. 
By strategically selecting a subset of these components, we achieve a substantial reduction in dimensionality while retaining a significant portion of data variability. 
This reduction not only facilitates improved interpretability but also proves vital in enhancing FDR control. 
Furthermore, PCA's versatility extends beyond dimensionality reduction, providing additional benefits such as data compression, noise reduction, and feature extraction. 
Its application in exploratory data analysis offers valuable insights into inherent patterns and structures within complex datasets. 
In the realm of FDR control, our proposed approach opens new avenues for researchers seeking robust methods in multiple testing scenarios. 
By harnessing the power of PCA, we not only address the challenges associated with high-dimensional covariates but also pave the way for more effective FDR control strategies.

Suppose we have observations $ ( {\bf x}_i,  p_i)$ for $1\leq i \leq m$ where ${\bf x}_i$ is the $d$-dimensional column vector.
Let us define ${\bf X} = ({\bf x}_1,\ldots, {\bf x}_m)^T$.  
We then apply the PCA to ${\bf X}$ and obtain PC scores corresponding to $p$ principal components, 
denoted by  $ {\bf PC} = (pc_{ij})_{1\leq i \leq m, 1\leq j \leq d} = ( {\bf pc}_1, \ldots, {\bf pc}_d)$ 
where ${\bf pc}_j = (pc_{1j},pc_{2j},\ldots, pc_{mj})^T$ for $1\leq j \leq d$
is the principal component scores corresponding to $j$th principal component.  
We select the best principal component which obtains the maximum number of rejections among all principal components. 
The following algorithm presents this procedure. 
\begin{algorithm}
    \caption{FDR controlling method with PCA}\label{alg:FDR_PCA}
    \begin{algorithmic}
    \State [1] Apply the PCA to $ {\bf X}$ and obtain pc score matrix, ${\bf PC}$. 
    \State [2] For $j$th principle component ($1\leq j \leq d$), apply the method to $(y_i, {pc}_{ij})$ for $1\leq i \leq m$ 
    and then obtain the rejections. 
    \State [3] Repeat the Step [2] for all principle components $1\leq j \leq d$. 
    \State [4] Select the principle component with the maximum rejections in the Step [3]. 
    \end{algorithmic}
\end{algorithm}

We applied the IHW and Boca-Leek methods to this algorithm \ref{alg:FDR_PCA}, respectively.
Note that unlike the IHW, the Boca-Leek method can utilize more than a single covariate.
However, we do not explore the optimal combination of multiple covariates in our paper. 
Rather than this issue, we focus on the structure that each of the methods can detect more efficiently. 
We demonstrate these properties in the following sections on simulations and real data analysis, and then we present some insight on the structure of our GWAS data by comparing simulation studies and the results of the analysis of GWAS data. 


\section{Simulation}\label{sec4}
 {
In this section, we outline the design of our simulation studies, specifically crafted to emulate real GWAS datasets and their characteristics. We aim to provide a robust framework to evaluate the performance of FDR controlling procedures in a context similar to actual GWAS data.}  
We define basic notations as follows. We test $m$ hypotheses where the $i$th test consists of the null and the alternative hypotheses, denoted by $H_i=0$ and $H_i=1$, respectively.  
We set an indicator function $\delta_i = \mathbb{I}(H_{i} \mbox{ is rejected})$ 
to denote whether hypothesis $H_i$ has been rejected ($i = 1, \ldots, m$). Then we can define
\begin{eqnarray*}
    P = FP+TP   &=& \sum_{i=1}^m \delta_i, \\
    TP          &=& \sum_{i=1}^m \delta_i H_i, \\ 
    FP          &=& \sum_{i=1}^m \delta_i (1-H_i), \\
    m_1 = FN+TP &=& \sum_{i=1}^m H_i.   
\end{eqnarray*}
We calculated the averages of the true positive rate (TPR) and the FDR, given by
\begin{align*}
    \text{TPR} = \mathbb{E} \left[\frac{TP}{m_1 \vee 1} \right],  \quad 
    \text{FDR} = \mathbb{E} \left[\frac{FP}{P \vee 1} \right].
\end{align*}
Throughout our simulation studies, we set $d=30$ and $m=2 \times 10^4$, meaning the number of covariates is $30$ and the number of hypotheses is $2 \times 10^4$.

\subsection{Scenario 1: Null Distribution Proportion Estimation}
In the first simulation, we employed the setting in \cite{Boca-Leek} to estimate the null proportion when there are covariates associated with the $p$-values.
To evaluate the efficacy of these methods in concurrently maintaining high statistical power and controlling a given level of FDR, we conduct 100 simulation runs emulating real GWAS datasets. 
The actual FDR was computed as the ratio of the number of rejected null hypotheses to the total number of rejected hypotheses. Furthermore, 
we determined the number of rejected hypotheses (discoveries) for each method and compared these results to the FDR values, 
assessing the extent of their statistical power in comparison. 
The simulation settings are outlined below:

\subsubsection{Generation of Covatiate Matrix X}

 {To mirror the characteristics of real GWAS covariates, we generate a covariate matrix  $\mathbf{X}$ that is sparse and high-dimensional. The generation process begins with creating a 30-dimensional variable $\mathbf{z} = (z_1, z_2, \ldots, z_{30})^T$ from a multivariate normal distribution:}

\begin{eqnarray}
    \mathbf{z} \sim N_{30} ( \mathbf{0}, AA^T).
    \label{eqn:z}
\end{eqnarray}
where $\mathbf{0}=(0,0,\ldots, 0)^T$ and $A$ is a random 30$\times$30 matrix. 
Following this, we generate multivariate binary data $\mathbf{y} = (y_1, y_2, \ldots, y_{30})^T$ using the algorithm from \cite{MVB}:
\begin{eqnarray}
    \mathbf{y} \sim \rm{multivariate~Binary}(\mathbf{p}, \delta)
    \label{eqn:y}
\end{eqnarray}
Here, $\mathbf{p}$ is the probability vector of $\mathbf{y}=\mathbf{1}$, 
i.e., $\mathbb{E}[\mathbf{y} |\mathbf{p}] = \mathbf{p}$, with each component set to 0.1.
 {We specifically set $\mathbf{p}$ to 0.1 to mimic the sparsity observed in real GWAS data, where the LD score matrix contains many zeros, and the average proportion of non-zero entries per column is approximately 0.1.}
The parameter $\delta$ represents the correlation between different components of $\mathbf{y}$:

\begin{eqnarray}
    corr(\mathbf{y}) := 
    \delta = 
    \begin{bmatrix} 
    1        & 0.2      & 0.2^2    & \ldots & 0.2^{29} \\
    0.2      & 1        & 0.2      & \ldots & 0.2^{28} \\
    \vdots   & \vdots   & \vdots   & \ddots & \vdots   \\
    0.2^{29} & 0.2^{28} & 0.2^{27} & \ldots & 1        \\
    \end{bmatrix}.
    \label{eqn:corr_mat}
\end{eqnarray}
The detailed method of generating $\mathbf{y}$ using the correlation matrix was implemented by employing the function provided at \url{https://github.com/shz9/mvbin}.
Using $\mathbf{z}$ and $\mathbf{y}$, we obtain final covariate $\mathbf{x} = (x_1, x_2, \ldots, x_{30})^T$:
\begin{eqnarray*}
    \mathbf{x} = (x_1,x_2,\ldots, x_{30})^T &=&       |\mathbf{z}| \circ \mathbf{y} \\
                                            &\equiv&  (|z_1|y_1, |z_2|y_2, \ldots, |z_{30}|y_{30})^T,
\end{eqnarray*}
where $\circ$ denotes elementwise multiplication. We generate $m$ such vectors to form the covariate matrix $\mathbf{X}$:
\begin{eqnarray*}
    \mathbf{X}=(\mathbf{x}_1,\mathbf{x}_2,\ldots, \mathbf{x}_m)^T.
\end{eqnarray*}
we obtained a  $m\times d = (2 \times 10^4) \times 30$ covariate matrix $\mathbf{X} = (x_1, x_2, \ldots, x_{30})^T$ 
through $2 \times 10^4$ iterations of generating $\mathbf{x}_i$ for $1\leq i \leq 2 \times 10^4$.

\subsubsection{Modeling $p$-values}
 {
In GWAS, $p$-values are often derived from the association between SNPs and traits, typically through contingency tables. 
This process involves several key steps. First, SNP data is prepared with each SNP characterized by two alleles (A1 and A2). Trait data, such as BMI, is collected for study participants. A $2 \times 2$ contingency table is constructed for each SNP, representing the relationship between alleles and the trait, including frequency counts of individuals with different combinations of alleles and trait values. Statistical tests, such as the chi-squared test, are performed on the contingency table to evaluate the association between the SNP and the trait, yielding a chi-squared statistic ($\chi^2$) that measures the strength of the association. This statistic is then transformed into a z-value using the relationship  $z = \text{sign}(\beta) \times \sqrt{\chi^2}$, where $\beta$ is the regression coefficient representing the effect size of the SNP on the trait. The $z$-values are subsequently used to compute $p$-values, indicating the statistical significance of the association. Smaller $p$-values suggest a stronger association between the SNP and the trait. (\cite{SNP-statistic})
In our simulation, we simplify this process by directly modeling $p$-values in a way that maintains the relationship between covariates and $p$-values.
}

We assume that the null probability $\pi_0$ is dependent on a specific principal component of $\mathbf{X}$. We define $x^*$ as a function of the third principle component:

\begin{eqnarray*}
    x^*_i  
    =  {\rm logit} (\mathbf{e}_3 \cdot \mathbf{x}_i) 
    =  \frac{e^{\sum_{j=1}^{30} e_{3j} x_{ij}}}{1+e^{\sum_{j=1}^{30} e_{3j} x_{ij}}},
\end{eqnarray*}

where $\mathbf{e}_3$ is the third eigenvector of $ \mathbf{X}$ corresponding to the 3rd largest eigenvalue and 
the covariate $\mathbf{x}_i$ affects the null probability $\pi_0$ only through $x_i^*$, 
i.e.,  $\pi_0(\mathbf{x}_i) = \pi_0(x_i^*)$.  
 {Figure \ref{fig::pi0_plot} depicting the function $\pi_0(x^*)$ 
is attached in the online supplementary material.}
The null probability $\pi_0$ depends on $x^*$:
$$
\pi_0(x^*) = f_1(x^*) + 1.5 \cdot f_2(x^*) + 0.9 \cdot f_3(x^*),
$$
where $f_1(x)$ is defined as:
$$
f_1(x) = 
\begin{cases} 
- a \cdot (x - a_1) \cdot (x - a_2)     & \text{if } x < 0.7 \\
- a \cdot (0.7 - a_1) \cdot (0.7 - a_2) & \text{if } x \geq 0.7 \\
1                                       & \text{if } x \leq \frac{a_1 + a_2}{2}
\end{cases}
$$
and $f_2(x)$ is defined as:
$$
f_2(x) = 
\begin{cases} 
-2.5 \cdot (x - 0.7)^2  & \text{if } x \geq 0.7 \\
0                       & \text{otherwise}
\end{cases}
$$
Lastly, $f_3(x)$ is defined as:
$$
f_3(x) = 
\begin{cases} 
-(x - 0.1)^2 & \text{if } x < 0.7 \\
-(0.7 - 0.1)^2 & \text{if } x \geq 0.7 \\
0 & \text{if } x \leq 0.1
\end{cases}
$$
Here, $a$ is a constant defined as $a = \frac{4}{(a_1 - a_2)^2}$, $a_1 = 1.2$, $a_2 = -0.2$, and $x$ ranges from 0 to 1.
This function is referenced from the paper \cite{Boca-Leek} available at the GitHub repository \url{https://github.com/SiminaB/Fdr-regression/tree/master}.

\subsubsection{Generating $p$-values}

The $p$-values are generated from a mixture distribution:
\begin{eqnarray}
    f(p|\mathbf{x}) = \pi_0(x^*) I_{(0,1)}(p) + (1-\pi_0(x^*)) Beta(a,b),  
    \label{eqn:p_distribution}
\end{eqnarray}
where $Beta(a,b)$ is a beta distribution with parameters $a \sim U(0.2,0.4)$ and $b = 4$. 
 {Figure \ref{fig::pval_plot} in the online supplementary material shows the density of the null and alternative distribution of $p$-values showing that this mixture distribution captures the behavior of $p$-values under the null and alternative hypotheses, with the alternative distribution typically having smaller $p$-values.}

 {By structuring our simulation in this way, we ensure that the design is closely aligned with the characteristics of real GWAS data. This approach highlights the impact of covariates, particularly LD scores, on the $p$-values, thereby providing a meaningful context for evaluating FDR control methods.}

\subsection{Scenario 1 Results}
In each simulation run($n = 1, 2, \ldots, 100$), 
we applied various FDR control methods to the previously created dataframe and conducted comparative analyses. 
 {For Bonferroni method, BH and Storey $q$-value methods, we identified significant features based solely on $p$-values.}
In the case of the covariate matrix, after Principal Component (PC) transformation, 
we applied each PC component to IHW and Boca-Leek.
After completing $100$ simulation runs, 
we computed the averages of the simulation results to ascertain the final counts of discoveries and FDR.
The simulation results revealed that the average number of actual alternative features was 3819.84 out of 20,000 features. 
 {Using the Bonferroni method, we observed a TPR of 0.0425 and an FDR of 0.0002. The BH method yielded a TPR of 0.2936 with an associated FDR of 0.0413. When applying the Storey q-value method, the TPR increased to 0.3184, with an FDR of 0.0501.}
The outcomes of applying PC-transformed covariate values to each method are illustrated in Figure \ref{fig::simul_pca_result_TPR} and \ref{fig::simul_pca_result_FDR} below. 
The x-axis denotes the type of PC axis applied to each method, 
while the y-axis in (a) indicates the TPR and in (b) represents the FDR. 
IHW results are depicted by the solid green line, and Boca-Leek results are represented by the solid yellow line. 
In (b), the nominal $\alpha$ = 0.05 is denoted by the red line.

\begin{figure}[ht]
    \centering
    \includegraphics[width=0.7\textwidth]{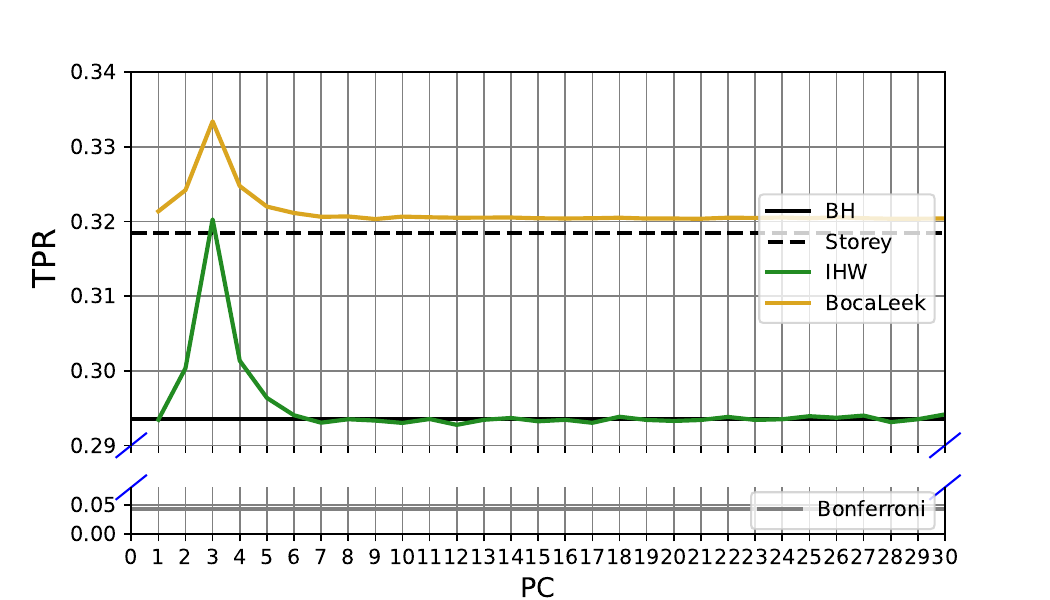}
    \caption{The TPR of simulation result on scenario 1}
    \label{fig::simul_pca_result_TPR}
\end{figure}

\begin{figure}[ht]
    \centering
    \includegraphics[width=0.75\textwidth]{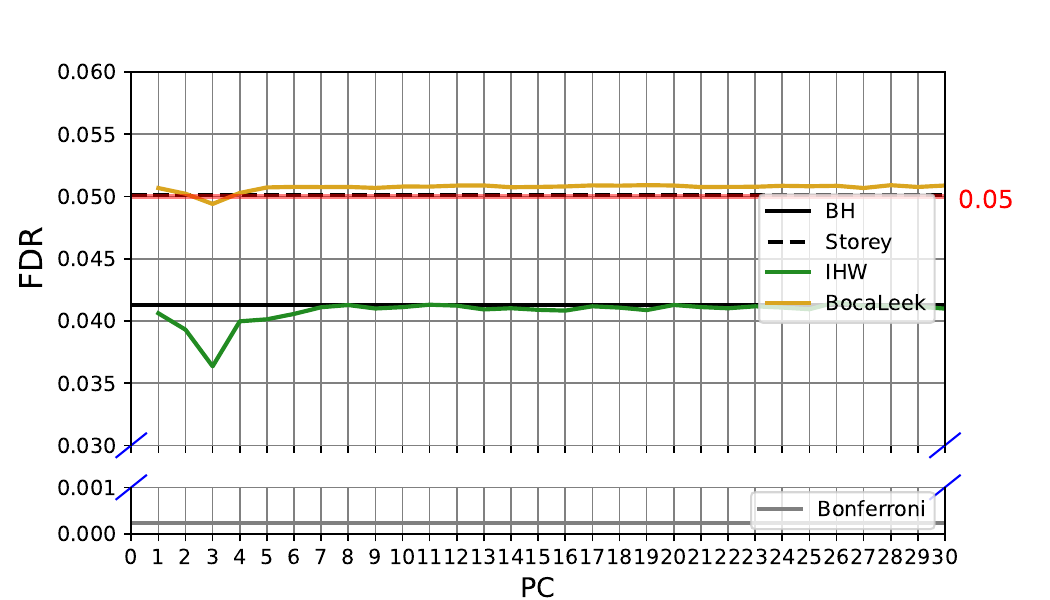}
    \caption{The FDR of simulation result on scenario 1}
    \label{fig::simul_pca_result_FDR}
\end{figure}

In summary, based on the findings presented in Figure \ref{fig::simul_pca_result_TPR}, 
the TPR values in Boca-Leek was conclusively comparable to or slightly higher than those identified using the Storey $q$-value method. 
Particularly noteworthy was the higher tendency in discoveries on PC3, which is associated with the distribution of $p$-values.
Regarding IHW, the overall number of discovered SNPs closely mirrored the results obtained through the BH method, 
with the highest number of SNP rejections concentrated near PC3. 
Notably, IHW exhibited a higher sensitivity to the real associated covariate PC3 compared to Boca-Leek.
Figure \ref{fig::simul_pca_result_FDR} displays the calculated actual FDR values for each scenario, indicating an overall control at the 0.05 threshold. Specifically, IHW demonstrated lower FDR values near PC3. 
While Boca-Leek occasionally surpassed the 0.05 threshold slightly, 
these instances did not involve values significantly exceeding 0.05.

\subsection{Scenario 2: Size Investing}
In the second simulation, we implemented a size investing strategy with the objective of applying the IHW method. 
 {
The goal is to further mimic actual GWAS data in our simulations. We begin by generating the covariate matrix $\mathbf{X}$ using the same approach as in Scenario 1.
}
To define the mean value under the alternative distribution, we used the first principal component ($e_1$) of covariance matrix of $\mathbf{x}$ ($\delta$):
$$
\mu_i = (2\Phi(e_1 \cdot \mathbf{x}_i)+1)H_i, \; where \; e_1 \cdot \mathbf{x}_i =\sum_{j=1}^d e_{1j} x_{ij}.
$$. 
The simulation proceeded with the size-investing methodology:
\begin{align*}
    H_i     &\sim   \text{Bernoulli}(\pi_1), \\
    W_i|H_i &\sim   N(\mu_i, \; 1), \\
    P_i     &=      1 - \Phi(W_i).
\end{align*}
From the definition of $\mu_i$, we have 
$W_i \sim N(0,1)$ under $H_i=0$, leading to the uniform distribution of $p_i$ in $(0,1)$, and $W_i \sim N(\mu_i, 1) $ under $H_i=1$, implying a normal distribution with $\mu_i>0$.
Note that $\pi_1$ does not depend on the covariate $\mathbf{x}$ and $p_i$ depends on the covariate $\mathbf{x}_i$ through the $\mu_i$.
The alternative hypothesis proportion $\pi_1$ was set to $0.1$, and the nominal $\alpha$ was set to $0.05$. A total of 100 Monte Carlo replicates were employed for the number of hypotheses, $m = 20,000$. 

 {
By using this size-investing approach, we ensure that our simulation is robust and reflective of real-world scenarios encountered in GWAS, further demonstrating the applicability and efficacy of the IHW method in controlling the false discovery rate while incorporating covariate information.
}

\subsection{Scenario 2 Results}
Next are the results obtained by applying the generated variables to each FDR control method. 
For BH and Storey $q$-value, only $p$-values were utilized, while for IHW and Boca-Leek, both $p$-values and covariates were applied. 
Through a total of 100 repetitions, as in Scenario 1,
a comparison was made to determine which method among the four exhibited superior power.
 {Upon analyzing the simulation results, it was found that, on average, there were 1,996.43 actual alternative features among the total 20,000 features. For comparison, the most basic method, the Bonferroni method, yielded a TPR of 0.0284 and a FDR of 0.0002. Following this, the BH method produced a TPR of 0.2566 and an FDR of 0.0443. Lastly, the Storey $q$-value method resulted in an average TPR of 0.2668 and an FDR of 0.0490.}

\begin{figure}[ht]
    \centering
    \includegraphics[width=0.7\textwidth]{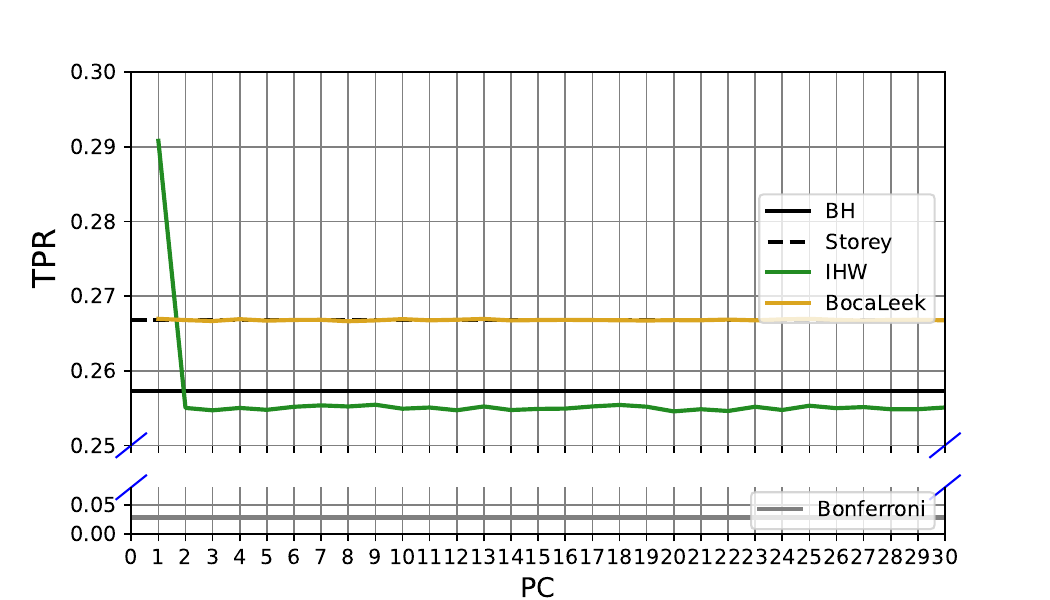}
    \caption{The TPR of simulation result on scenario 2}
    \label{fig::simul_ihw_pca_result_TPR}
\end{figure}

\begin{figure}[ht]
    \centering
    \includegraphics[width=0.75\textwidth]{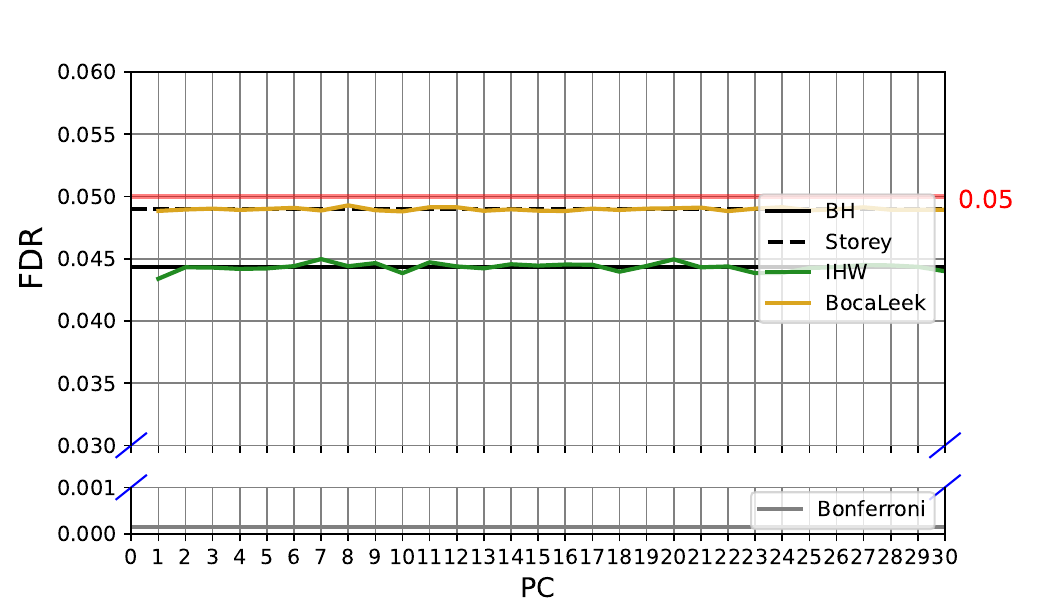}
    \caption{The FDR of simulation result on scenario 2}
    \label{fig::simul_ihw_pca_result_FDR}
\end{figure}

Figure \ref{fig::simul_ihw_pca_result_TPR} and \ref{fig::simul_ihw_pca_result_FDR} illustrates the outcomes of incorporating PC-transformed covariates into each method. 
The x-axis denotes the specific PC axis, while the y-axis in Figure \ref{fig::simul_ihw_pca_result_TPR} signifies the TPR, 
and in Figure \ref{fig::simul_ihw_pca_result_FDR}, it represents the FDR. 
The solid green line represents the results for IHW, while the solid yellow line represents Boca-Leek. 
In Figure \ref{fig::simul_ihw_pca_result_FDR}, the red line indicates the nominal $\alpha$ = 0.05.

In Figure \ref{fig::simul_ihw_pca_result_TPR}, the results deviated slightly from those of Scenario 1. 
In the case of IHW, utilizing $ e_1 \cdot \mathbf{x}_i$ as a covariate, 
which is associated with the distribution of $p$-values, yielded the highest TPR. 
However, for the Boca-Leek method, which determines the rejection region based on the estimation of the null distribution, 
using $ e_1 \cdot \mathbf{x}_i$ as a covariate showed little difference compared to using other unrelated covariates.
This suggests that in the context of size-investing simulations, IHW excels in identifying true features, 
whereas Boca-Leek does not demonstrate significant performance. 
Additionally, in Figure \ref{fig::simul_ihw_pca_result_FDR}, 
it was observed that all four methods effectively controlled the FDR within 0.05.

\section{Analysis of Real Data}\label{sec5}
In our study, we undertook a comprehensive examination of FDR control methods, 
including BH, Storey $q$-value, IHW and Boca-Leek, applied to a real-world dataset. 
In the practical application of these methods to real data, 
our objective was to assess their performance and comparative strengths. 
The utilization of diverse FDR control strategies allowed us to investigate how each method copes with the intricacies and nuances present in real-world datasets, 
where confounding factors can significantly impact the reliability of results.

\subsection{Motivating Data and Dataset Description}
GWAS provide valuable insights into the genetic basis of common diseases, with summary statistics readily available in online databases like the GIANT consortium. \cite{poly} highlights some characteristics of findings derived from GWAS. Notably, diseases such as BMI exhibit a polygenic genetic architecture involving the collective influence of numerous SNPs,  {each with a small individual effect,} rather than a single one. 
These SNPs, often located in non-coding regions with a minor allele frequency exceeding $0.05$, contribute to the complexity of genetic associations.
 {In particular, these non-coding DNA regions do not code for amino acids, however, they can serve functional roles, such as regulating gene expression. The identity of regulatory elements within non-coding areas is not yet fully understood \citep{encode}.}

To assess genetic effects within this polygenic framework, linkage disequilibrium score regression (LDSC) has been employed in \cite{LDSC1}. 
 {LD score was initially focused on population genetics and is now widely used to provide insights into evolutionary history and as a basis for gene mapping \citep{LD}. 
LDSC is known for its capacity to distinguish genetic effects from confounding factors such as population stratification, and for its computational efficiency achieved by excluding individual genotype data.
As a follow-up study, \cite{LDSC2} introduced stratified-LDSC by expanding the analysis to multiple groups of LD scores. These groups represent functional categories in the genome, including non-coding regions, with each element potentially contributing different roles to the disease.
By simultaneously considering functional groups of LD scores, stratified-LDSC leverages genome-wide information to explain diseases based on the functional roles of genomic regions.
However, the accuracy of both LDSC and stratified-LDSC diminishes with fewer available SNPs, and their reliance on certain assumptions may impact reliability. See \cite{LDSC3} for more detail. }

Here, we present a dataset focused on chromosome 3, comprising $p$-values and 30 covariates for 71,994 SNPs. The $p$-values were calculated based on BMI index data obtained from the GWAS 2010 BMI summary statistics dataset introduced in \cite{BMI-SNP}, reflecting the significance of each SNP's association with BMI. The covariates include minor allele frequency (MAF) and LD scores for 29 functional groups of SNPs. MAF, representing the frequency of a minor or recessive allele within a given population, helps to differentiate the common and rare SNPs. The functional groups are structured based on the role that the SNPs play in the comprehensive gene regulation mechanisms employed by cells to modulate the expression of specific gene products, including proteins or messenger RNA. These mechanisms entail both up-regulation and down-regulation of gene products, and eventually affect disease and other phenotypic traits. More information about these groups can be found in \cite{LDSC2}. To account for potential confounding factors and enhance the accuracy of identifying significant associations, we consider two approaches to incorporate covariates into FDR procedures, such as IHW and Boca-Leek. One approach involves using the original covariates directly, while the other incorporates dimension reduction on the 30 covariates through principal component analysis. This enables the identification of crucial principal components as substitutes for the original covariates and enhances interpretability.

\subsection{Result of Original Covariates}
In our analysis, we initially employed the Bonferroni method, renowned for controlling the FWER. 
We discuss the SNPs of chromosome 3 among the total 22 autosomal chromosomes. 
Out of the 71,994 SNPs tested, this method identified 27 SNPs as statistically significant. 
Subsequently, we applied the BH and Storey $q$-value methods for FDR control, 
leading to the identification of 49 significant SNPs for each method, 
with complete agreement between the results obtained from BH and Storey $q$-value methods.
The Bonferroni method, controlling FWER, is the most conservative in obtaining powers among all the methods we consider.

In the implementation of the IHW method, we systematically introduced covariates one by one into the FDR analysis.
Notably, the number of rejected SNPs varied depending on the specific covariate incorporated.
 {The results summarized for chromosome 3 are shown in Figure \ref{fig::SNP_result}, while comprehensive results for all chromosomes from 1 to 22 can be found in the online supplementary material, particularly in Figure \ref{all_result_SNP}.}
Specific covariates, including the LD score of ``Repressed\_Hoffman", ``H3K4me\_Trynka", and ``Intron\_UCSC" emerged as particularly noteworthy in our study. 
The incorporation of these covariates led to a substantially increased number of rejected SNPs compared to conventional FDR control methods like BH and Storey $q$-value. 
These outcomes underscore the importance of incorporating relevant covariates in FDR control methods to enhance the sensitivity of SNP detection.
 {
Specifically, the repressed region contains SNPs related to a subclass of transcription factors that negatively regulate gene transcription. For example, some transcriptional repressor proteins counteract the activity of positively acting transcription factors \citep{repressed}.} On the other hand,
in the realm of epigenetics, ``H3K4me1" signifies monomethylation of lysine 4 on histone H3, protein, and the roles of SNPs in this group are related to active and primed enhancers \citep{epig}. 
 {Additionally, \cite{encode} and \cite{intron} note that many SNPs discovered in GWAS are located in non-coding regions like introns, such as within or overlapping gene boundaries.}
Thus, our findings have a rationale from genetic and epigenetic points of view.

\begin{figure}[ht]
    \centering
    \includegraphics[width=0.99\columnwidth]{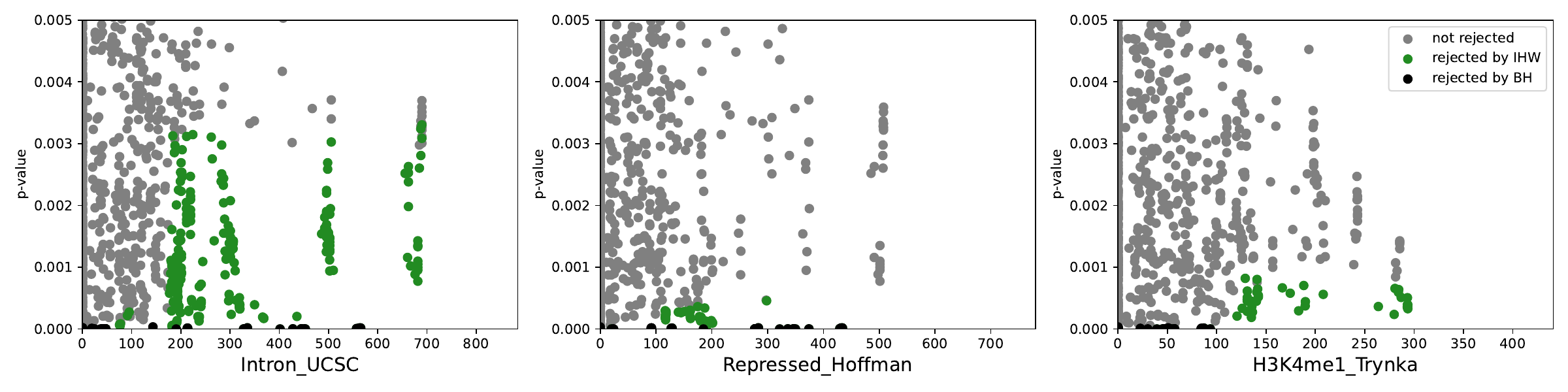}
    \caption{Scatterplot of SNPs with each covariate and $p$-value}
    \label{fig::SNP_scatter}
\end{figure}

The Figure \ref{fig::SNP_scatter} illustrate scatterplots for three covariates, which showed more significant results than the traditional FDR control methods BH and Storey $q$-value. 
The x-axis represents the values of each LD score, 
while the y-axis displays $p$-values for a total of 71,994 SNPs, denoted by grey dots. 
SNPs rejected by the IHW method are highlighted in green, and those rejected by the BH method are marked in black.
In each plot, a $p$-value range of 0 to 0.005 is chosen to offer a detailed view of the scatterplot for SNPs rejected by IHW. 
Specifically, in the left panel of Figure \ref{fig::SNP_scatter}, 230 SNPs were rejected using cov$\_$Intron$\_$UCSC, with 30 of these overlapping with those rejected by the BH method. 
Similarly, the panel in the middle of Figure \ref{fig::SNP_scatter} displays the scatterplot of the 79 SNPs rejected using cov$\_$Repressed$\_$Hoffman, of which 44 SNPs overlapped with those rejected by the BH method. 
Lastly, the right panel of Figure \ref{fig::SNP_scatter} presents the scatterplot of the 70 SNPs rejected using cov$\_$H3K4me1$\_$Trynka, with 32 SNPs overlapping with those rejected by the BH method.

We observe that the BH method tends to discover the significant SNPs when the covariates such as cov$\_$Repressed$\_$Hoffman or cov$\_$H3K4me1$\_$Trynka are relatively small. 
In the IHW method, these two covariates can be seen to affect the significance of the SNP as their values increase even, 
whereas the BH method primarily detects significant SNPs with fairly small $p$-values. 
On the other hand, in the case of cov$\_$Intron$\_$UCSC, relatively more SNPs are detected by the BH method when the values of this covariate are widely distributed, 
but it can be observed that a considerable number of new significant SNPs are discovered due to this covariate.
Moreover, it is noteworthy that when the LD score of ``Intron$\_$UCSC" was included as a covariate, 
almost three times as many SNPs were discovered compared to the previous two covariates. 
 {Although intron regions of DNA are typically considered unrelated to protein synthesis, \cite{encode} has highlighted the significance of non-coding regions, such as intron, in the study of human biology and disease for elucidate systematic analyses of transcripts and gene regulatory mechanisms.}
Given the ongoing discoveries of biologically significant mechanisms in non-coding DNA regions, 
a thorough examination of this covariate is warranted.
To summarize, the BH and Storey $q$-value method detects only the SNPs with small $p$-values while the IHW method finds the significant SNPs even with relatively large $p$-values and large values of corresponding covariates. 

In contrast to the IHW method, which exhibited varying results depending on the included covariate, the Boca-Leek method demonstrated consistent outcomes across different covariate settings. Specifically, when using cov\_baseL2, it showed 50 discoveries, while for the remaining 29 covariates, each exhibited 49 discoveries, aligning with results comparable to conventional methods like BH or Storey $q$-value. This trend remains consistent with the simulation findings, suggesting that the Boca-Leek method, displayed lower sensitivity to true covariates associated with the distribution of $p$-values compared to IHW.

\subsection{Result of PC Transformed Covariates}
In this section, we opted for transformed covariates using PCA as an alternative to the original matrix which are presented in Algorithm \ref{alg:FDR_PCA}. 
PCA serves to capture essential information more efficiently through dimensionality reduction and enhance interpretability by revealing underlying patterns in the covariate data. 
Overall, we aim to identify true associations between covariates and SNP significance, 
providing a more nuanced and interpretable perspective on the underlying biological relationships.

\begin{figure}[ht]
    \centering
    \includegraphics[width=0.8\columnwidth]{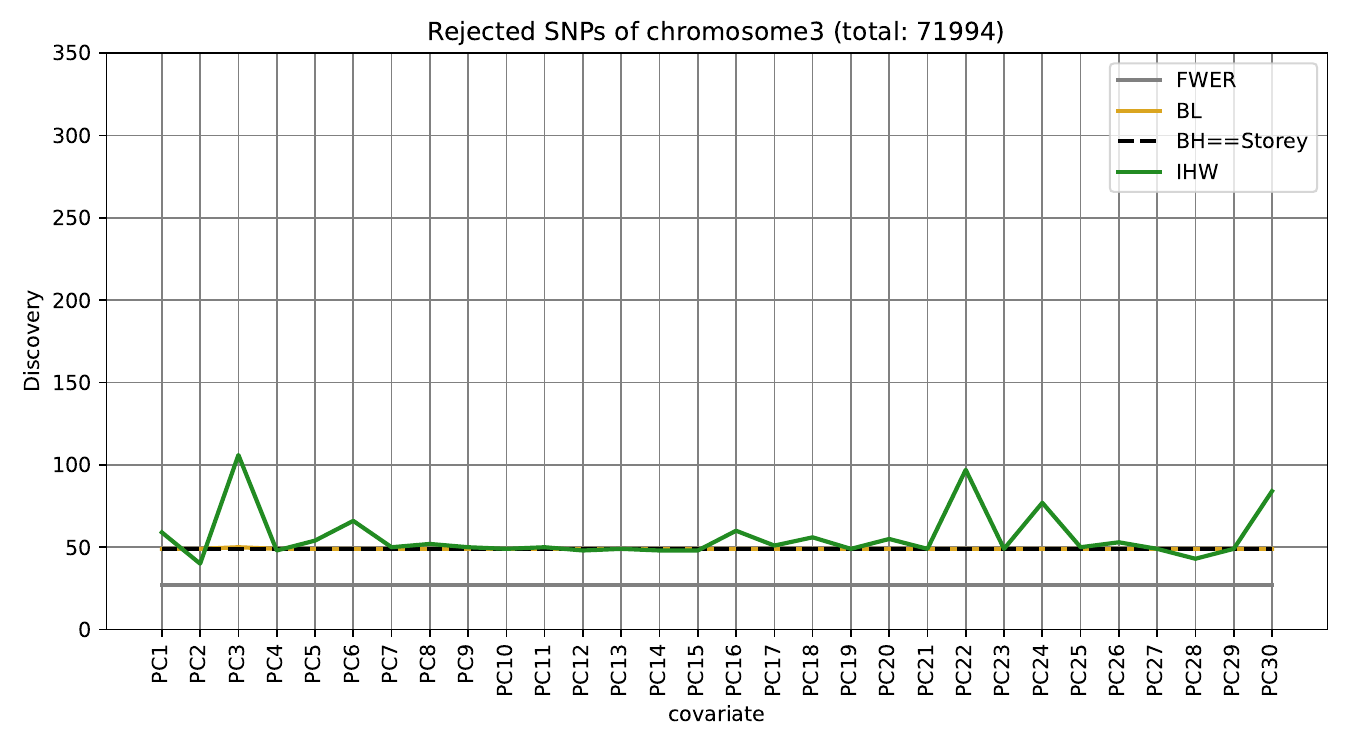}
    \caption{The discovery of each method on PC-axis}
    \label{fig::SNP_PCA_result}
\end{figure}

Applying the IHW method, we incorporated a PC-transformed covariate one by one into the FDR analysis, revealing variations in the number of rejected SNPs depending on the chosen PC-transformed covariate. 
 {While Figure \ref{fig::SNP_PCA_result} presents the summarized findings for chromosome 3, detailed results for chromosomes 1 through 22 are included in the online supplementary material, specifically in Figure \ref{all_result_SNP_PCA} 
in the online supplementary material.}
Notably, when utilizing PC3 as a covariate in the IHW method, we identified 136 significant SNPs, and with PC6, we identified 106 significant SNPs. This suggests that using PC-transformed covariates resulted in rejecting approximately twice as many SNPs compared to when using other covariates prior to transformation. In contrast, when employing the Boca-Leek method, we observed consistent results regardless of the PC axis used. Specifically, the number of discoveries remained constant at 49, mirroring the outcomes obtained when applying the original methods such as BH and Storey $q$-value.

\begin{figure}[ht]
    \centering
    \includegraphics[width=0.9\columnwidth]{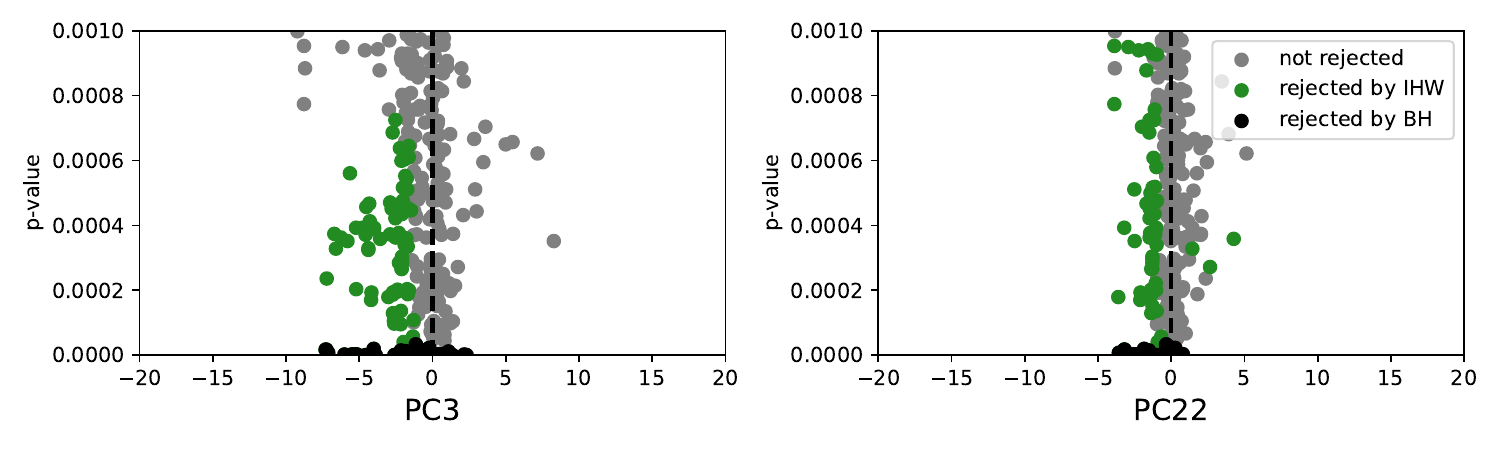}
    \caption{The discovery of each method on PC-axis}
    \label{fig::SNP_PCA_scatter}
\end{figure}

The Figure \ref{fig::SNP_PCA_scatter} displays scatterplots for PC3 and PC22, which showed more significant results in IHW method than other PC-axes. The x-axis represents the values of PC scores, while the y-axis represents $p$-values for a total of 71,994 SNPs. Similar to Figure \ref{fig::SNP_scatter}, SNPs identified by BH are represented in black, those discovered by IHW are shown in green, and other non-significant SNPs are depicted as gray dots.

To explore the significance of the two significant axes, PC3 and PC22, we computed the PC loading values for each PC. The outcomes are depicted in Table \ref{table::pc_loadings} in the online supplementary material, with the PC loading values rounded to the fourth decimal place. Particularly, we sorted the loading values of PC3 in descending order. Following PC transformation of the covariate matrix, first of all, the coding covariate exhibits a positive PC loading, whereas the intron, representing a non-coding covariate, shows a negative PC loading.
Specifically, we observe that the covariates representing LD scores for promoters, transcription-initiated sites, and enhancers show positive values for PC3, while those representing LD scores for repressed region display negative values.
From a gene expression regulatory point of view, promoters are regions of DNA located upstream where RNA polymerase binds to initiate transcription, the process of synthesizing RNA from DNA.
Promoters play a crucial role in determining transcription start sites (TSS) and are frequently enriched with transcription factor binding sites (TFBS).
Enhancers regulate gene expression by interacting with promoters, and during transcription initiation, the binding becomes more permissive.
On the other hand, a repressed covariate is associated with the inhibition of gene expression. This occurs when a repressor protein binds to the promoter region, preventing the synthesis of messenger RNA.
This observed grouping suggests that covariates with similar functions are coalescing in the same direction.

Following a similar analysis for PC22, we find it challenging to precisely interpret the grouping of covariates based on the sign of their loading values. However, given that PC22 has less explanatory power in the original data compared to PC3, we opt to bypass an in-depth exploration of its implications.

\begin{figure}[ht]
    \centering
    \includegraphics[width=0.5\columnwidth]{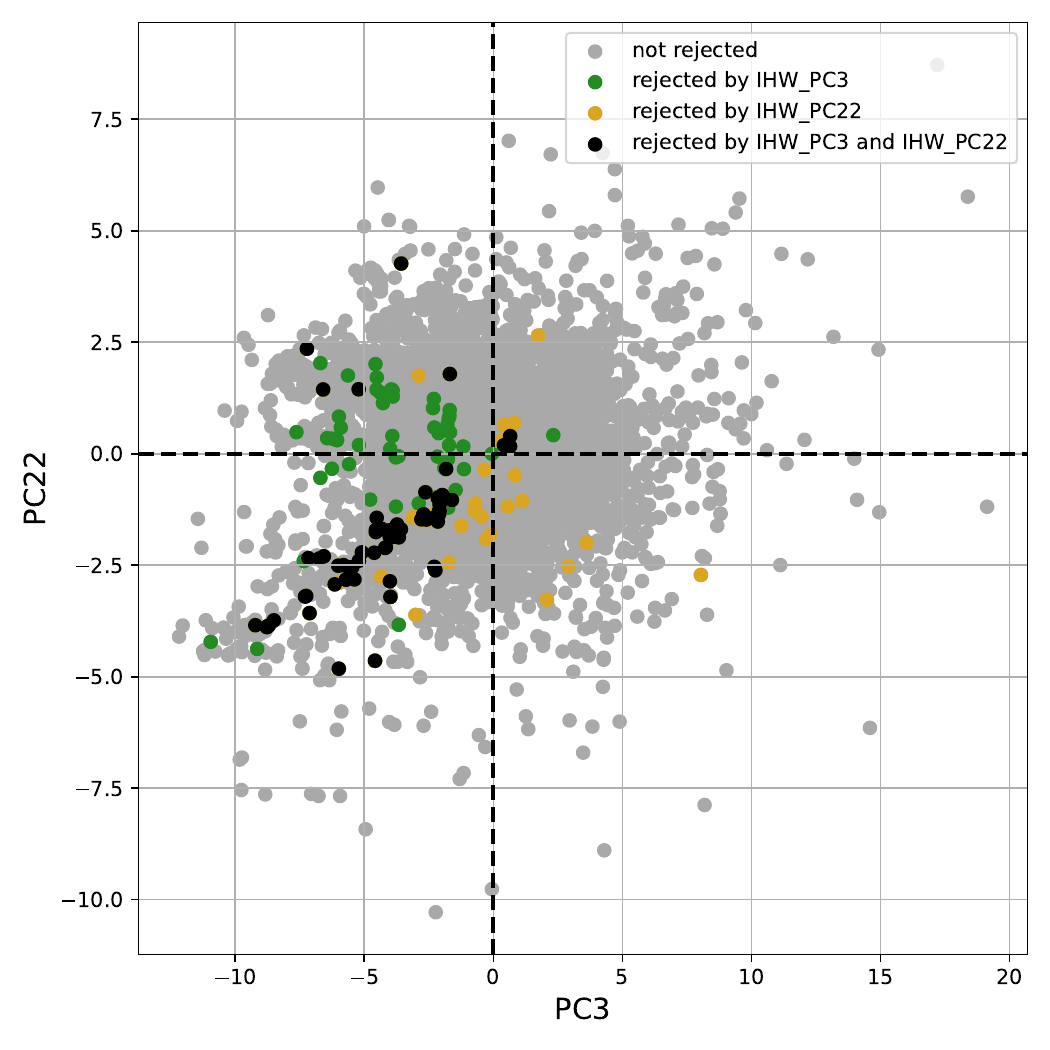}
    \caption{Scatterplot of PC scores for PC3 and PC22}
    \label{fig::pc_score}
\end{figure}

We further transformed the existing covariate values into principal component scores for 71,994 SNPs, and plotted the resulting values on a scatter plot with PC3 on the x-axis and PC22 on the y-axis. As shown in Figure \ref{fig::pc_score}, SNPs that were determined to be significant only by PC3 are denoted as green dots, SNPs significant only by PC22 are denoted as yellow dots, and SNPs significant by both PC3 and PC22 are denoted as black dots.
In this figure, we can observe groups represented by the same color gathering together on the scatterplot. Specifically, when we consider PC3 as the $x$-axis and PC22 as the $y$-axis, we can see that significant SNPs determined only by PC3 are located above some line, for example   $y = 0.5x$, while significant SNPs determined by PC22 are located below the line. SNPs with black dots 
that are located  along with the line, $ y=0.5x$,  
are declared to be significant by the IHW for 
both PC3 and PC22.

\section{Conclusion}\label{sec6}
This study has provided an advancing FDR control methods in statistical hypothesis testing, 
especially when incorporating covariate information. 
We conducted a comparative analysis of traditional FDR control methods using only $p$-values, 
such as the BH and Storey $q$-value methods, 
against methods utilizing covariate information like IHW and the Boca-Leek approach. 
Recognizing the challenges posed by high-dimensional covariate data, 
including multicollinearity and the limitations inherent in existing methods, we proposed two strategies.
The first involved the meticulous selection of a subset of meaningful covariates through individual application, 
while the second leveraged PCA on the original covariates. 
The application of PCA not only streamlined the complexity of the data but also significantly enhanced the interpretability and reliability of our statistical inferences.

Simulation studies revealed that transforming high-dimensional covariates using Principal PCA effectively enhances the performance of FDR control methods for each PC axis.
When the data were structured using the approach of estimating the null distribution, 
both IHW and Boca-Leek methods, utilizing covariates in addition to $p$-values, 
demonstrated a higher capacity to detect significant features, particularly on specific PC axis directly associated with $p$-values.  

Moving to real data applications,
our analyses revealed significant rejections of SNPs associated with BMI indices when utilizing LD scores of specific functional regions, such as Repressed, H3K4me1, and Intron, as covariates within an epigenetic context. 
Notably, employing PC-transformed covariates, specifically PC3 and PC22, resulted in a greater number of SNP discoveries. 
The interpretation of these principal components enriched our understanding of the genomic landscape and its relevance to BMI-related traits has been notably enriched.
This observed significant rejection of SNPs was predominantly attributed to the efficacy of the IHW method. 
It is noteworthy that Boca-Leek consistently showed no significant impact, irrespective of the covariate used, 
and yielded similar rejection counts to those obtained with the conventional BH method. 
This observation aligns with our simulation results, where Boca-Leek exhibited limited effectiveness compared to IHW, 
especially in scenarios where the data were structured using a size-investing strategy. 
The real-data application appears to have been designed following a size-investing approach, 
deviating from the null distribution proportion estimation scenario employed in the first simulation. 
This fundamental difference in data construction strategies helps explain why Boca-Leek did not demonstrate notable effects, corroborating the insights gained from our simulation findings.

In conclusion, our study introduces strategies to enhance both FDR control and statistical power in the presence of covariates, addressing limitations present in existing methods. 
Importantly, we positioned the proposed covariate-based methods within the broader landscape of classical FDR control procedures. By comparing their performance against Bonferroni correction, the Benjamini-Hochberg procedure, and Storey’s q-value method in both simulation and real data analyses, we have provided a comprehensive evaluation of where these approaches excel or fall short under different data structures. This comparative framework directly addresses concerns regarding how our current methods relate to widely used traditional approaches and highlights the additional utility gained by incorporating informative covariates into the multiple testing correction process. 
Through rigorous simulations and real data applications, we demonstrated the superior performance of IHW and Boca-Leek in capturing true associations by effectively utilizing a meaningful subset obtained through covariate dimensionality reduction. 
This study contributes to advancing statistical methodologies in genomics,
offering sophisticated and effective FDR control strategies in the era of high-dimensional data. 
Future research avenues include refining LD score interpretations, exploring Boca-Leek's multi-covariate potential, and unraveling the grouping patterns observed in PC-transformed covariate analyses.



\section*{Acknowledgement}
This work was supported by the National Research Foundation of Korea (NRF) grant 601 funded by the Korea government (MSIT) (RS-2025-00556575).


\bibliographystyle{plainnat}
\bibliography{bibliography}

\appendix

\section{Figures}

\begin{figure}[ht]
    \centering
    \includegraphics[width=0.7\textwidth]{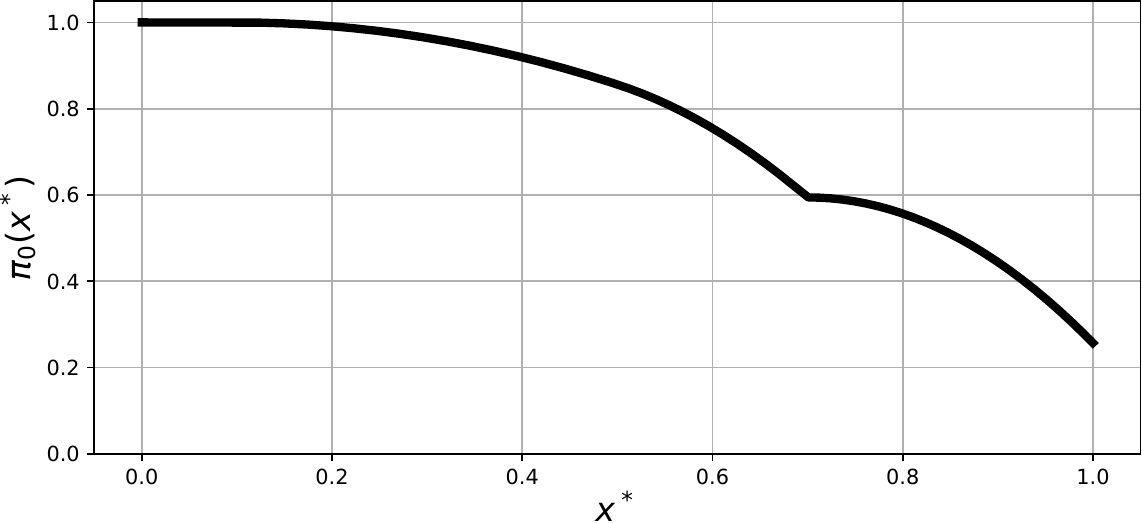}
    \caption{The proportion of null hypotheses depending on specific variables}
    \label{fig::pi0_plot}
\end{figure}

We depict the function $\pi_0(x^*)$ in the Figure \ref{fig::pi0_plot}, 
showing that it is a decreasing function with respect to $x^*$. 

\begin{figure}[ht]
    \centering
    \includegraphics[width=0.7\textwidth]{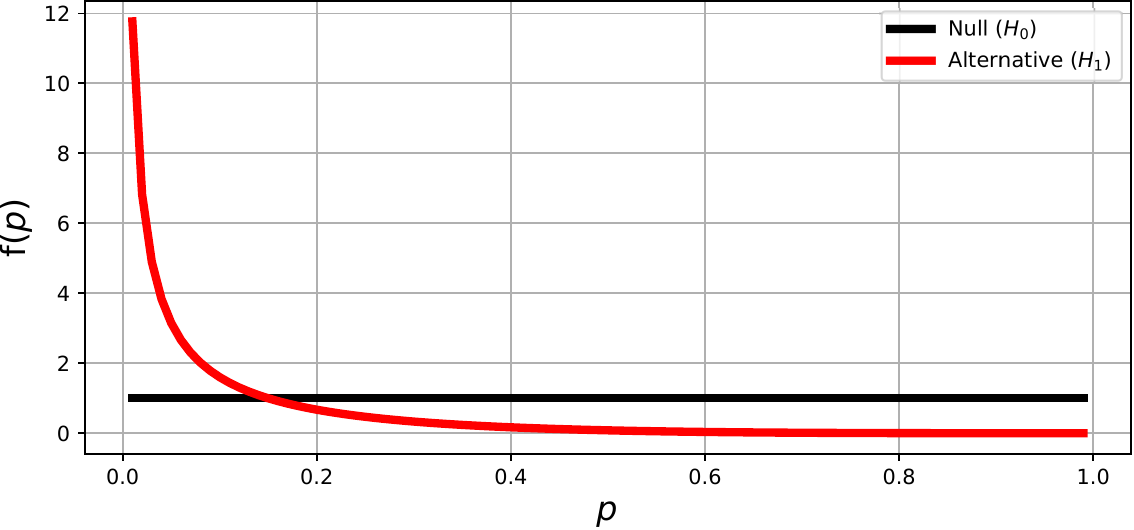}
    \caption{The distribution of $p$-value depending on each hypothesis}
    \label{fig::pval_plot}
\end{figure}


    
    

    
    

\section{Real Data Analysis}

For all 22 chromosomes, We add the following graphs on the number of rejections for different methods. 
As in the case of chromosome3 in our main text,  
we see  at most hundreds of SNPs are discovered among tens of  thousands SNPs. The IHW discovers the largest numbers of SNPs for some covariate or PC except chromosome21 which is the case that extremely few SNPs are rejected.

{
\centering
\begin{longtable}{cc}
    \includegraphics[width=.49\textwidth]{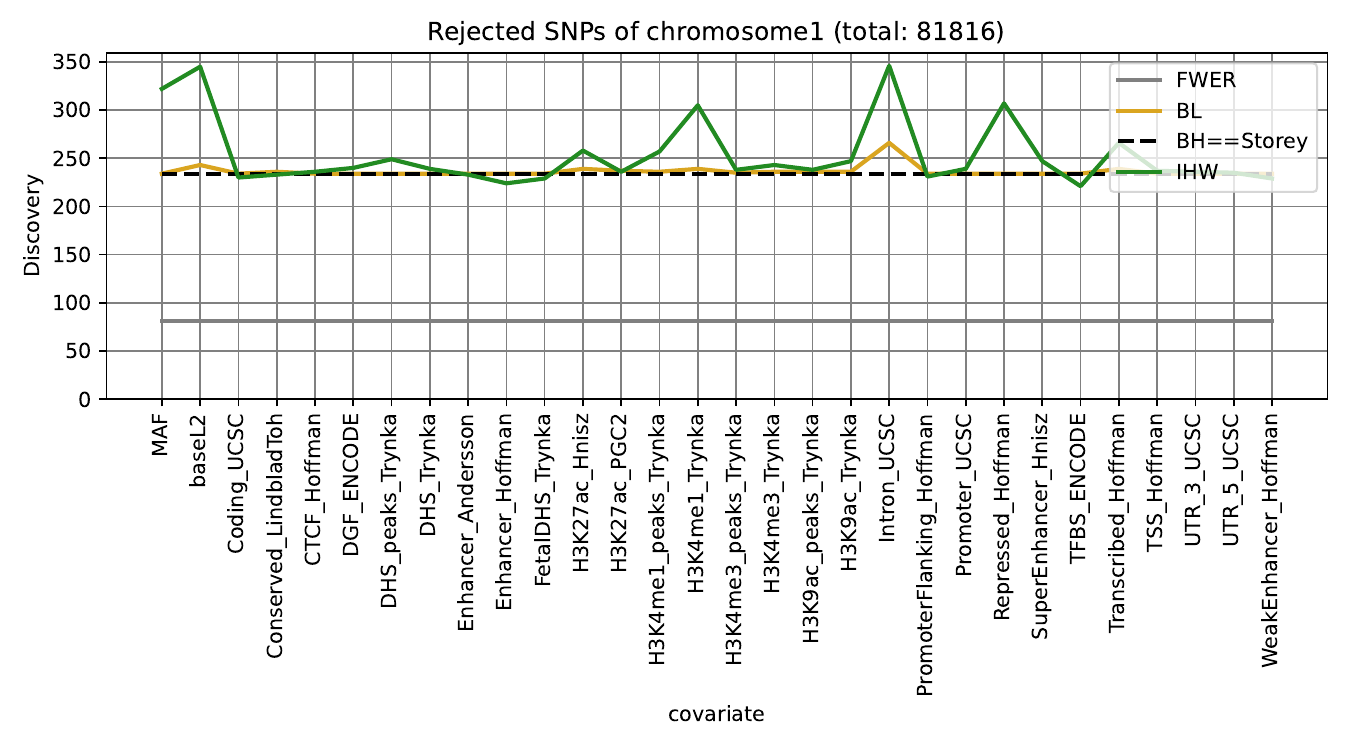} &
    \includegraphics[width=.49\textwidth]{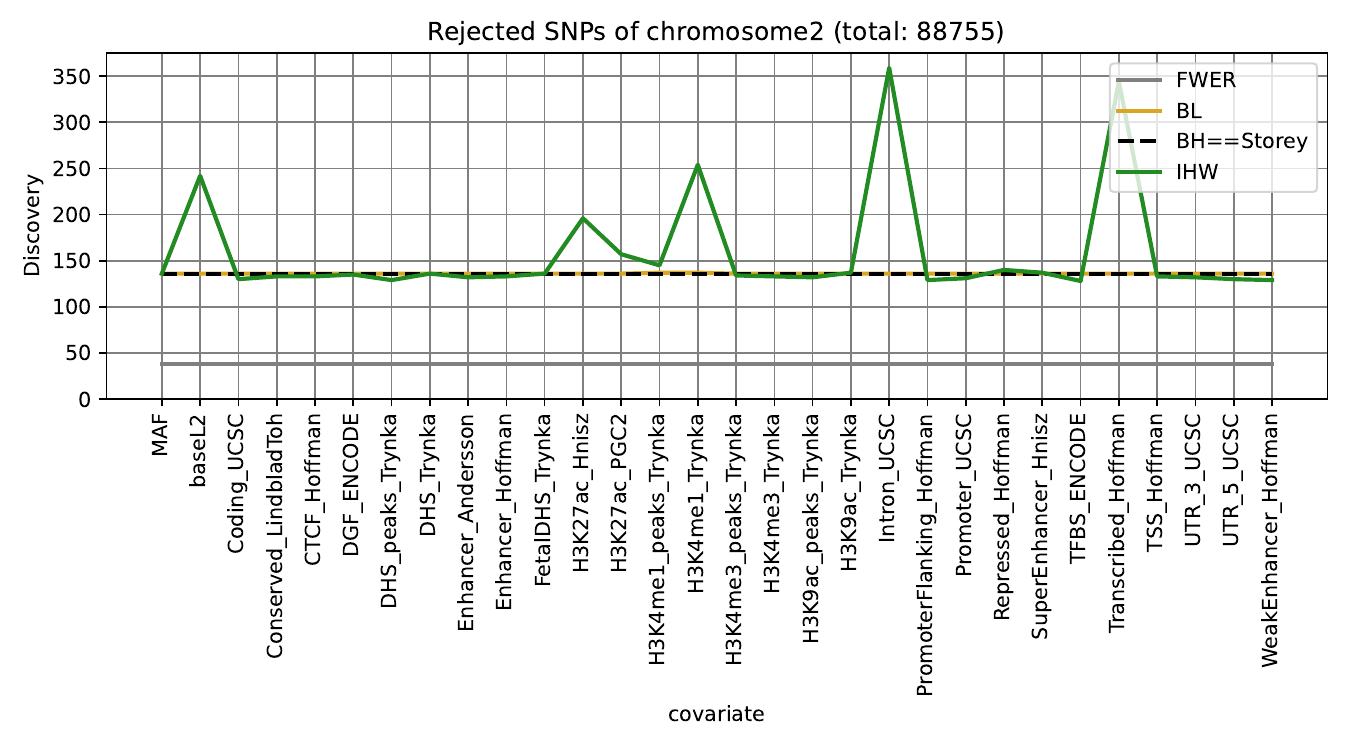} \\

    \includegraphics[width=.49\textwidth]{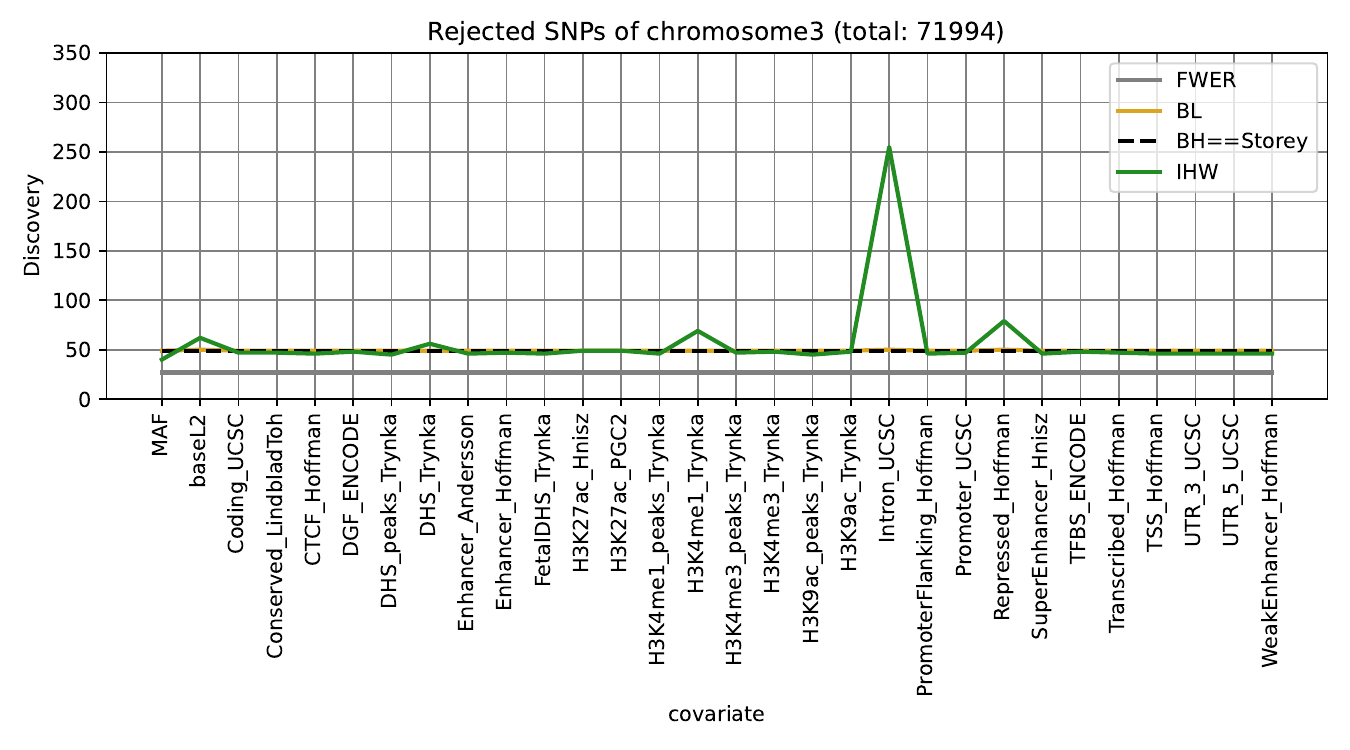} &
    \includegraphics[width=.49\textwidth]{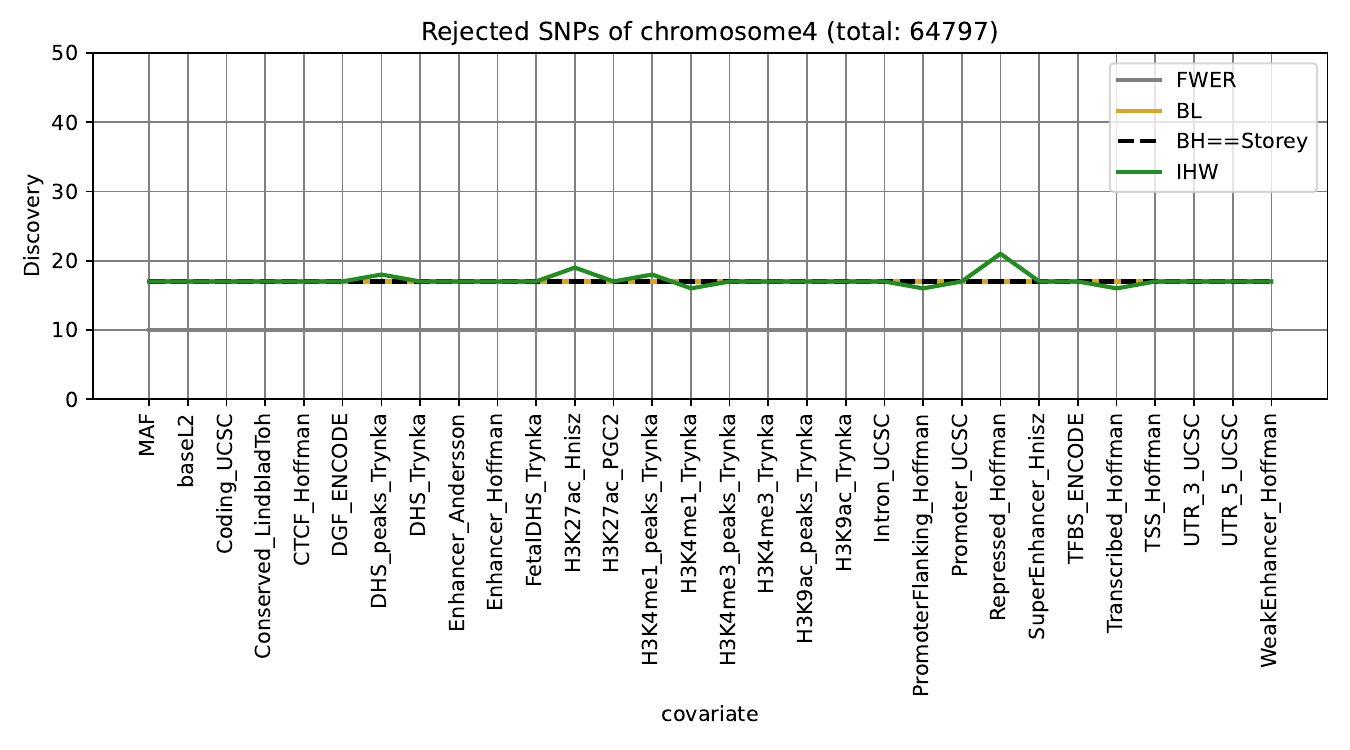} \\

    \includegraphics[width=.49\textwidth]{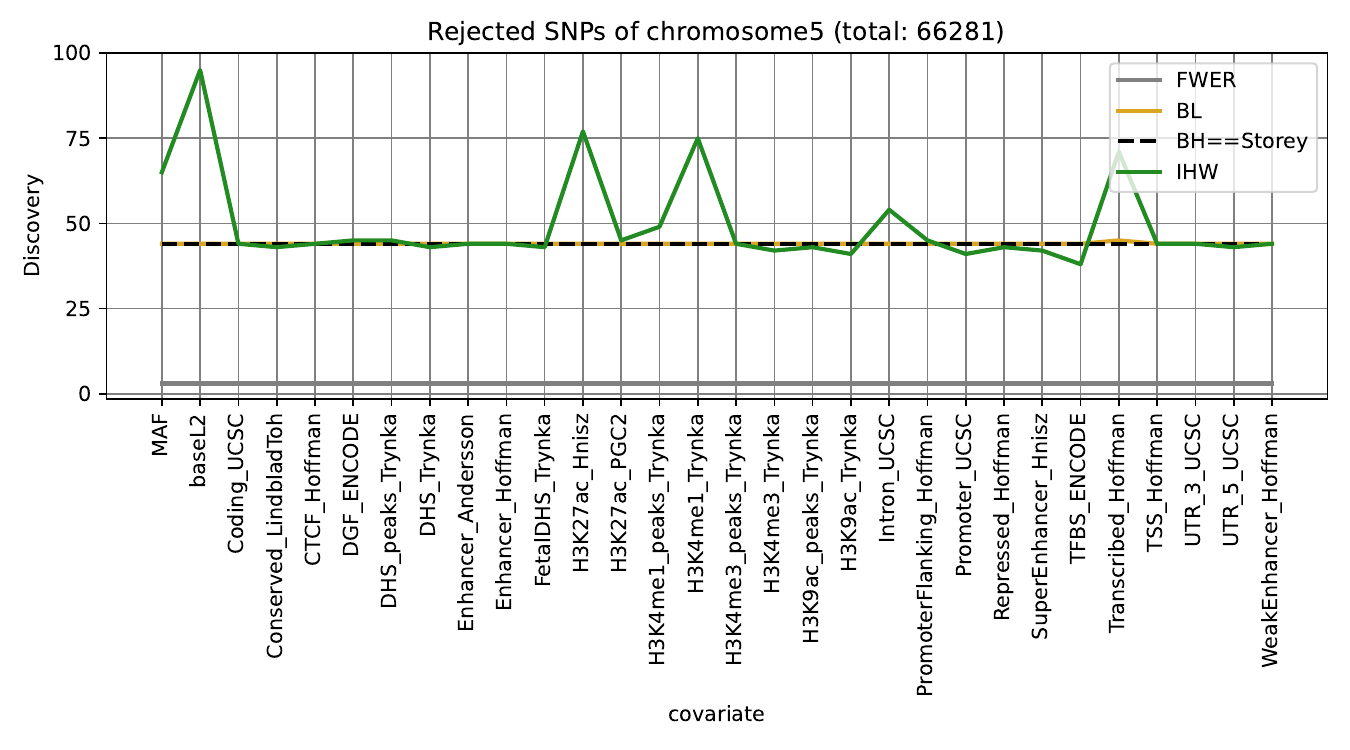} &
    \includegraphics[width=.49\textwidth]{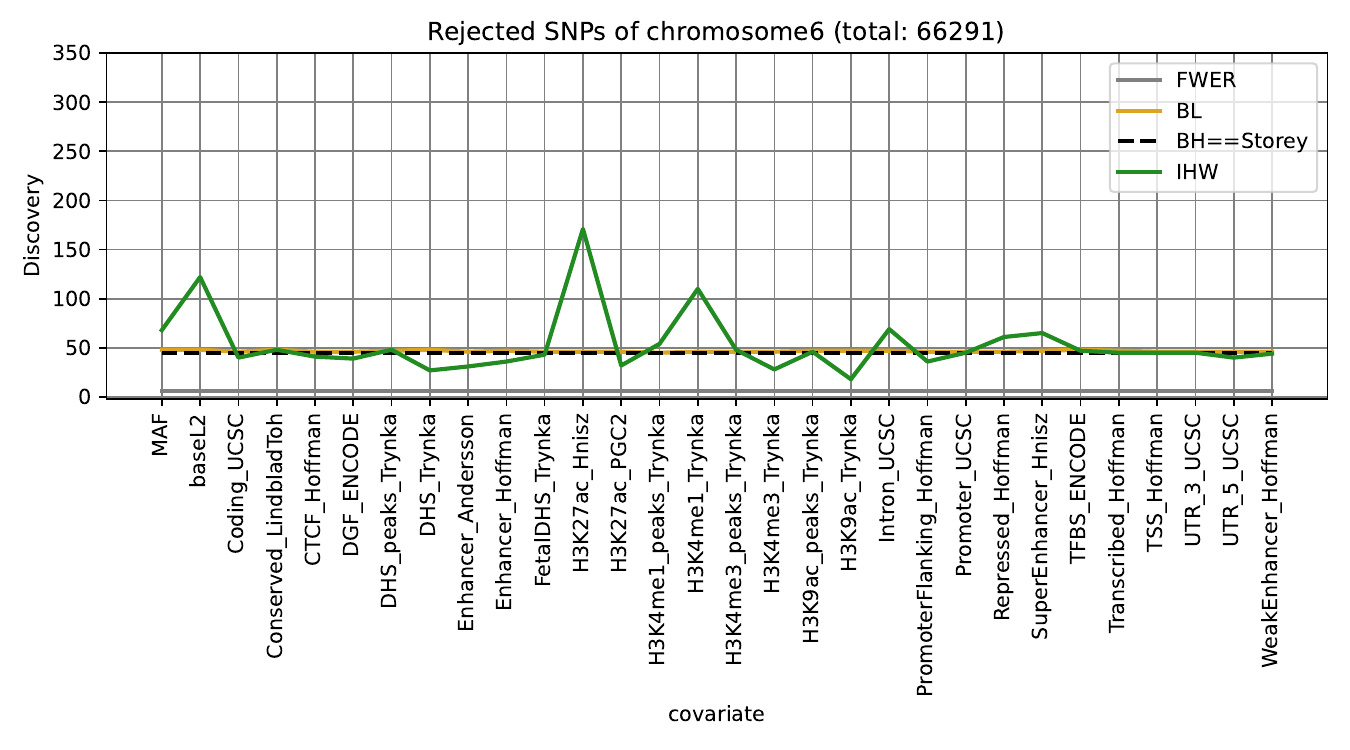} \\

    \includegraphics[width=.49\textwidth]{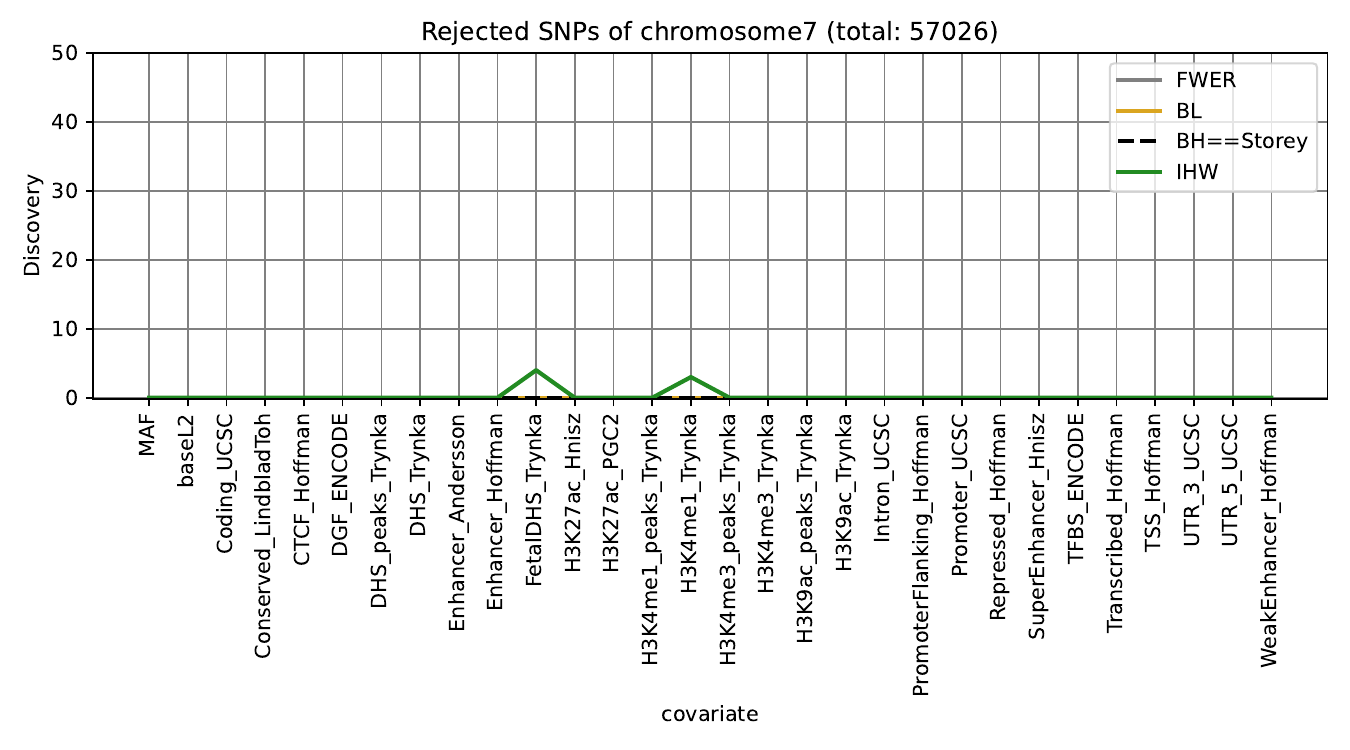} &
    \includegraphics[width=.49\textwidth]{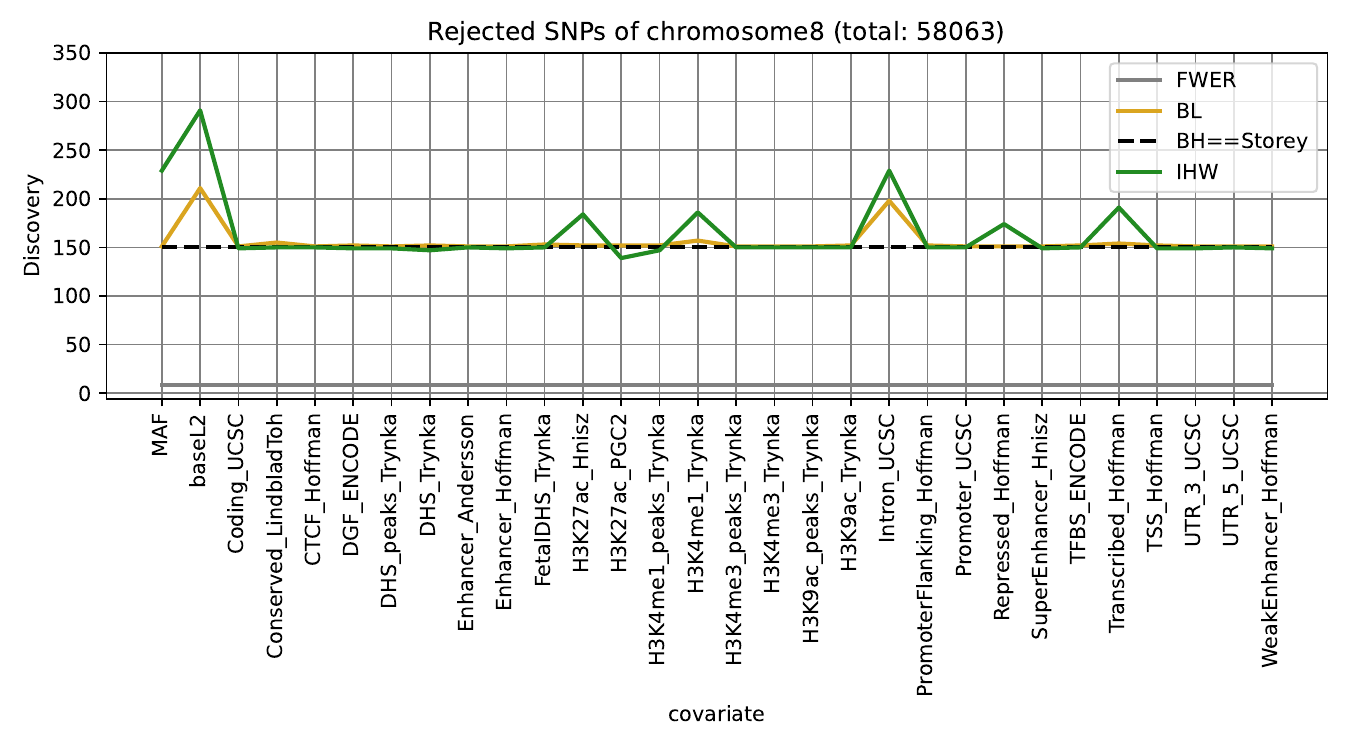} \\

    \includegraphics[width=.49\textwidth]{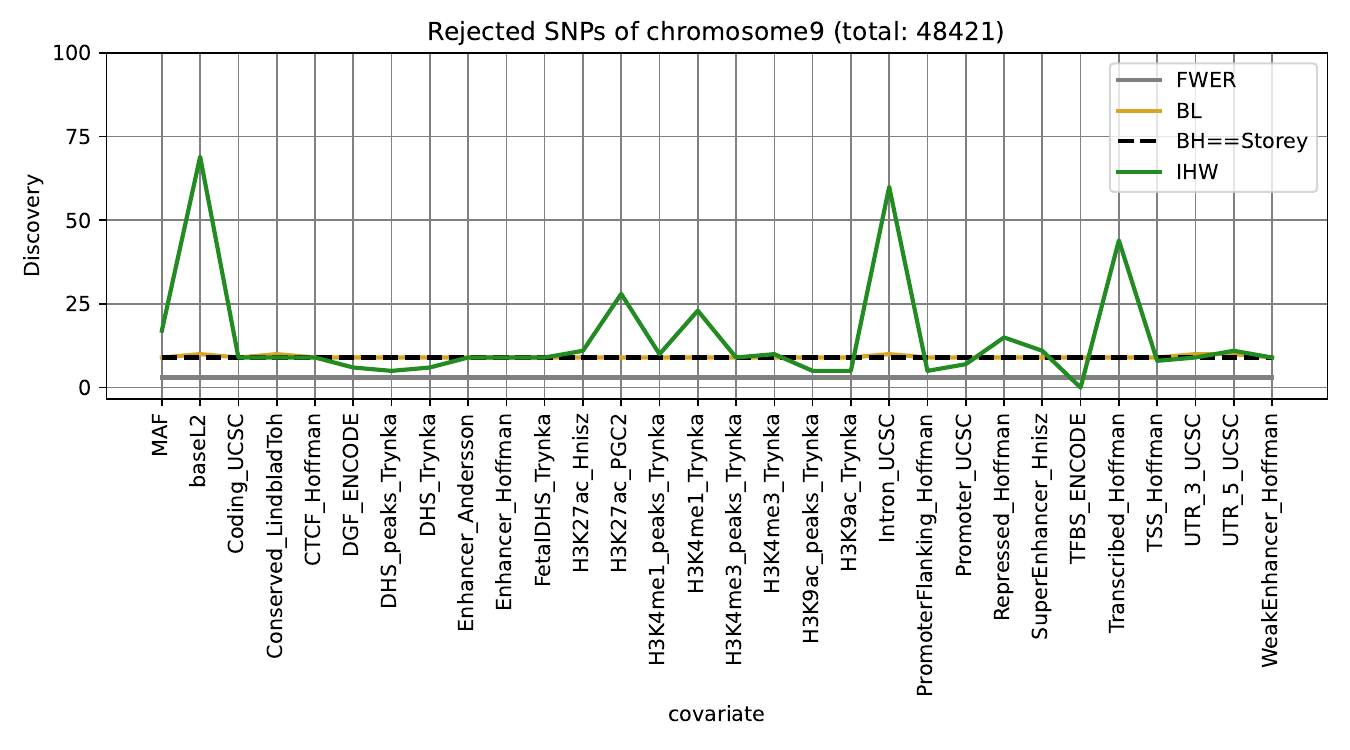} &
    \includegraphics[width=.49\textwidth]{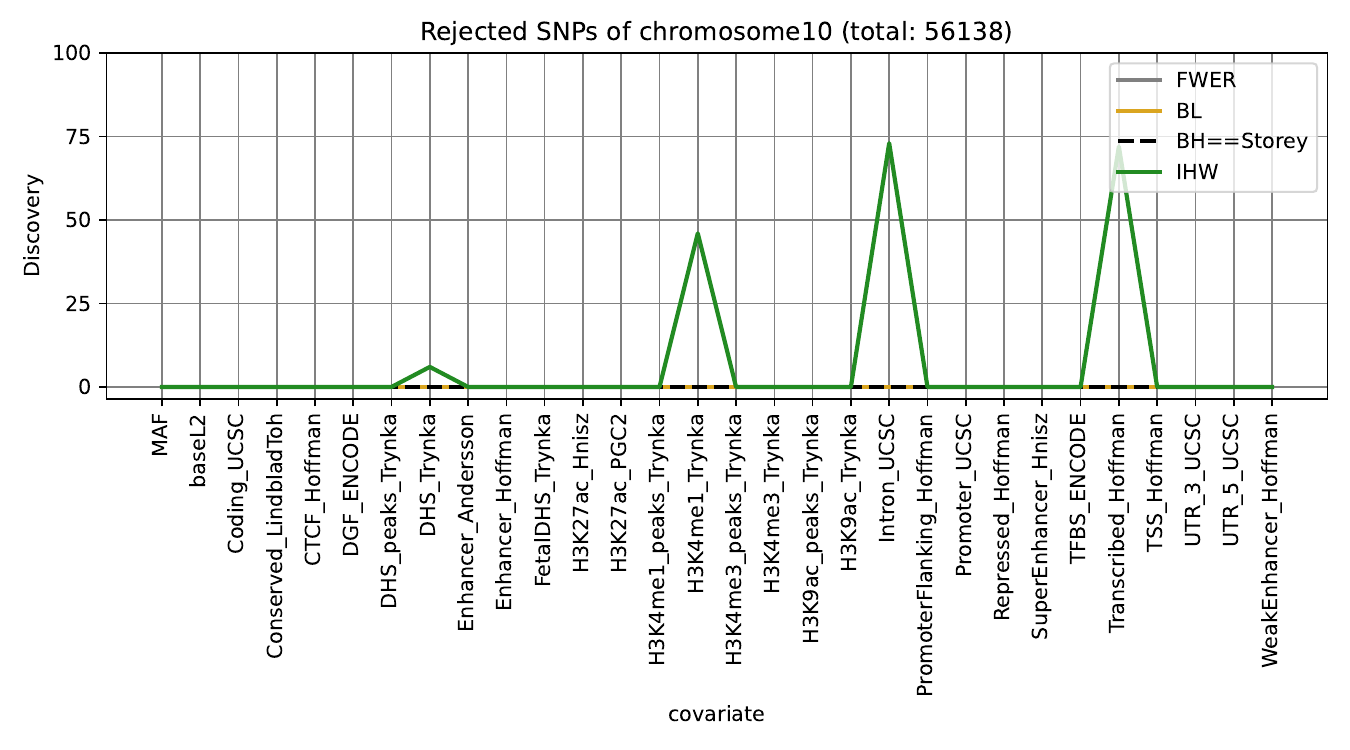} \\

    \includegraphics[width=.49\textwidth]{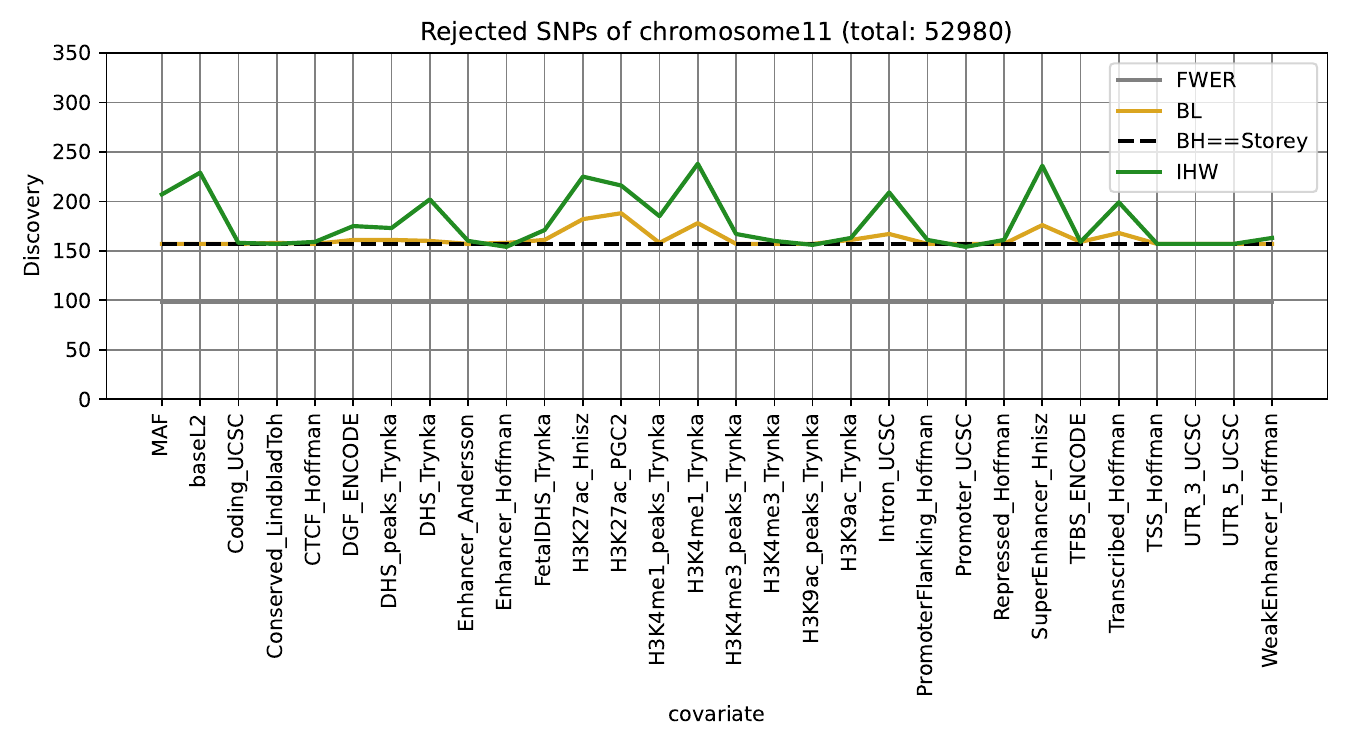} &
    \includegraphics[width=.49\textwidth]{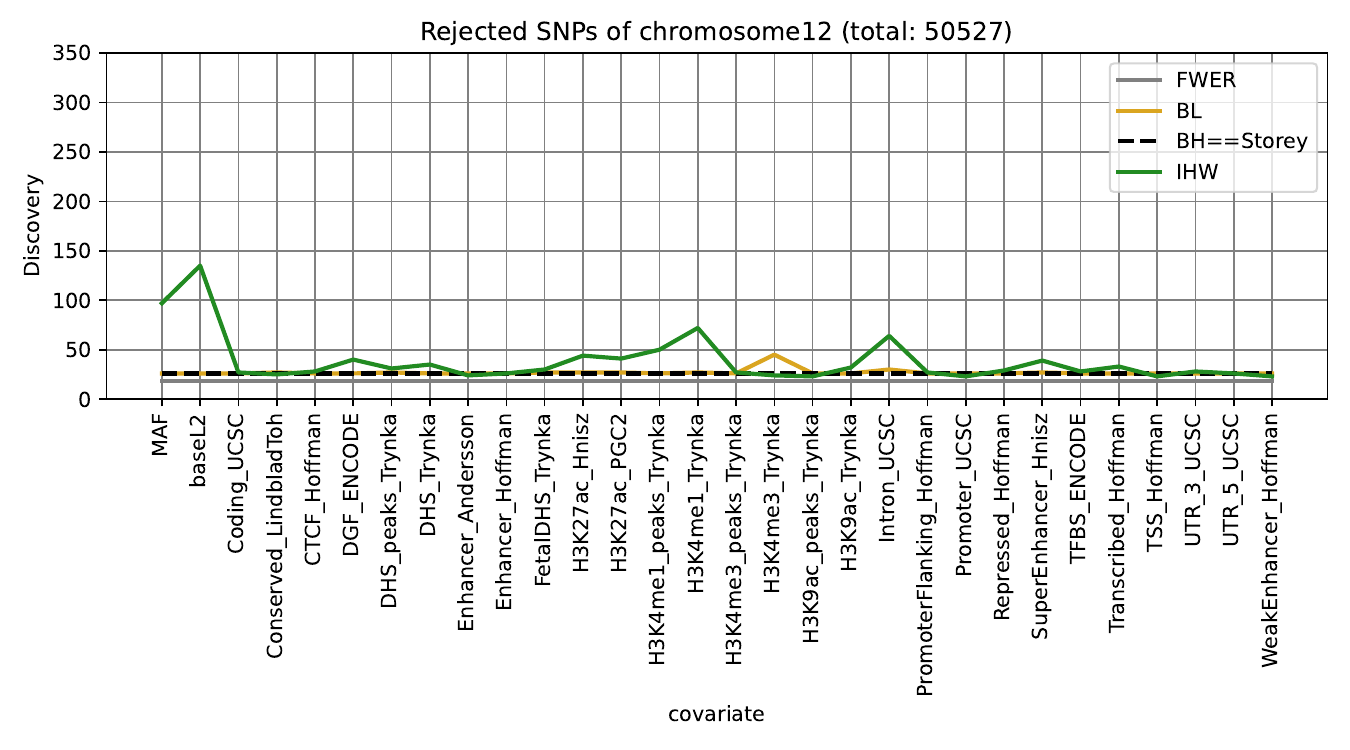} \\

    \includegraphics[width=.49\textwidth]{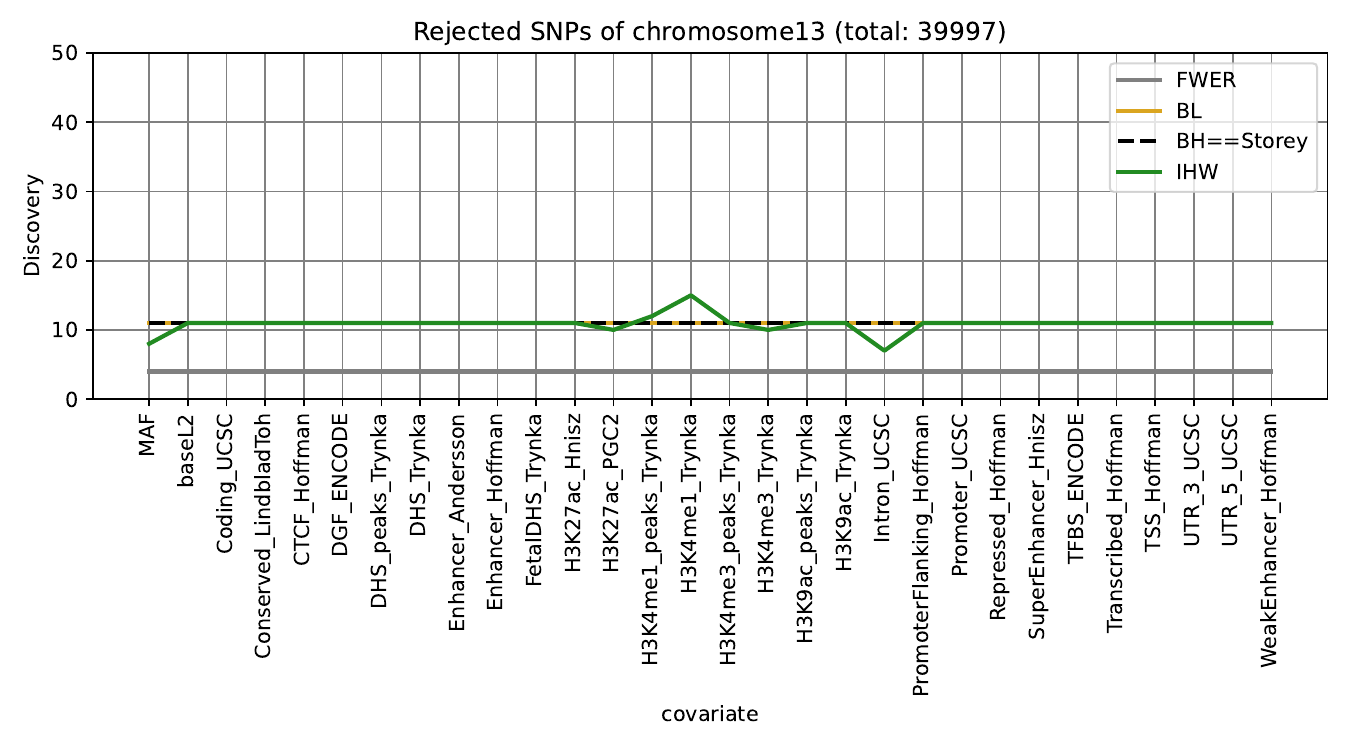} &
    \includegraphics[width=.49\textwidth]{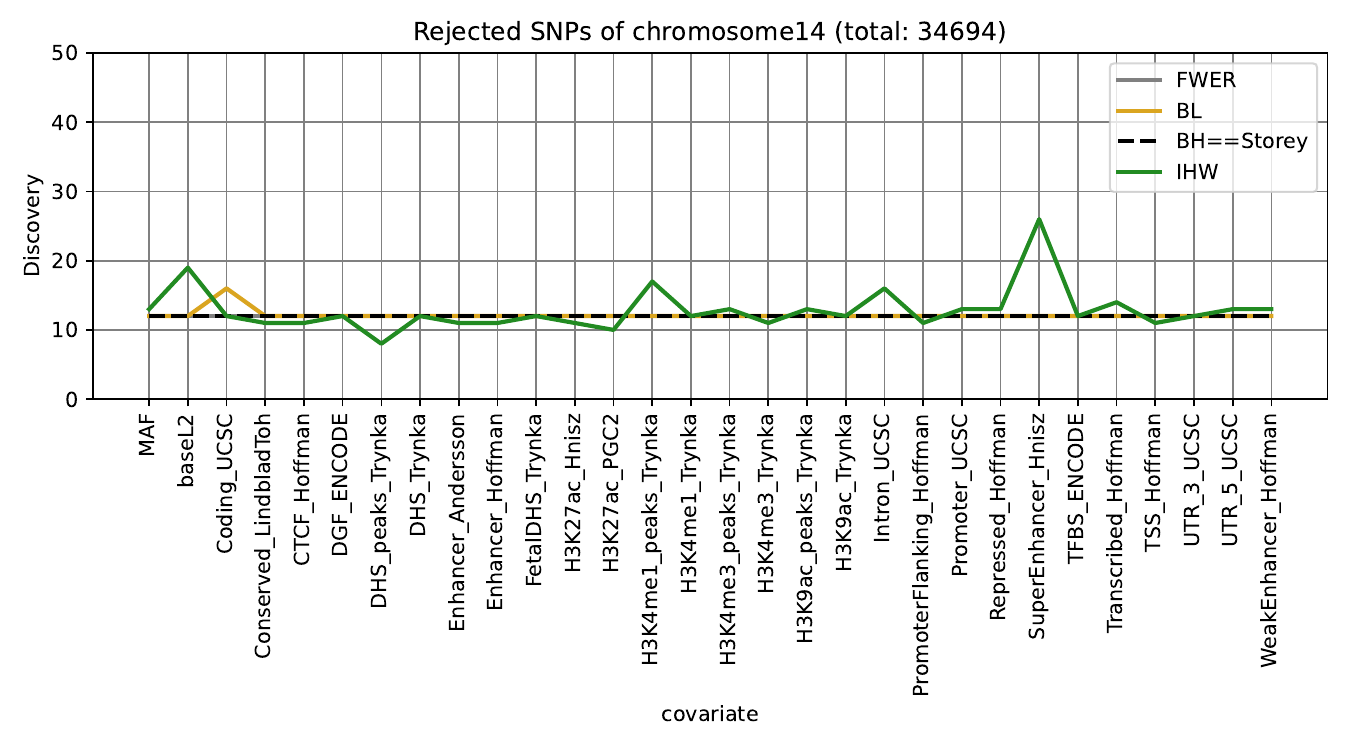} \\

    \includegraphics[width=.49\textwidth]{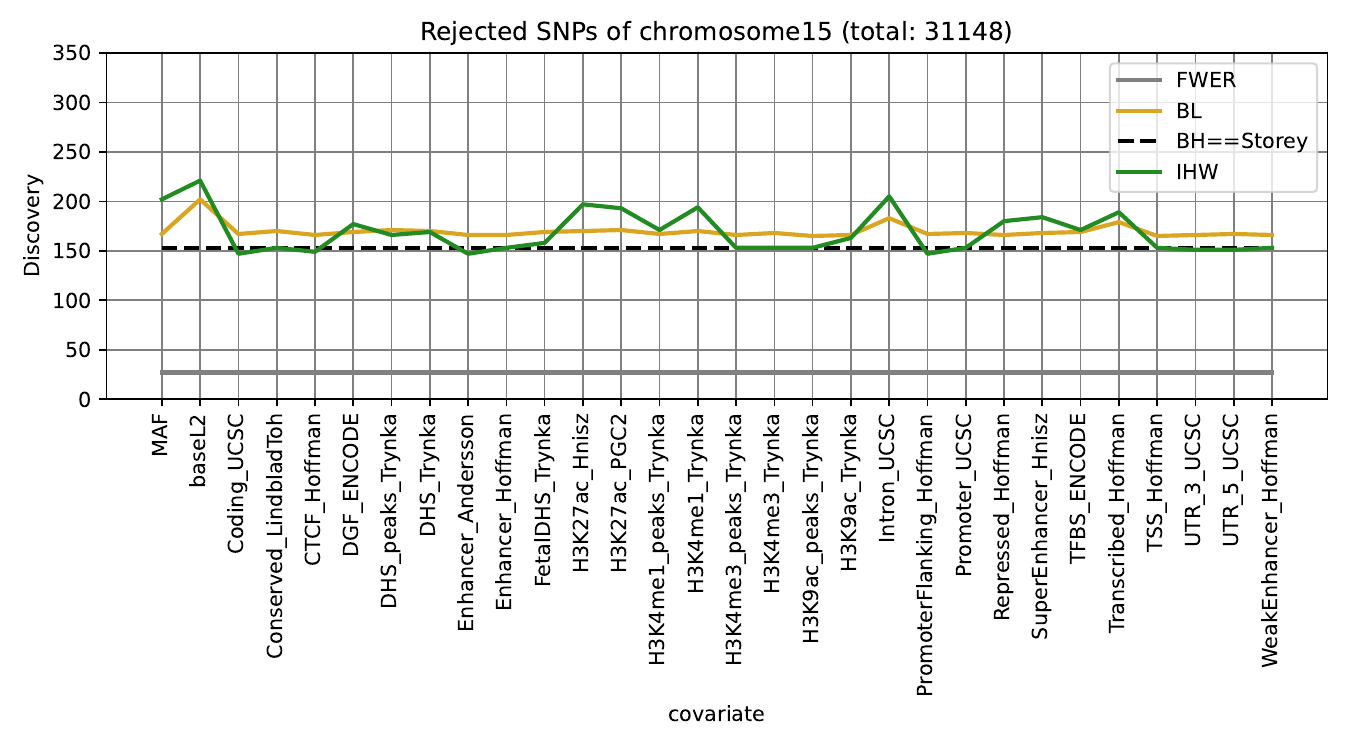} &
    \includegraphics[width=.49\textwidth]{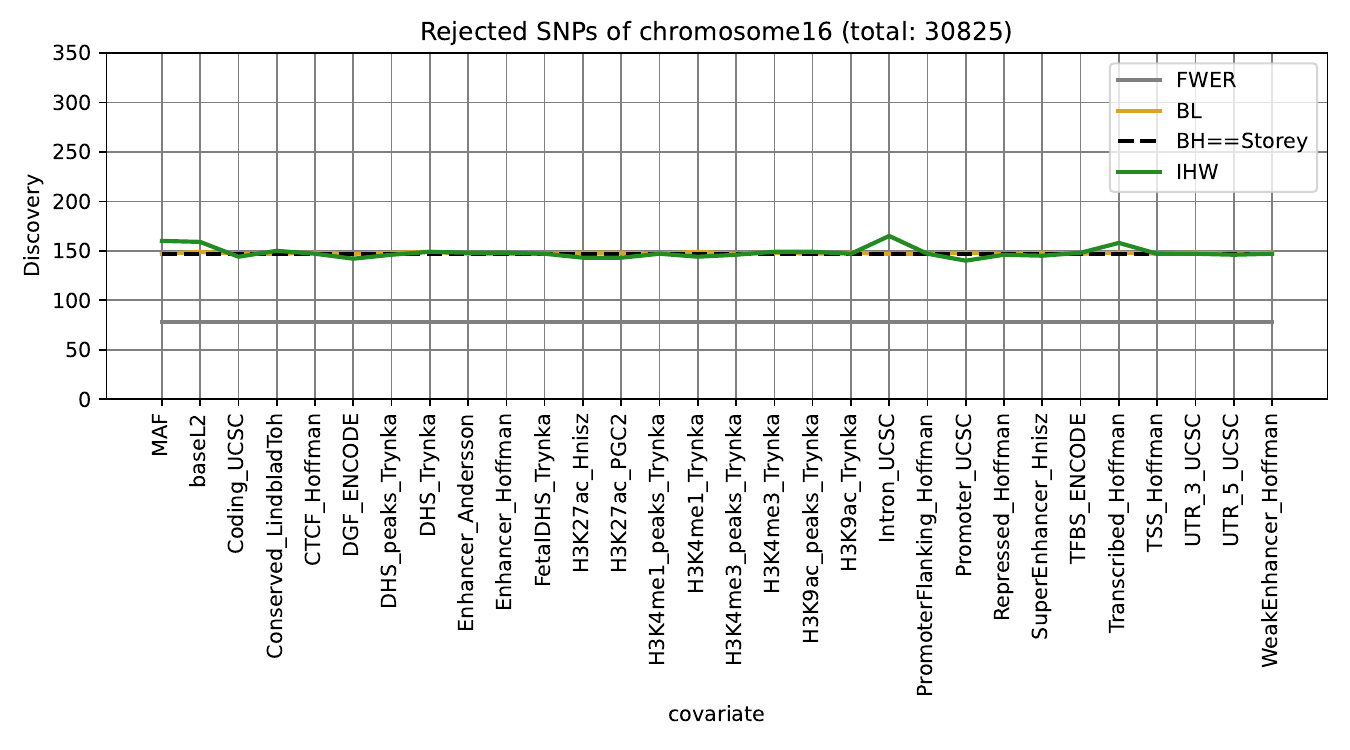} \\

    \includegraphics[width=.49\textwidth]{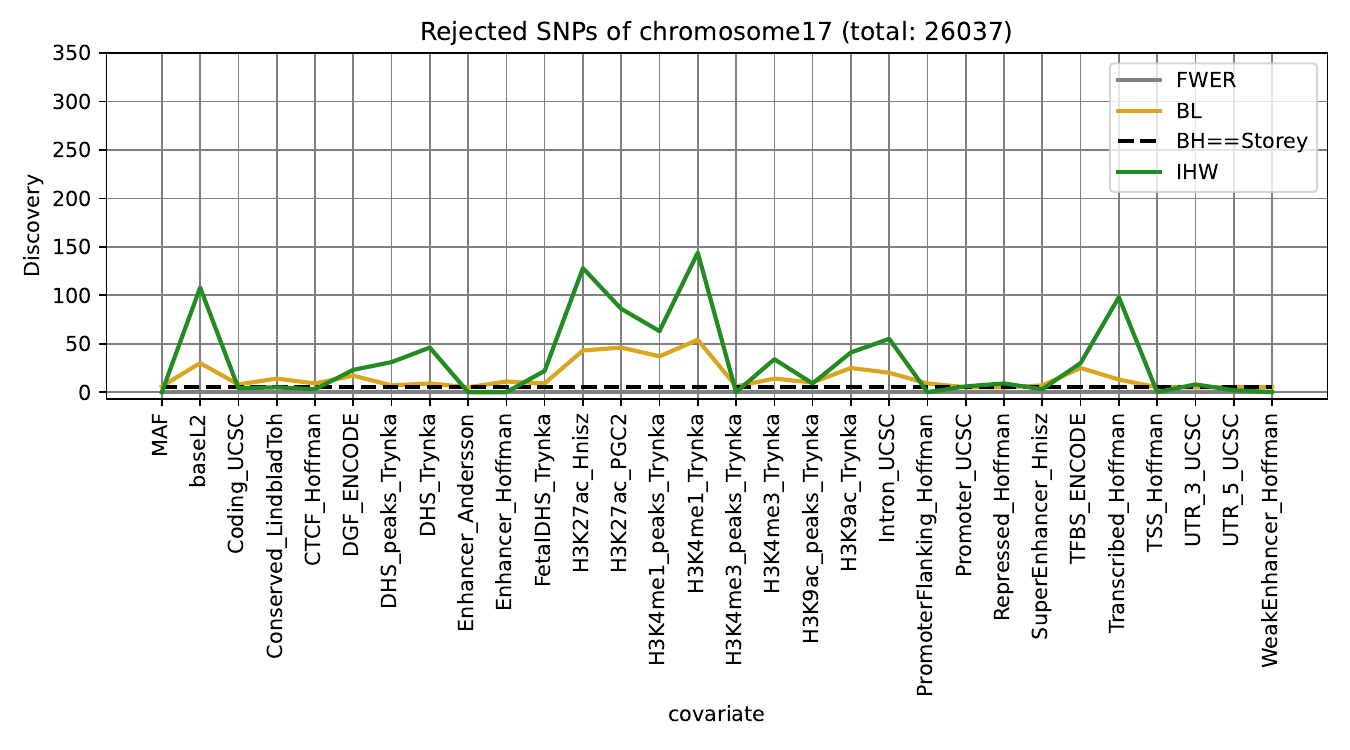} &
    \includegraphics[width=.49\textwidth]{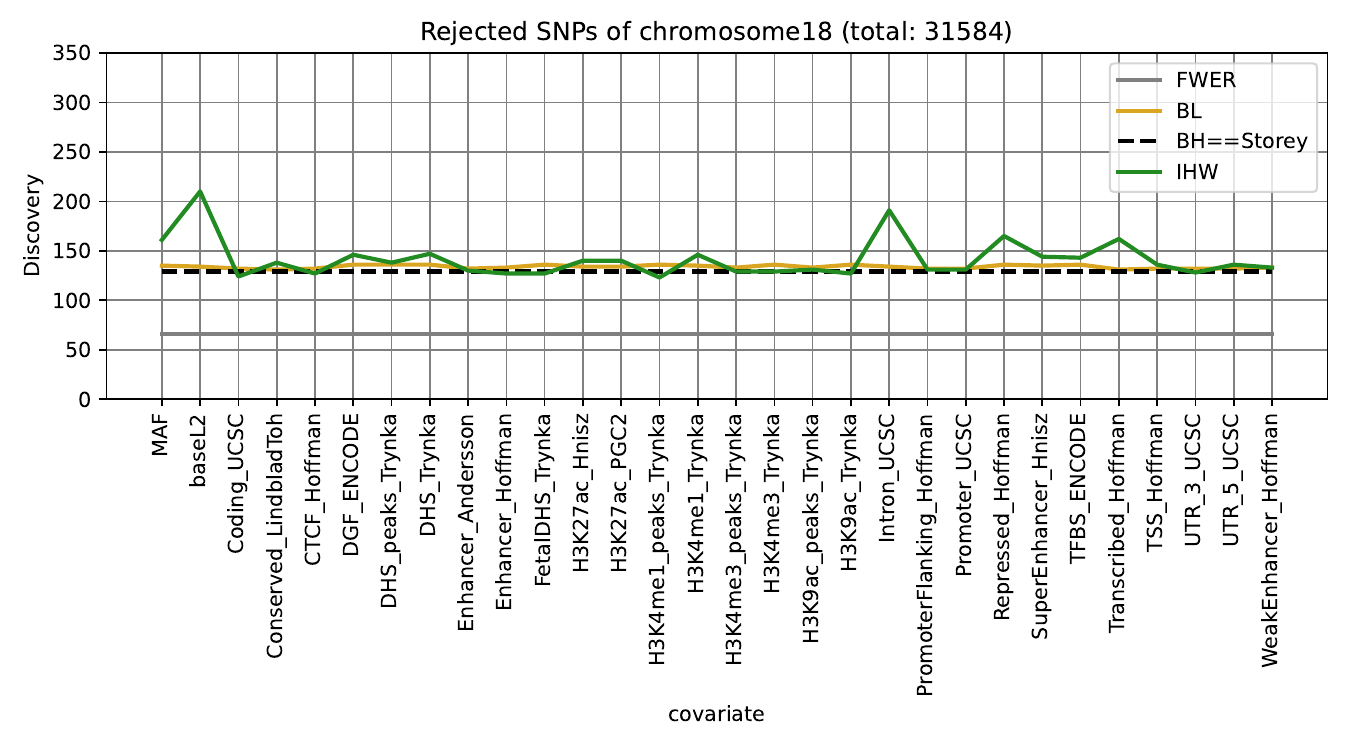} \\

    \includegraphics[width=.49\textwidth]{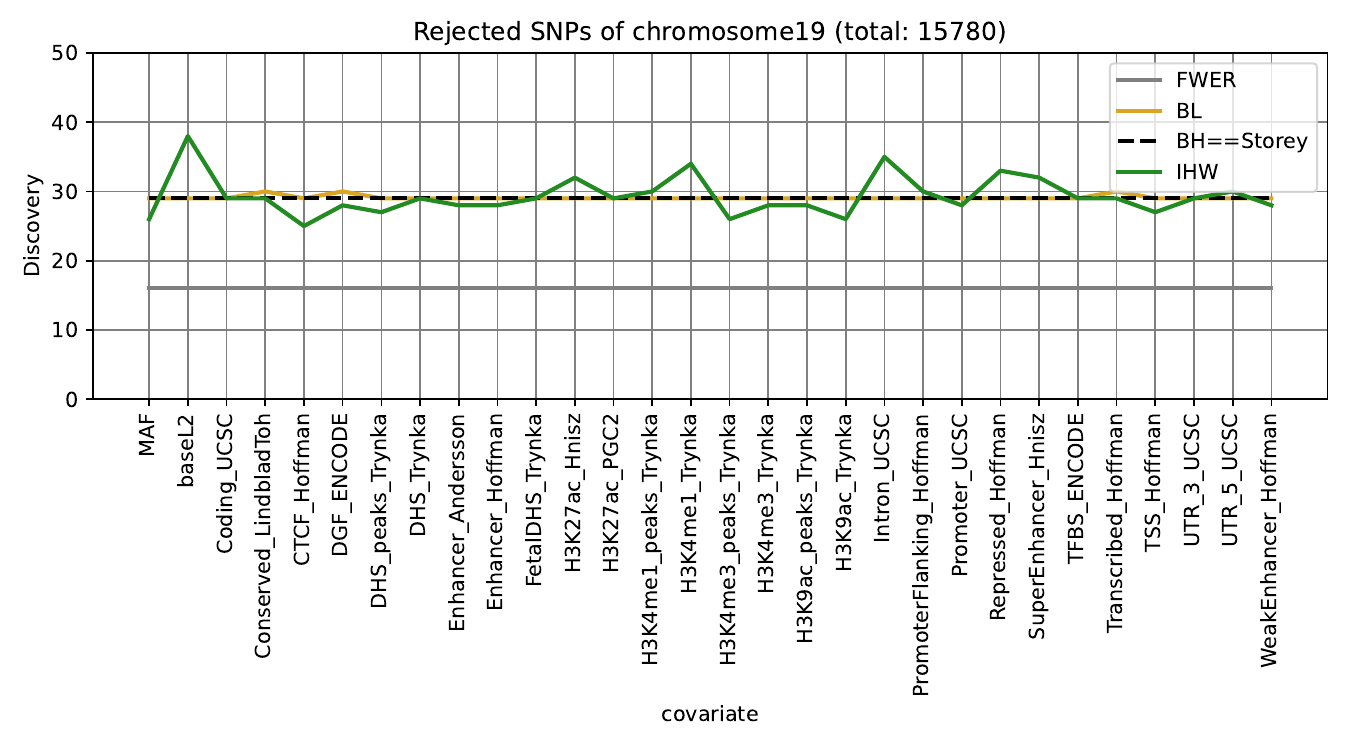} &
    \includegraphics[width=.49\textwidth]{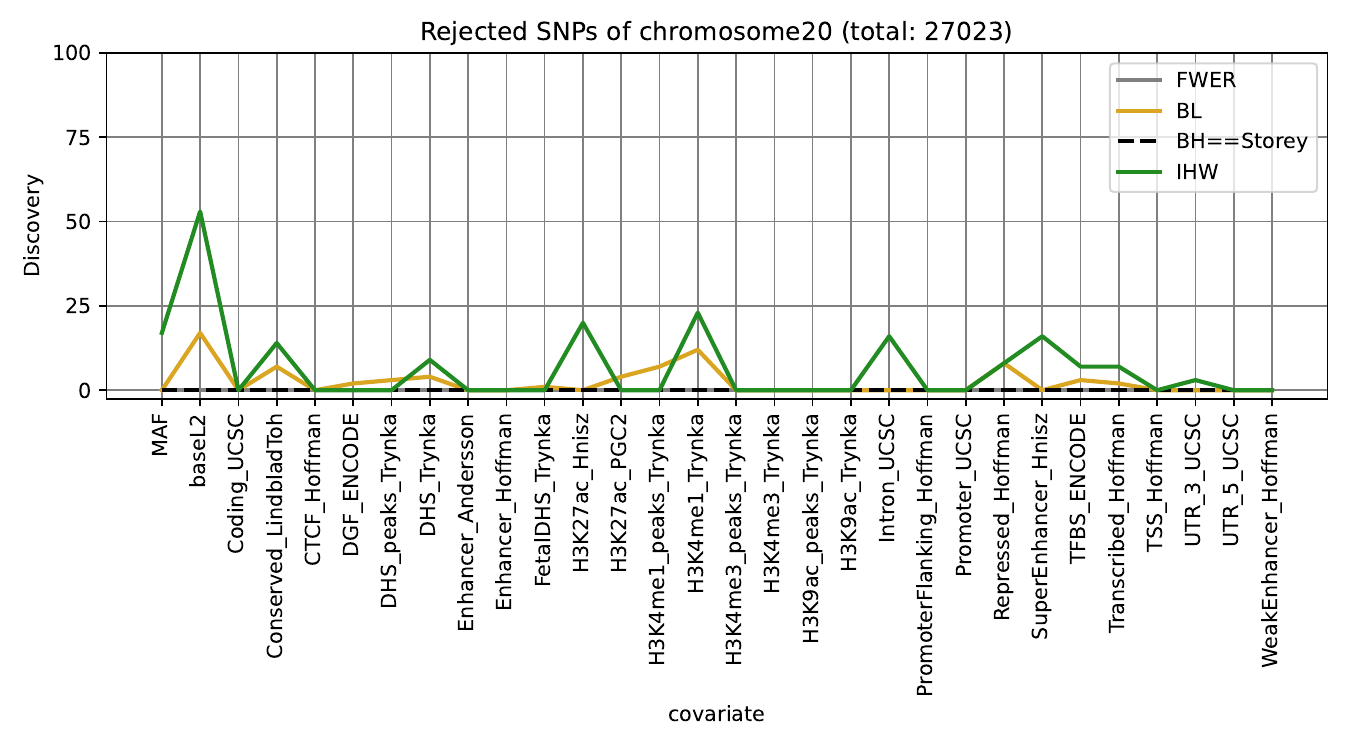} \\

    \includegraphics[width=.49\textwidth]{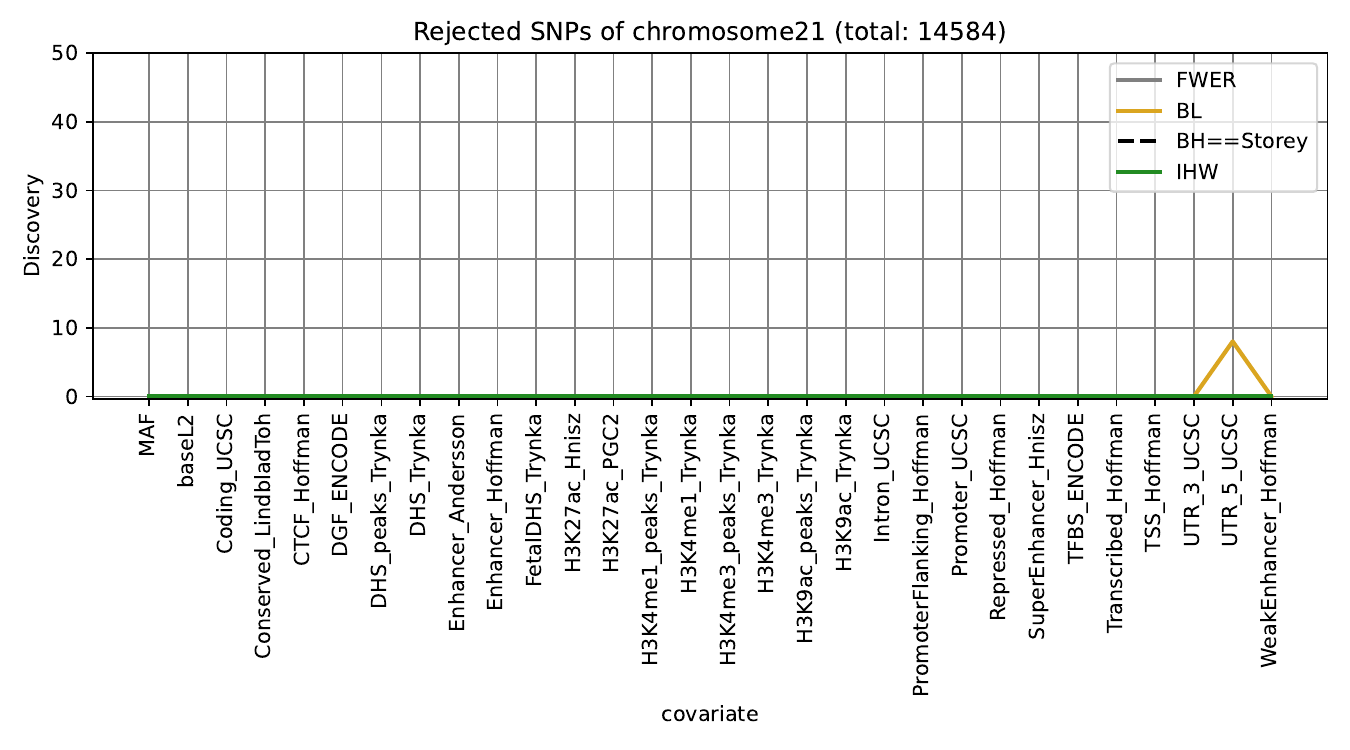} &
    \includegraphics[width=.49\textwidth]{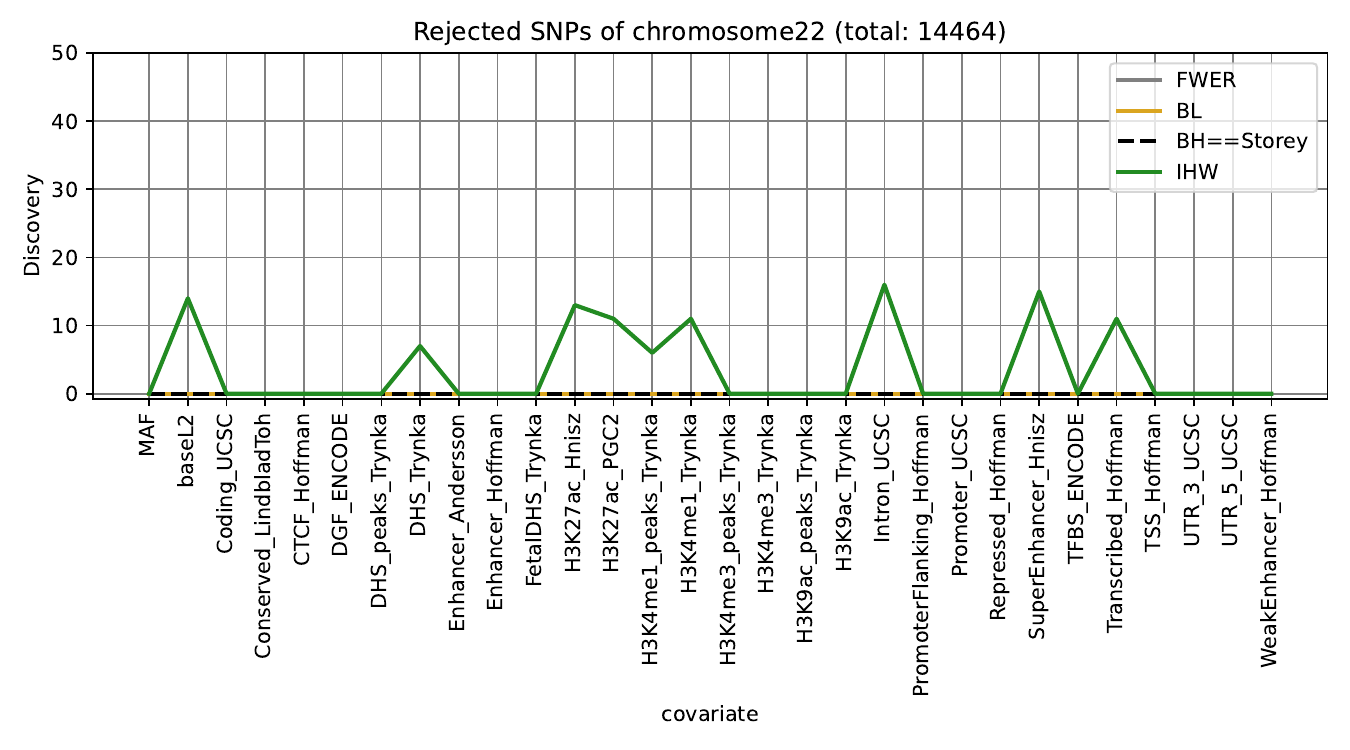} \\
\end{longtable}
\captionof{figure}{The discovery of each method on original covariate}
\label{all_result_SNP}
}

{
\centering
\begin{longtable}{c c}
    \includegraphics[width=.49\textwidth]{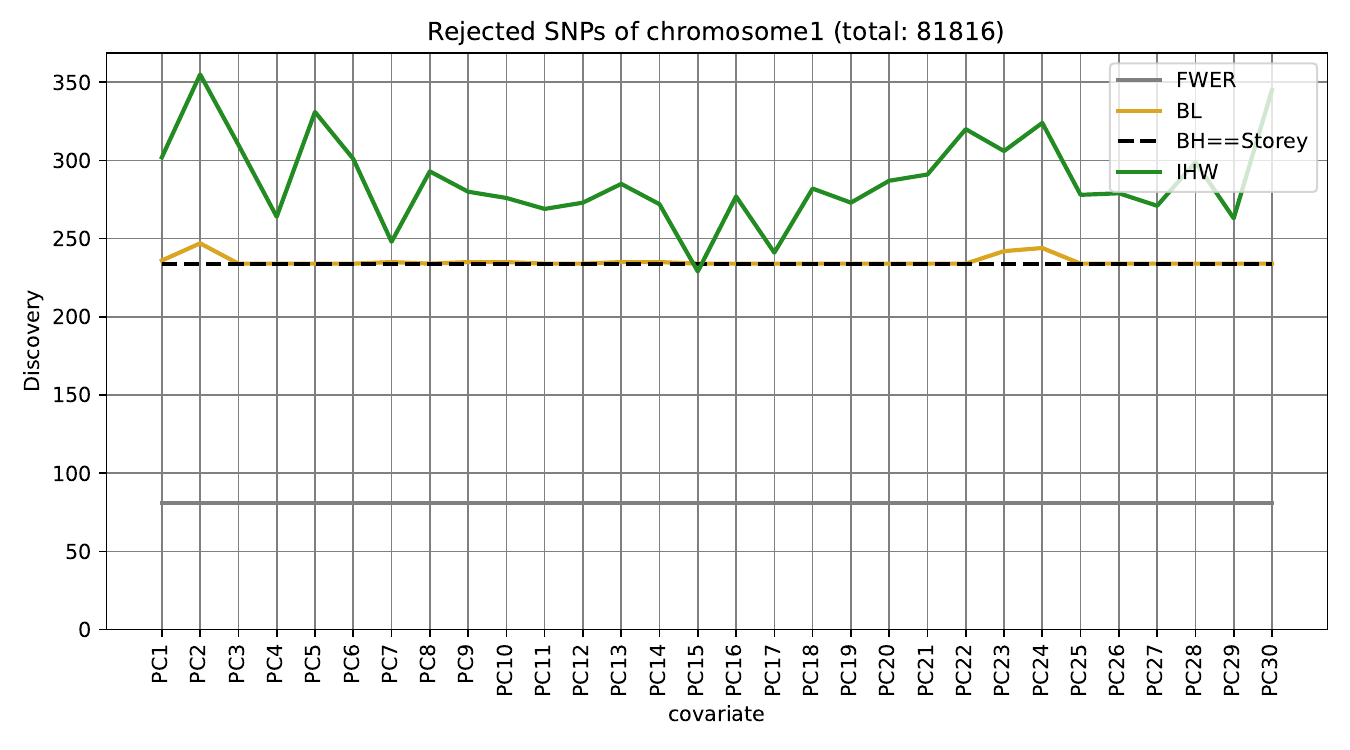} &
    \includegraphics[width=.49\textwidth]{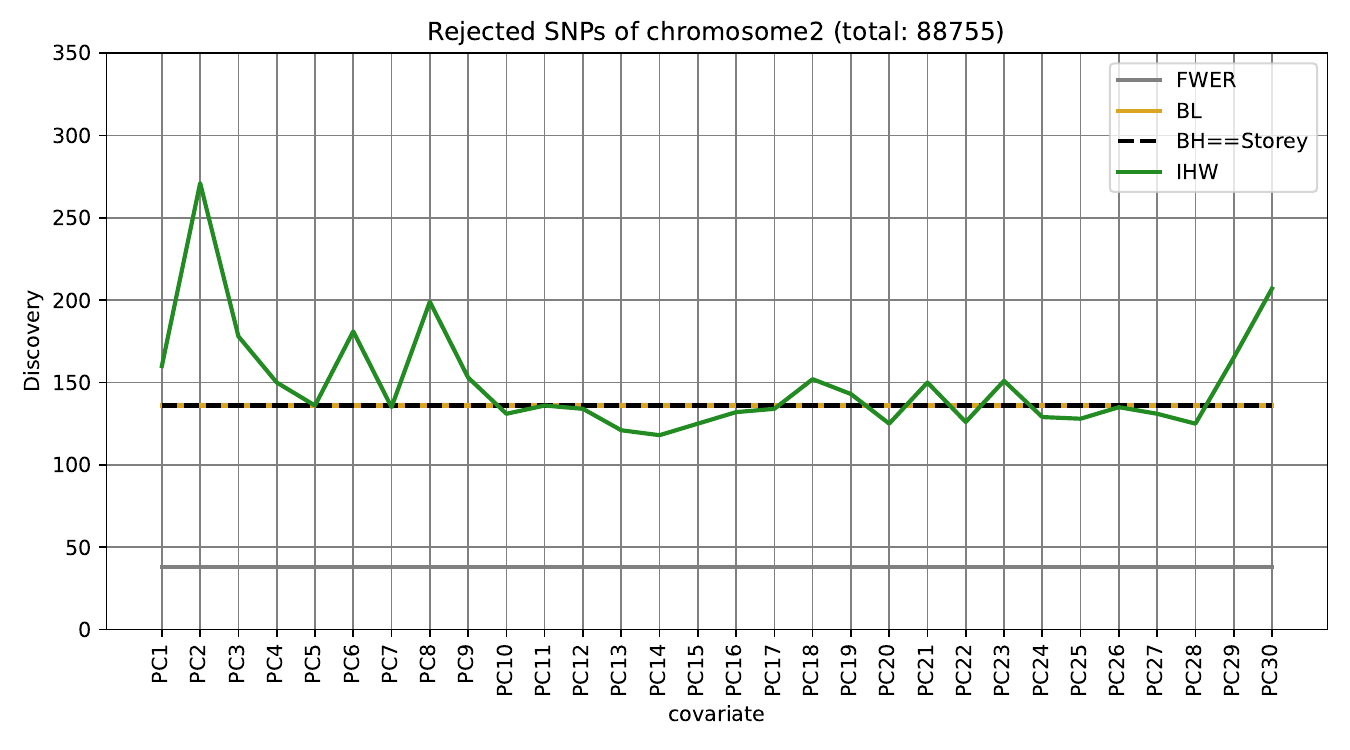} \\

    \includegraphics[width=.49\textwidth]{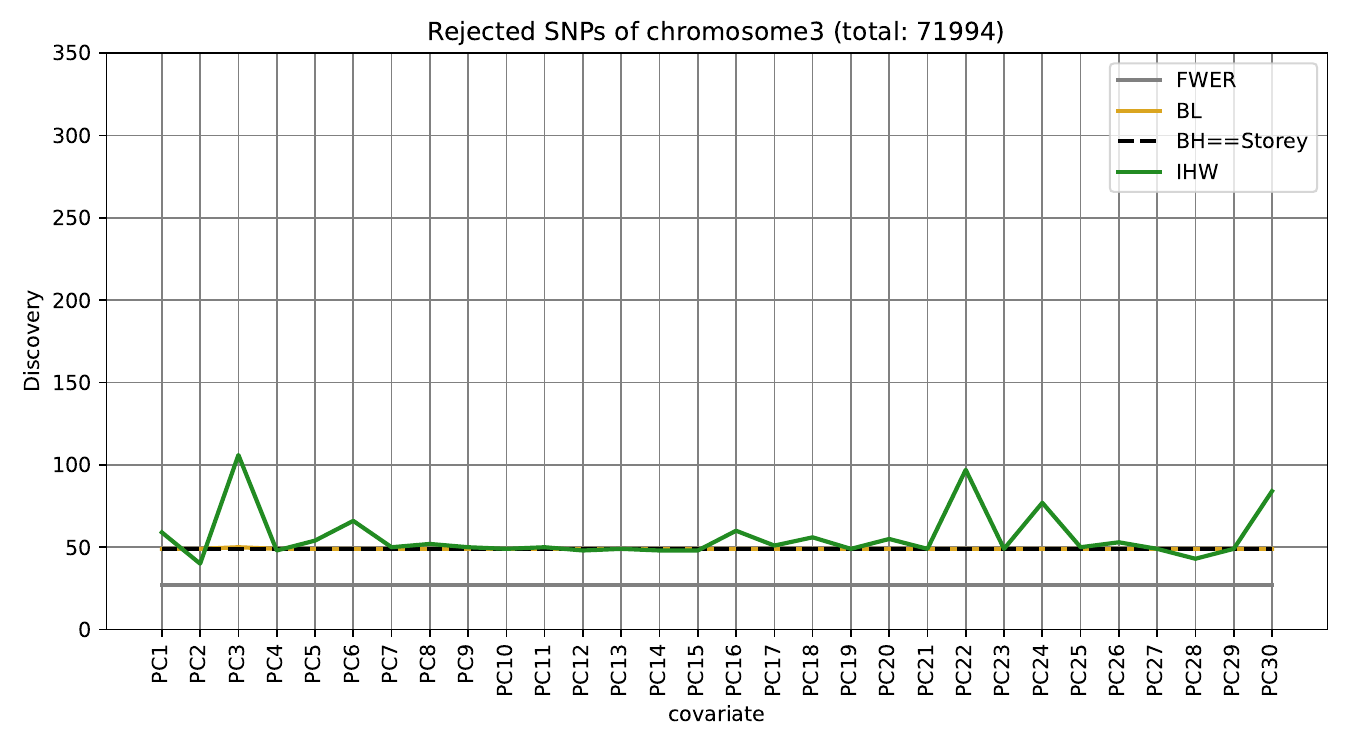} &
    \includegraphics[width=.49\textwidth]{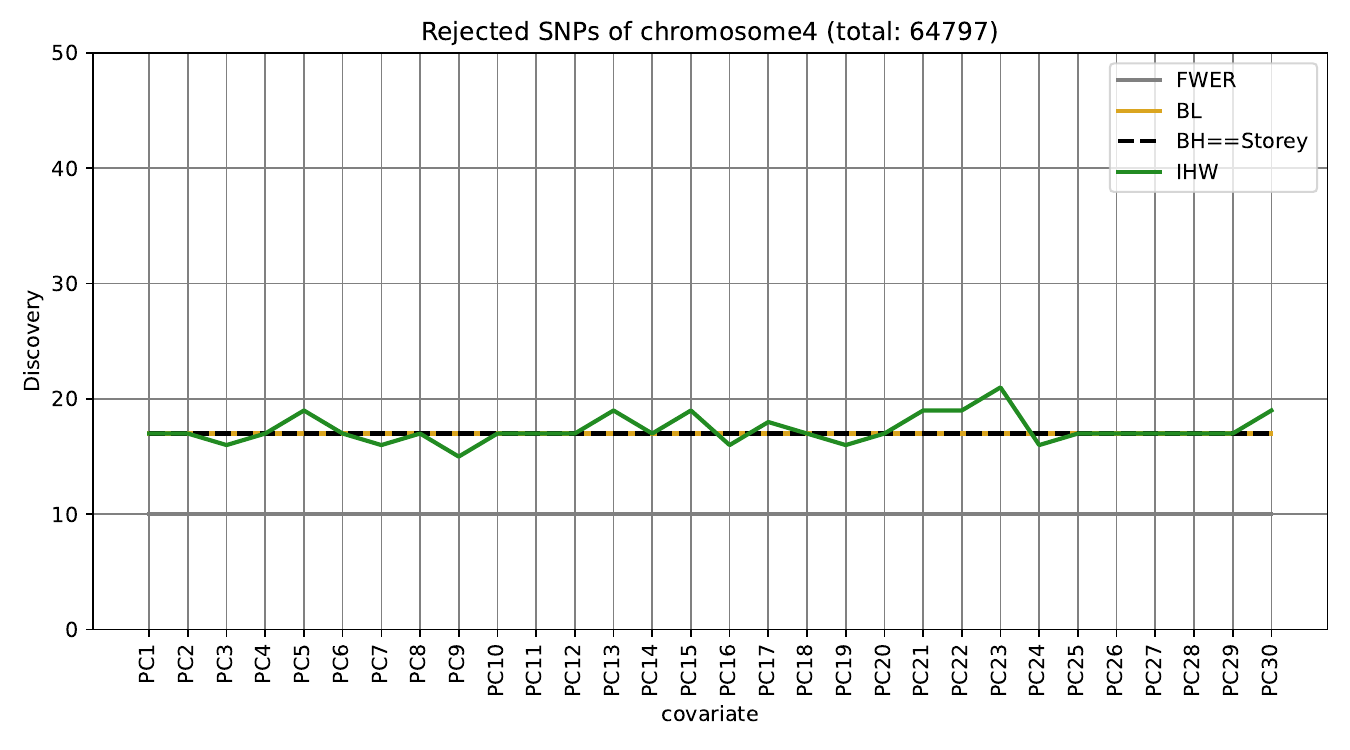} \\

    \includegraphics[width=.49\textwidth]{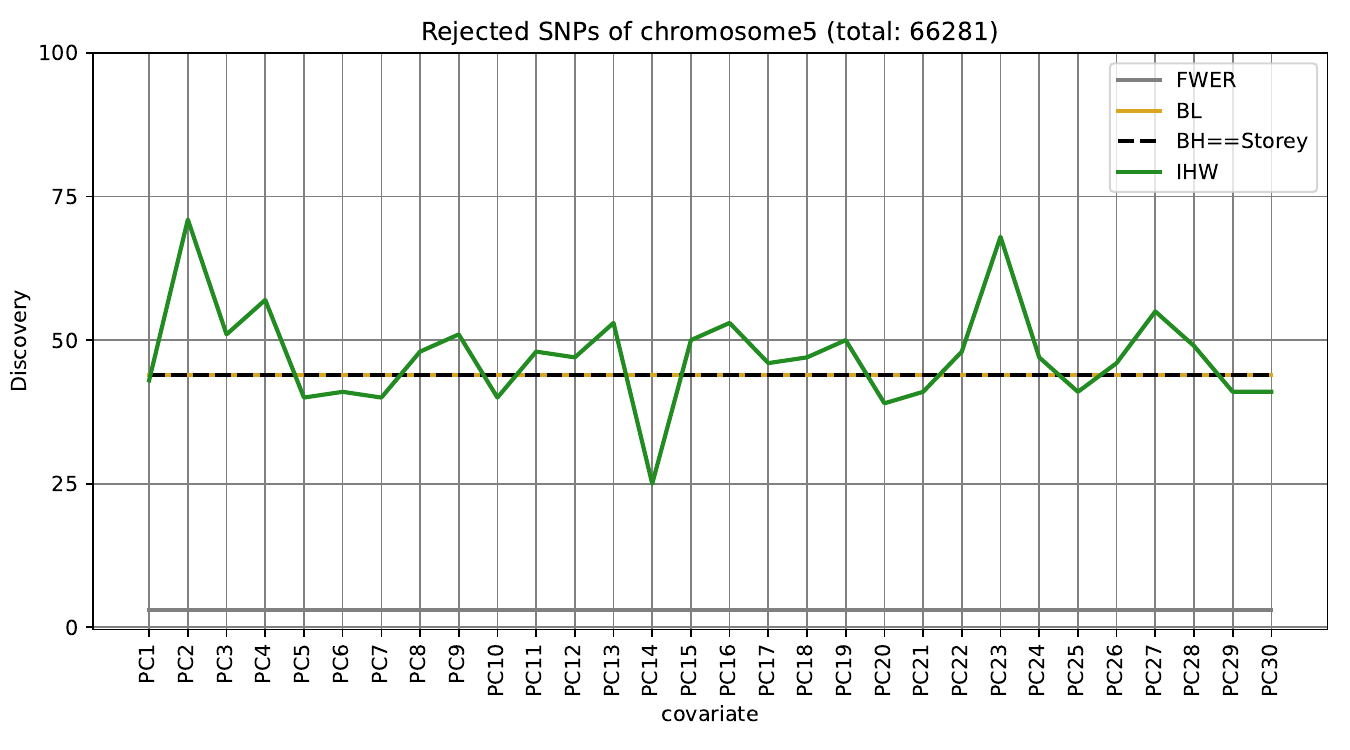} &
    \includegraphics[width=.49\textwidth]{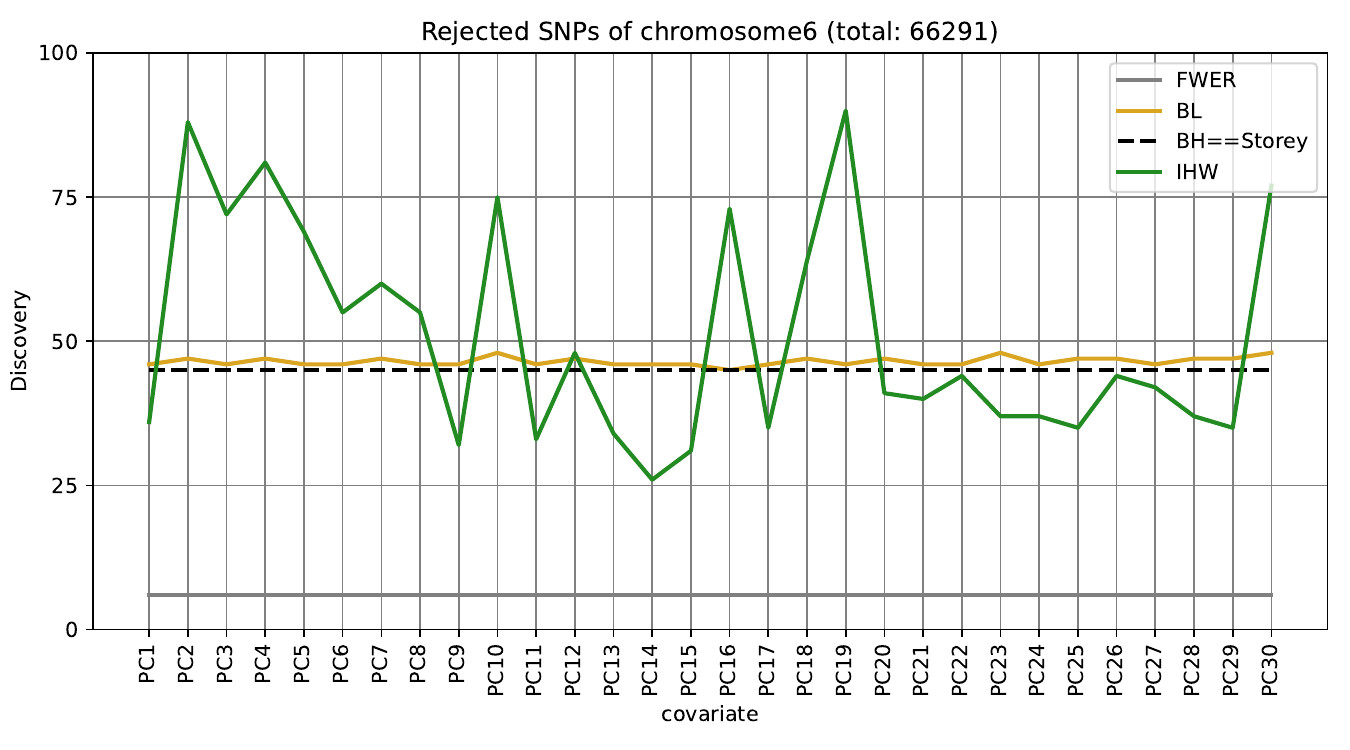} \\

    \includegraphics[width=.49\textwidth]{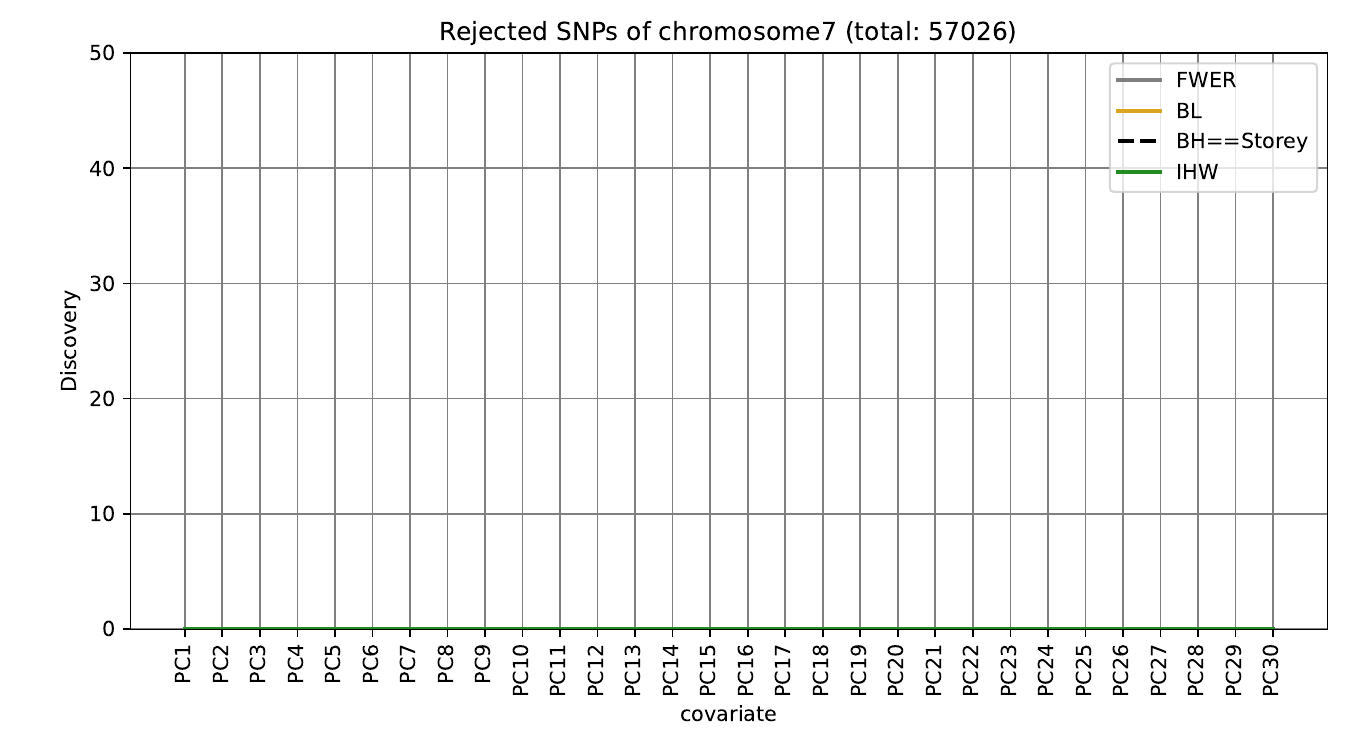} &
    \includegraphics[width=.49\textwidth]{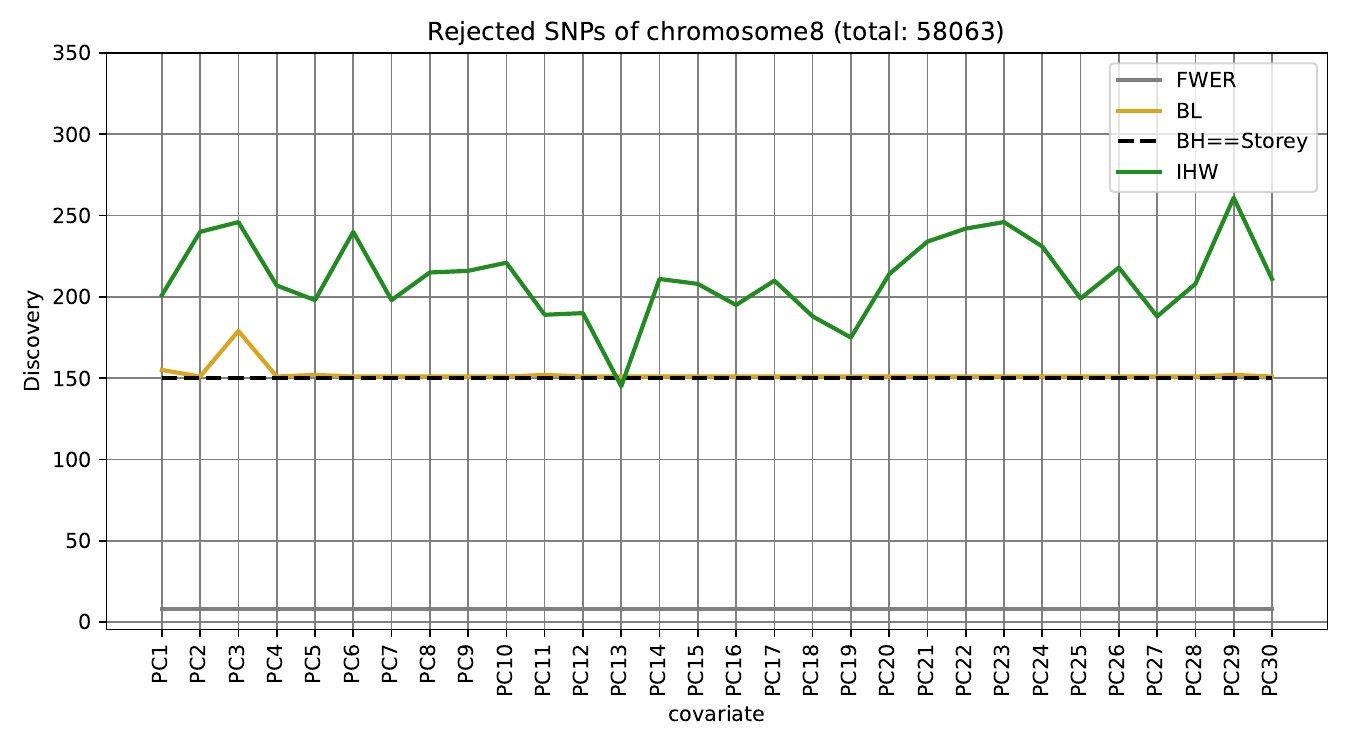} \\

    \includegraphics[width=.49\textwidth]{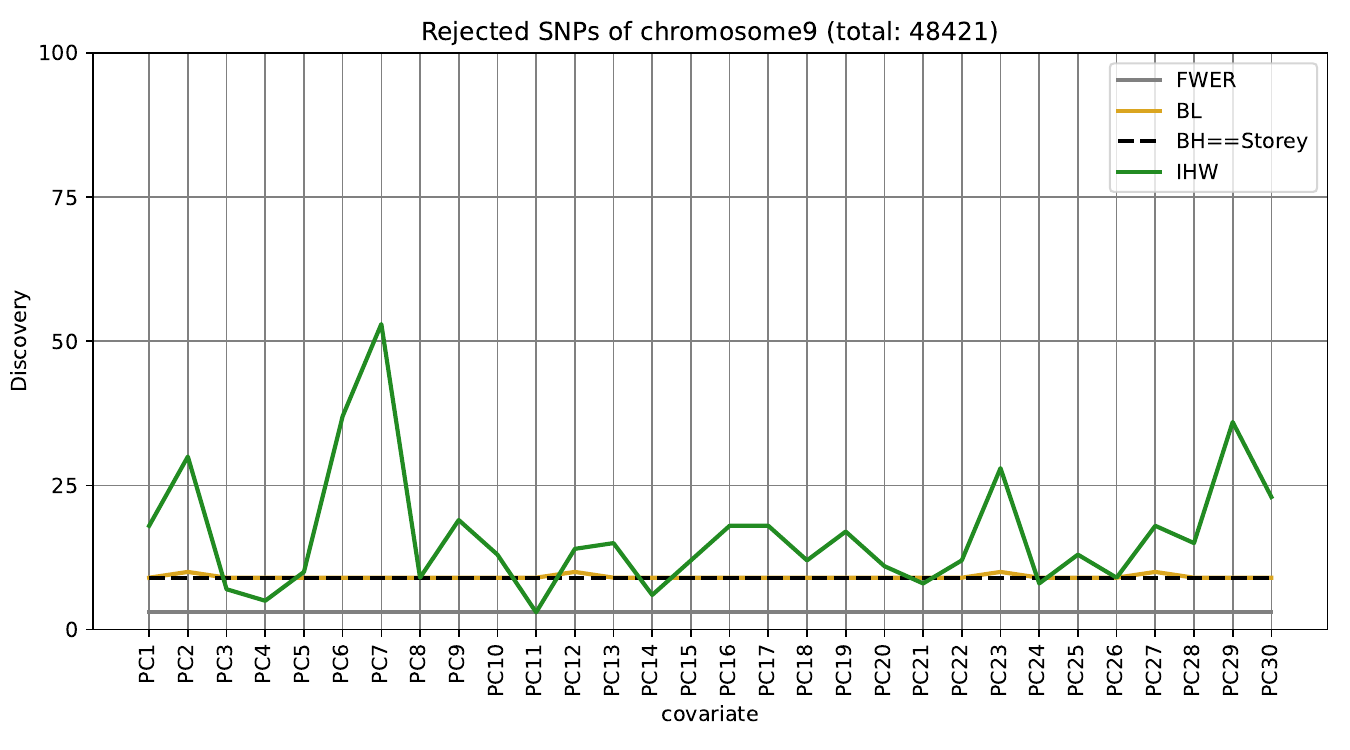} &
    \includegraphics[width=.49\textwidth]{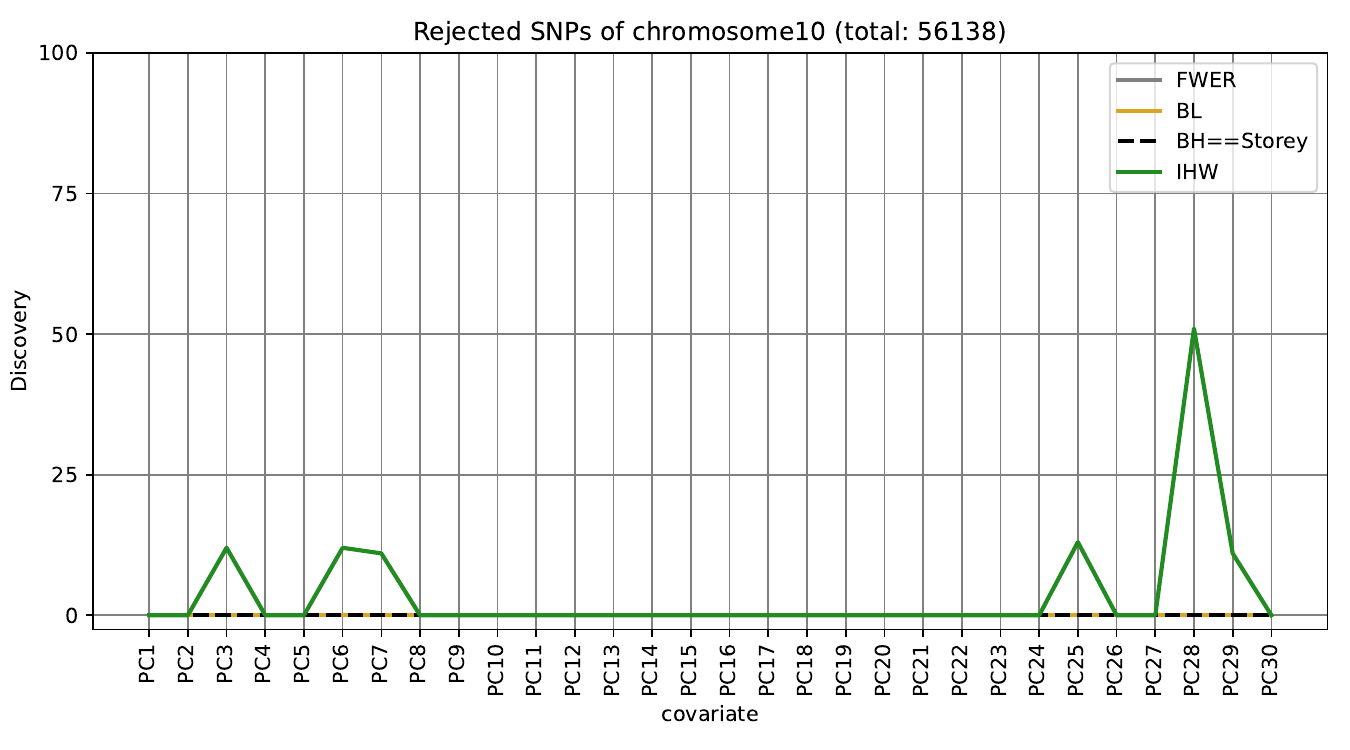} \\

    \includegraphics[width=.49\textwidth]{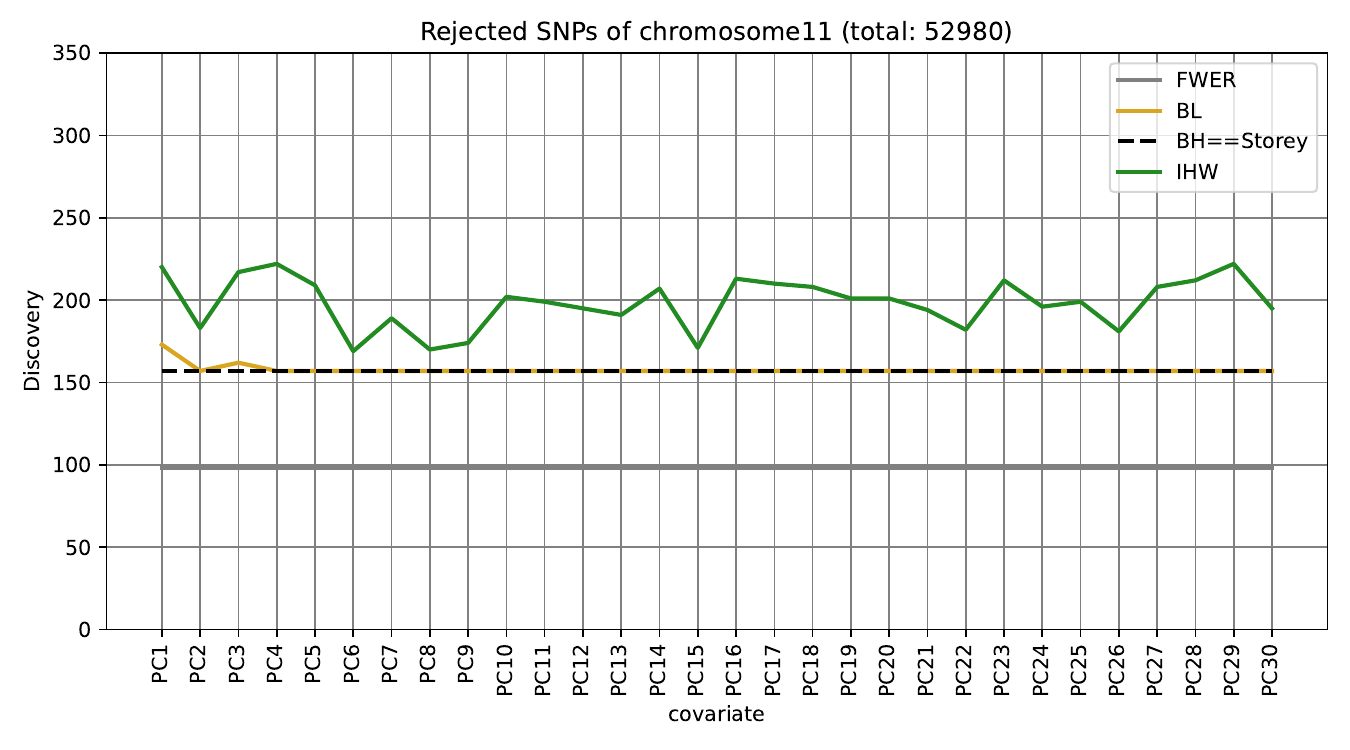} &
    \includegraphics[width=.49\textwidth]{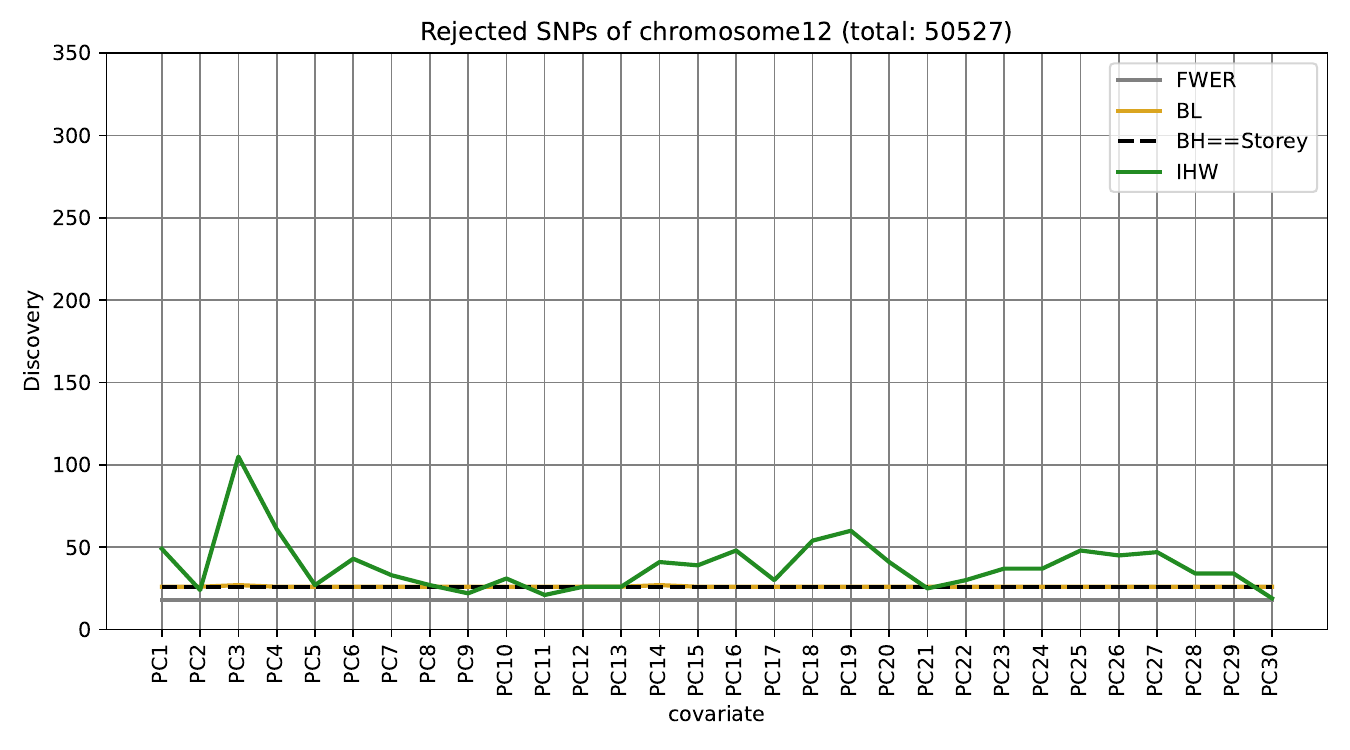} \\

    \includegraphics[width=.49\textwidth]{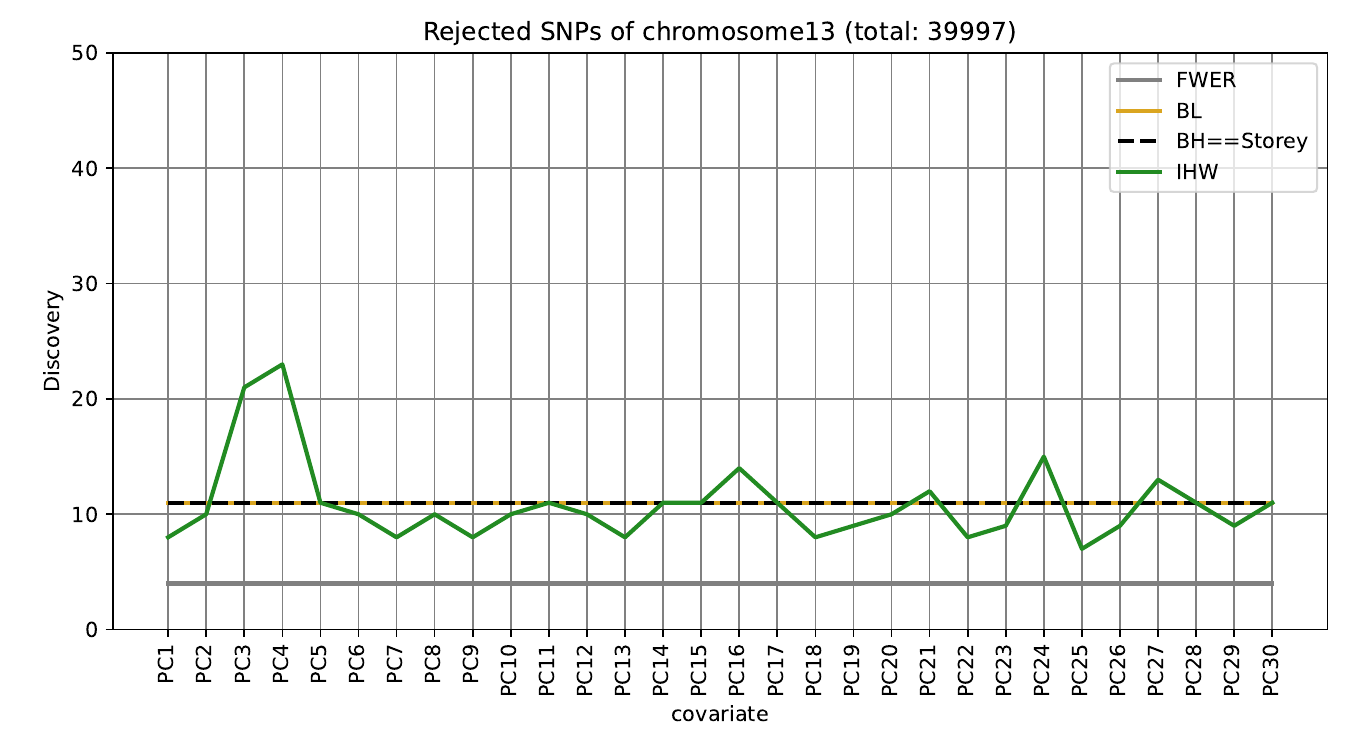} &
    \includegraphics[width=.49\textwidth]{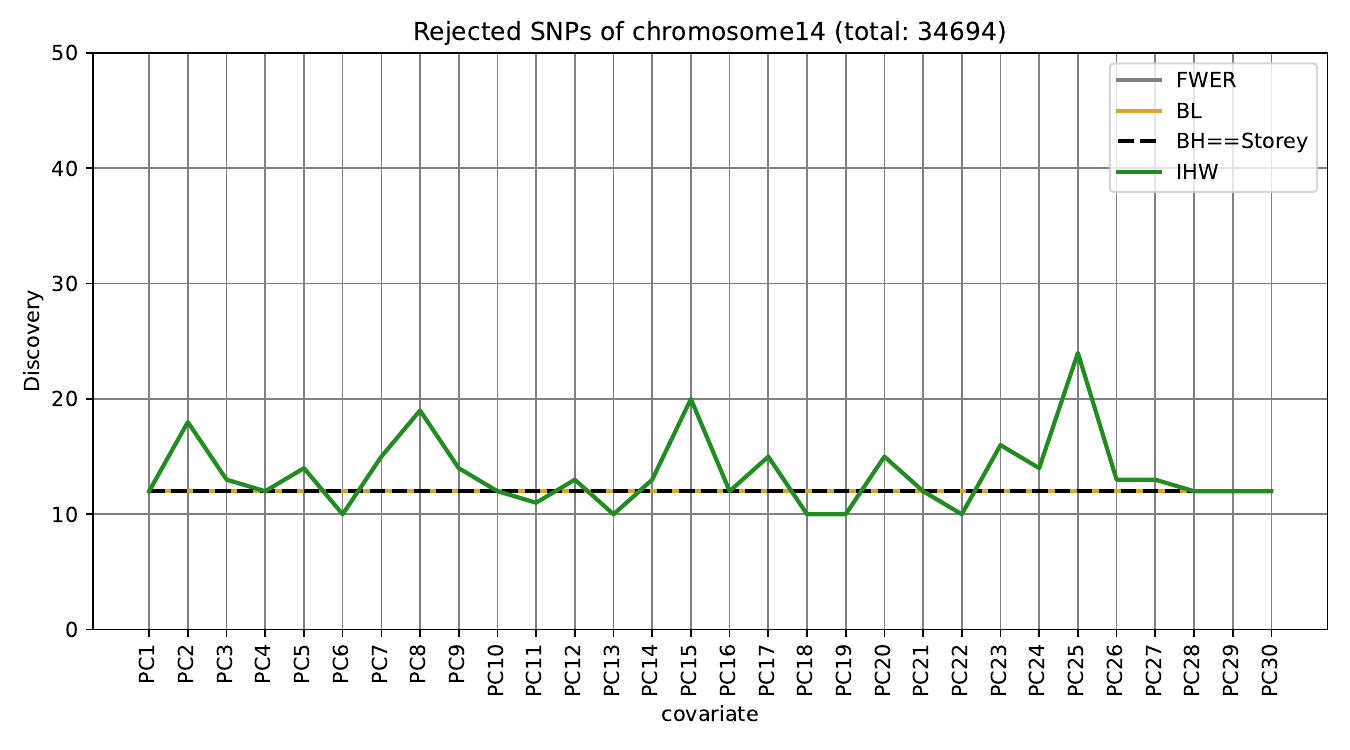} \\

    \includegraphics[width=.49\textwidth]{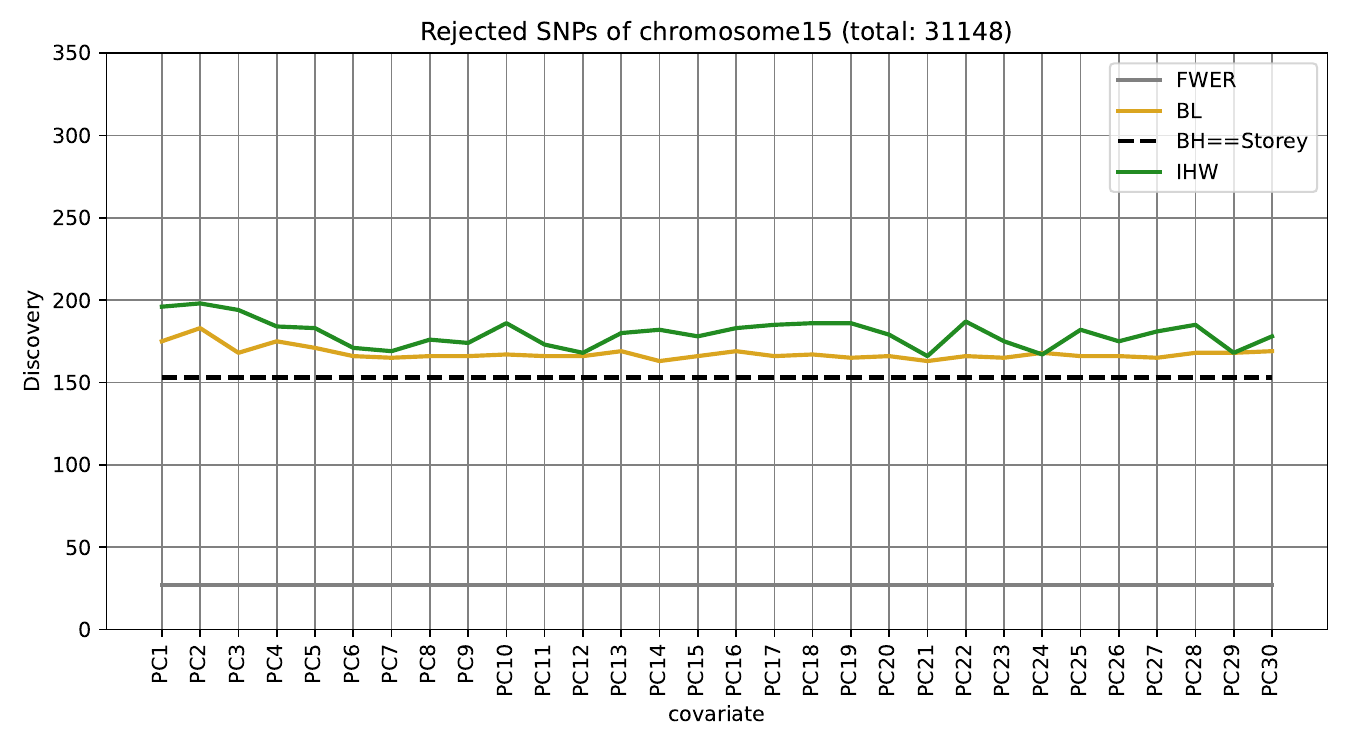} &
    \includegraphics[width=.49\textwidth]{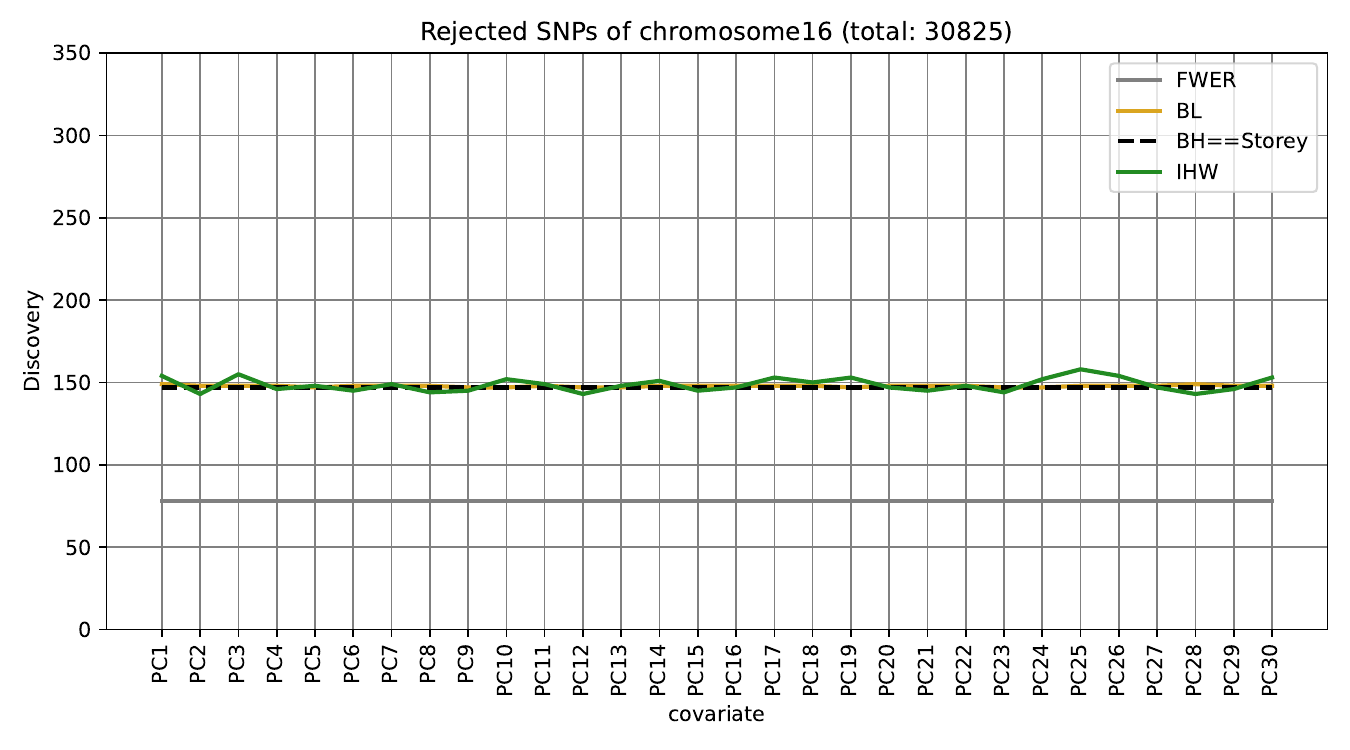} \\

    \includegraphics[width=.49\textwidth]{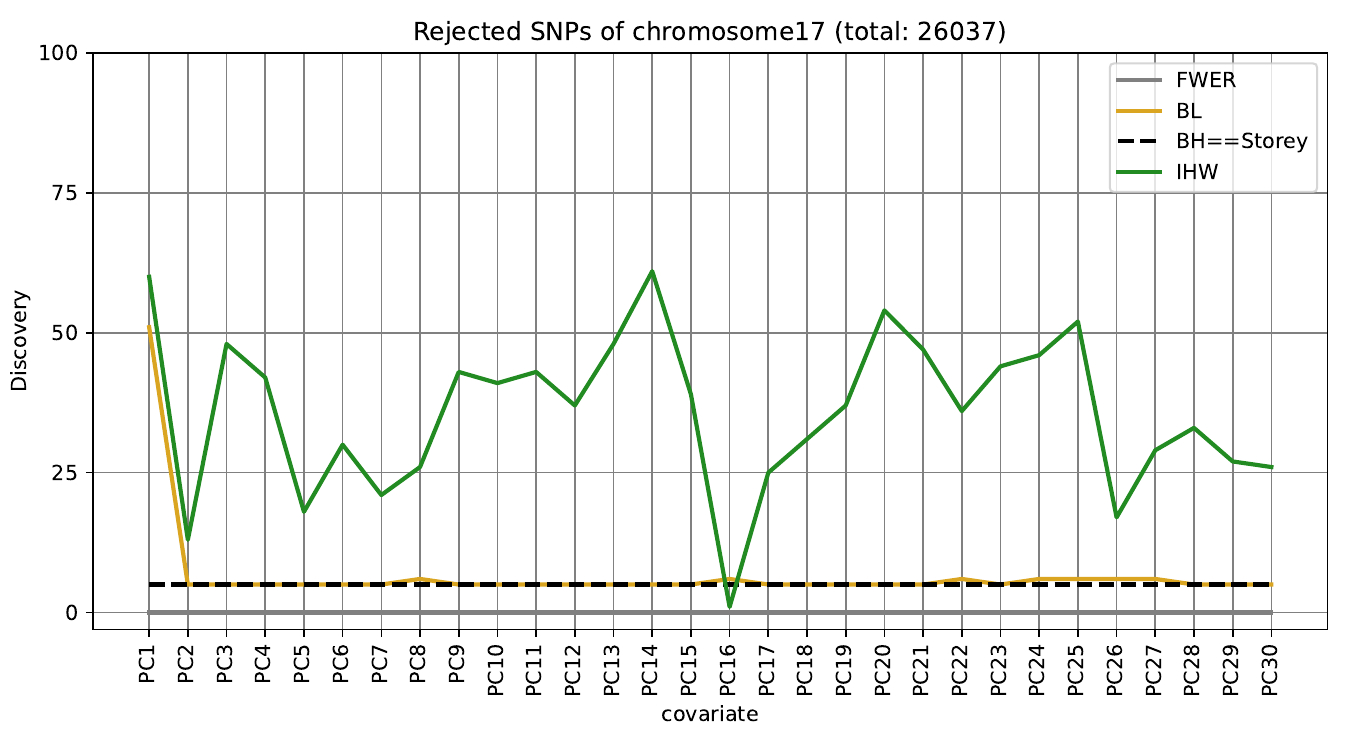} &
    \includegraphics[width=.49\textwidth]{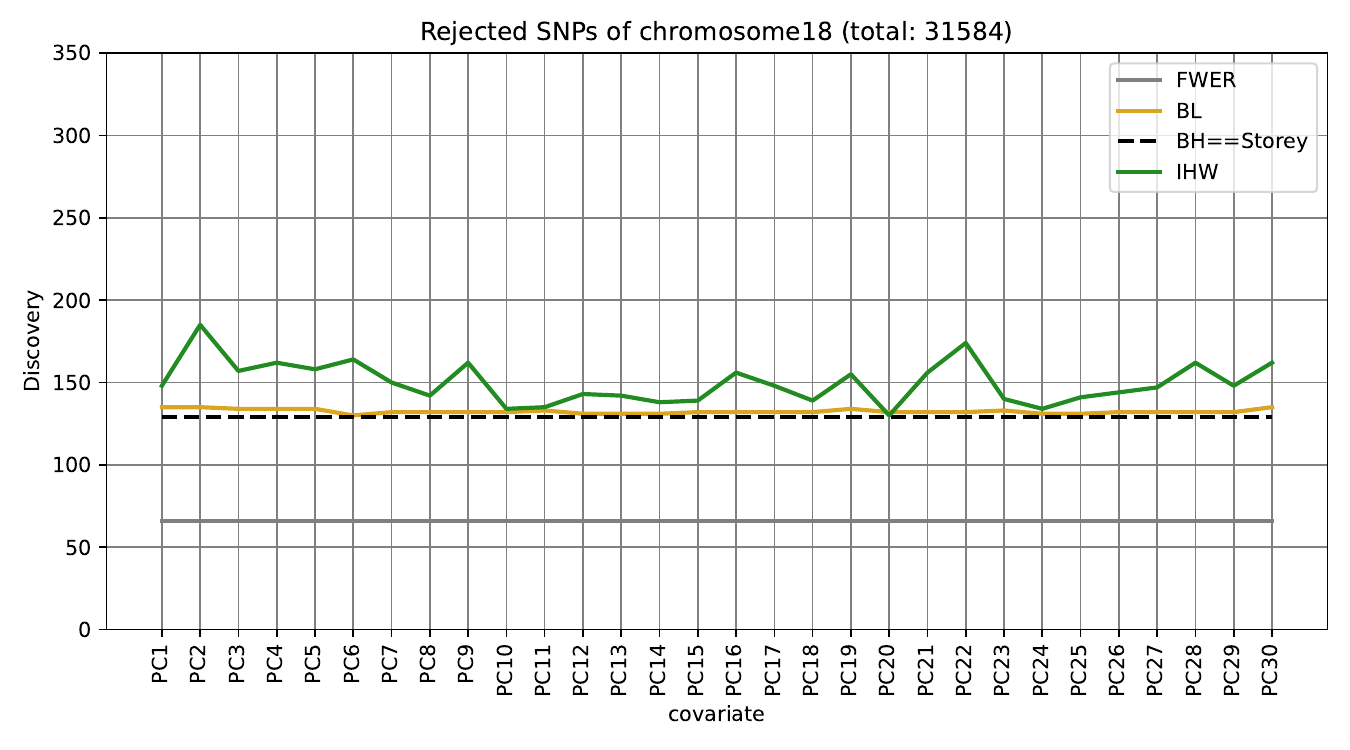} \\

    \includegraphics[width=.49\textwidth]{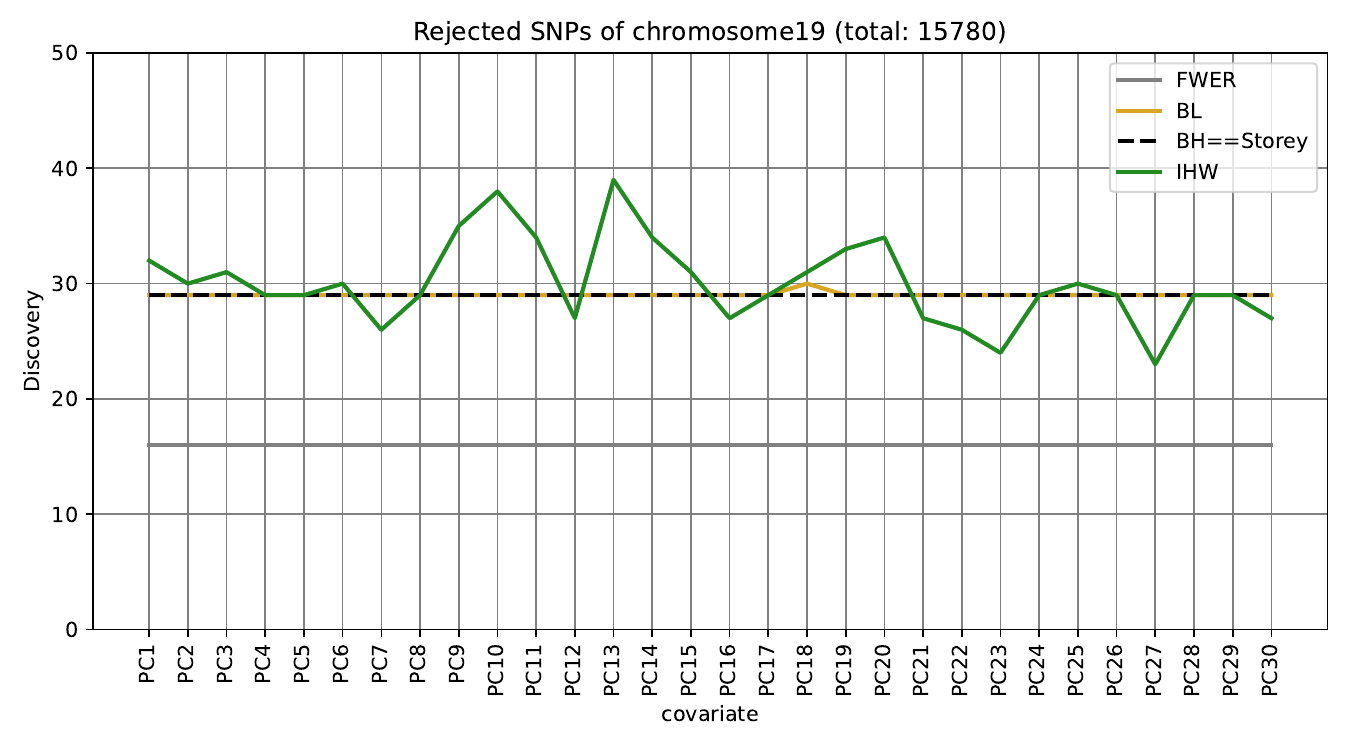} &
    \includegraphics[width=.49\textwidth]{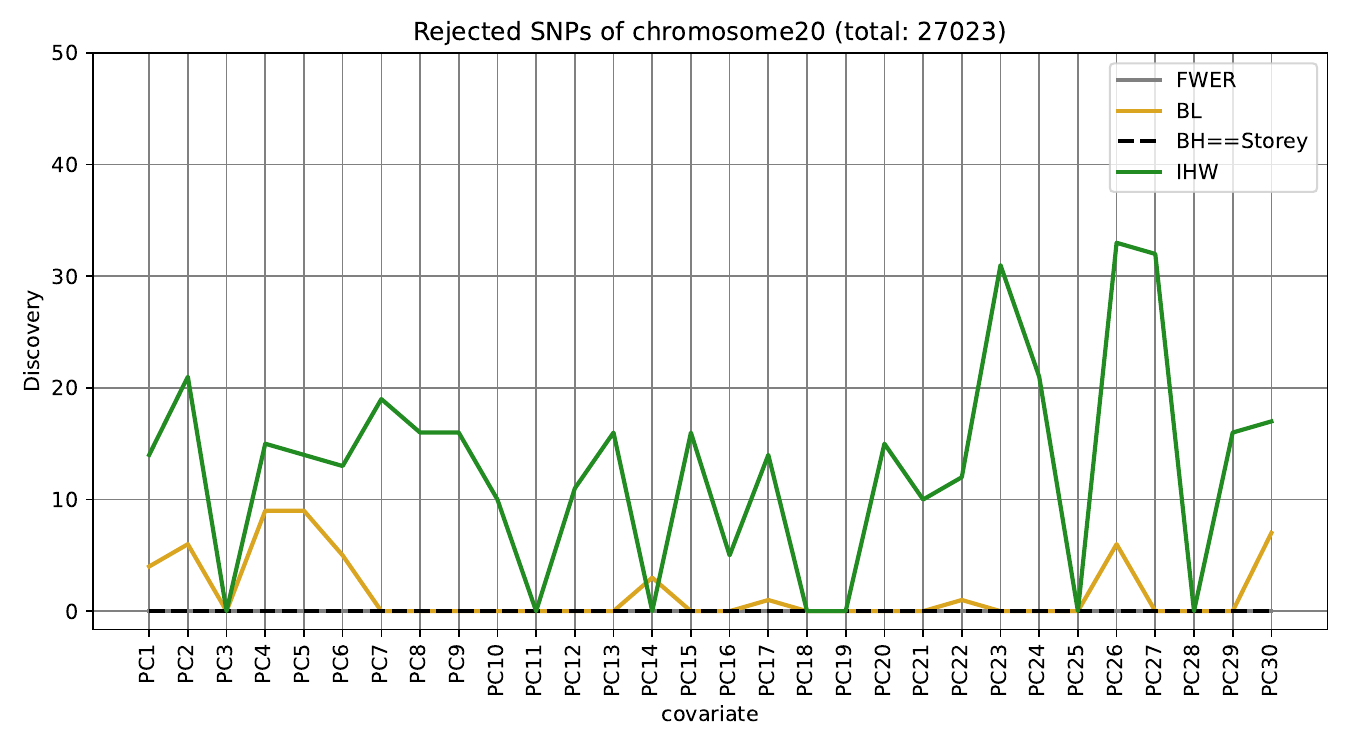} \\

    \includegraphics[width=.49\textwidth]{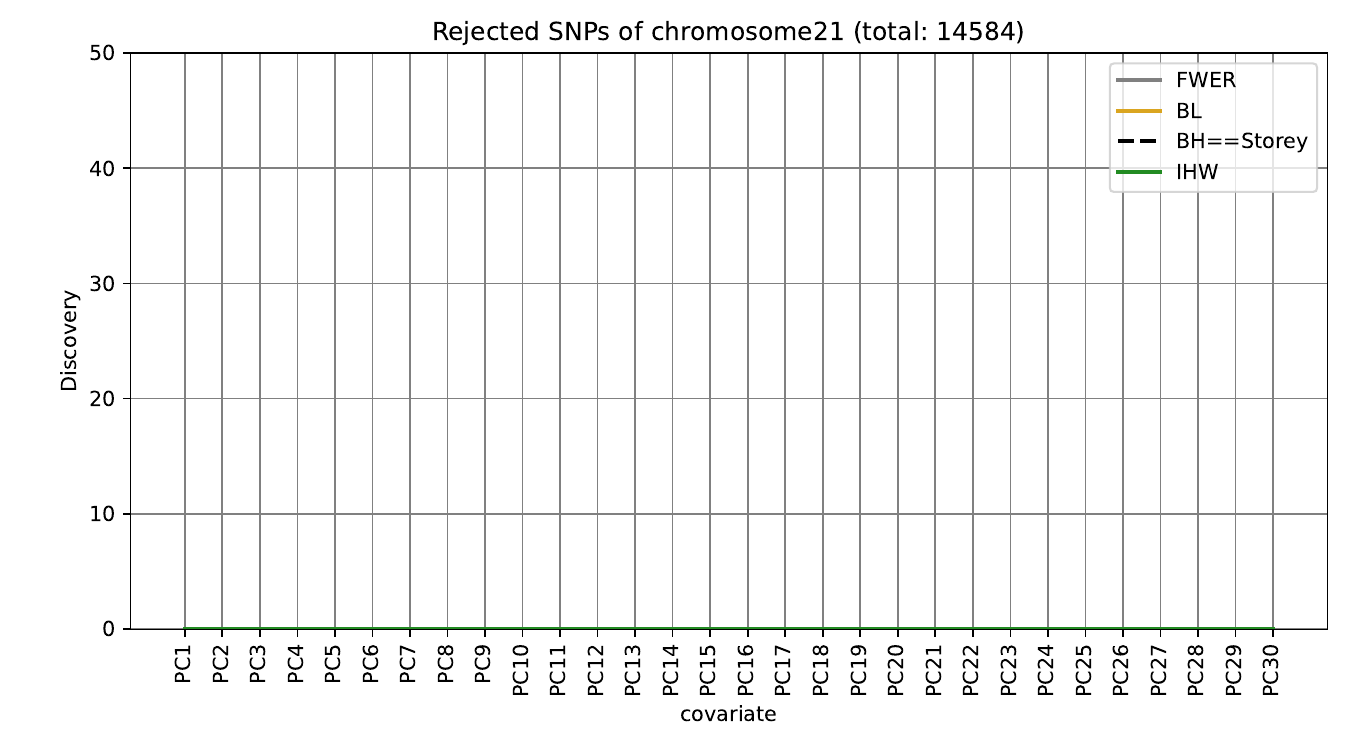} &
    \includegraphics[width=.49\textwidth]{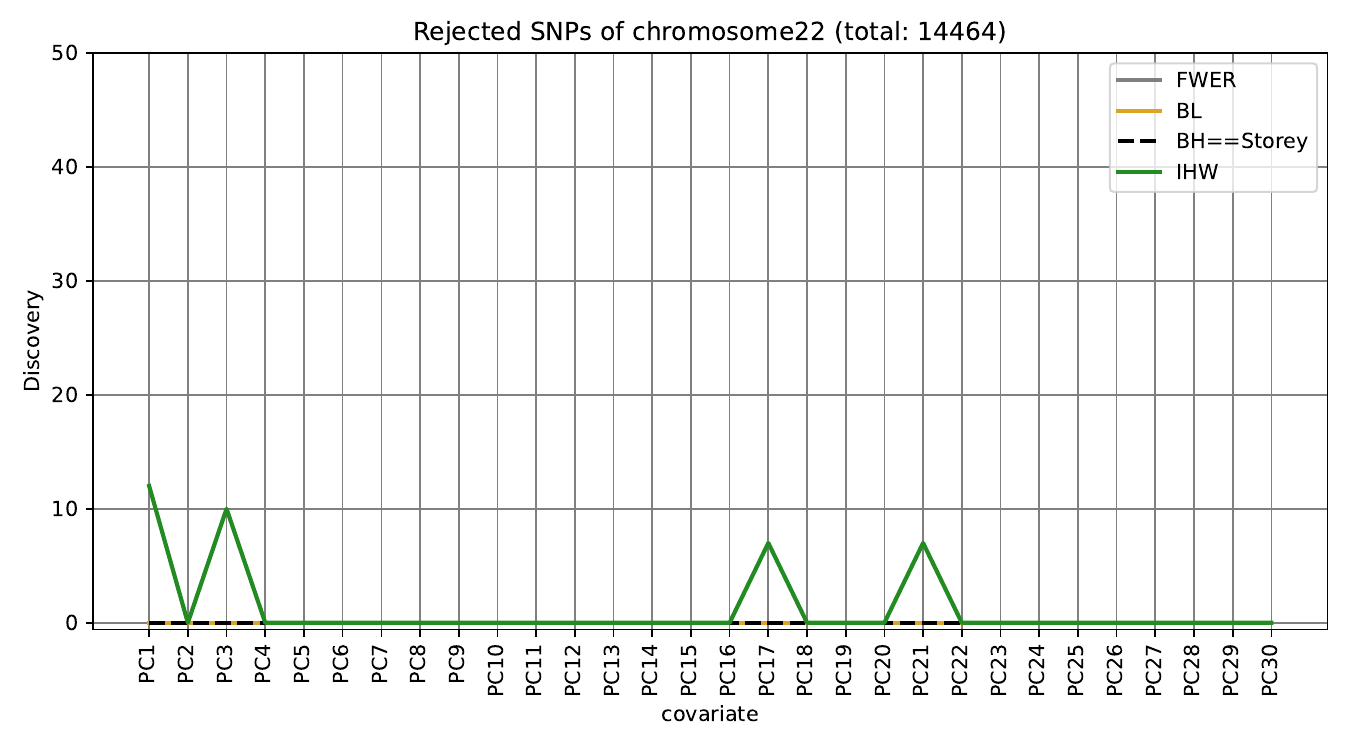} \\
\end{longtable}
\captionof{figure}{The discovery of each method on PC-axis}
\label{all_result_SNP_PCA}
}

\begin{figure}[ht]
    \centering
    \includegraphics[width=0.8\columnwidth]{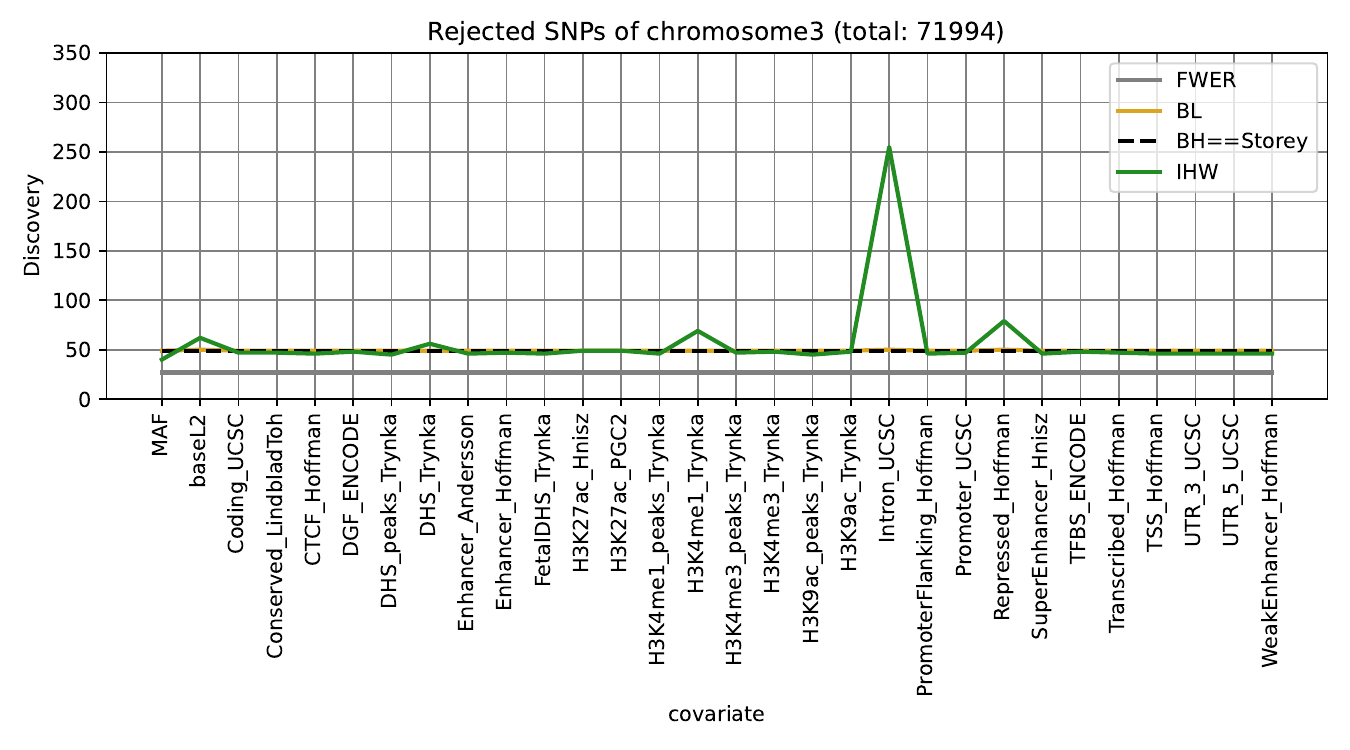}
    \caption{The discovery of each method on original covariate}
    \label{fig::SNP_result}
\end{figure}

\bigskip

\newpage 

\setcounter{table}{0}
\section{Table}
\begin{table}[ht]
    \centering
    \begin{tabular}{|l||l|l|}
    \hline
    \textbf{Covariate}               & \textbf{PC3}& \textbf{PC22}\\ \hline \hline
    Coding\_UCSC.bedL2               & 0.1993 & 0.0612  \\ \hline
    TSS\_Hoffman.bedL2               & 0.1939 & 0.1974  \\ \hline
    Promoter\_UCSC.bedL2             & 0.1796 & 0.0013  \\ \hline
    UTR\_3\_UCSC.bedL2               & 0.1671 &-0.1870  \\ \hline
    H3K4me3\_peaks\_Trynka.bedL2     & 0.1400 & 0.1824  \\ \hline
    H3K9ac\_peaks\_Trynka.bedL2      & 0.1349 & 0.1105  \\ \hline
    H3K4me3\_Trynka.bedL2            & 0.1270 &-0.2844  \\ \hline
    H3K9ac\_Trynka.bedL2             & 0.1259 &-0.1756  \\ \hline
    UTR\_5\_UCSC.bedL2               & 0.1143 &-0.0557  \\ \hline
    Enhancer\_Hoffman.bedL2          & 0.0815 & 0.1347  \\ \hline
    PromoterFlanking\_Hoffman.bedL2  & 0.0687 & 0.0569  \\ \hline
    SuperEnhancer\_Hnisz.bedL2       & 0.0628 & 0.0324  \\ \hline
    TFBS\_ENCODE.bedL2               & 0.0496 &-0.2629  \\ \hline
    Enhancer\_Andersson.bedL2        & 0.0322 & 0.0087  \\ \hline
    WeakEnhancer\_Hoffman.bedL2      & 0.0301 & 0.0625  \\ \hline
    H3K27ac\_PGC2.bedL2              & 0.0150 &-0.0382  \\ \hline
    CTCF\_Hoffman.bedL2              & 0.0139 & 0.1276  \\ \hline
    H3K4me1\_peaks\_Trynka.bedL2     & 0.0048 &-0.0088  \\ \hline
    Conserved\_LindbladToh.bedL2     & 0.0025 &-0.0184  \\ \hline
    H3K27ac\_Hnisz.bedL2             &-0.0358 & 0.1201  \\ \hline
    FetalDHS\_Trynka.bedL2           &-0.0489 & 0.0471  \\ \hline
    DGF\_ENCODE.bedL2                &-0.0529 & 0.0086  \\ \hline
    DHS\_peaks\_Trynka.bedL2         &-0.1046 & 0.0226  \\ \hline
    DHS\_Trynka.bedL2                &-0.1211 &-0.0014  \\ \hline
    H3K4me1\_Trynka.bedL2            &-0.1304 & 0.0116  \\ \hline
    MAF                              &-0.1921 &-0.0745  \\ \hline
    Transcribed\_Hoffman.bedL2       &-0.2279 & 0.4876  \\ \hline
    Repressed\_Hoffman.bedL2         &-0.3952 & 0.0061  \\ \hline
    Intron\_UCSC.bedL2               &-0.4150 &-0.5738  \\ \hline
    baseL2                           &-0.5414 & 0.2415  \\ \hline
    \end{tabular}
     \caption{PC loadings sorted by PC3}\label{table::pc_loadings}
\end{table}
\end{document}